\documentclass[acmsmall,screen]{acmart}
\setcopyright{rightsretained}
\copyrightyear{2025}
\acmYear{2026}
\acmDOI{XXXXXXX.XXXXXXX}

\usepackage{tikz}    
\usetikzlibrary{shapes.geometric, arrows.meta, positioning, fit}

\usepackage{xspace}
\usepackage{array}
\usepackage{url}
\usepackage{hyperref}
\usepackage{booktabs}
\usepackage{graphicx} 
\usepackage[table,xcdraw]{xcolor}
\usepackage{paralist}
\usepackage[size=tiny]{todonotes}
\usepackage{multirow}
\usepackage[stable]{footmisc}

\usepackage{pifont}          
\usepackage{amsmath}
\usepackage[scaled=.8]{beramono}
\usepackage{cleveref}
\usepackage{ifthen}
\usepackage[normalem]{ulem}
\usepackage{environ}

\newcommand{\etc}{etc.\xspace}
\newcommand{\etal}{et al.\xspace}
\newcommand{\ie}{i.e.\xspace}
\newcommand{\eg}{e.g.\xspace}

\newcommand{\secref}[1]{Section~\ref{#1}\xspace}
\newcommand{\figref}[1]{Figure~\ref{#1}\xspace}
\newcommand{\tabref}[1]{Table~\ref{#1}\xspace}

\newcommand{\versionflag}{diff} 
\renewcommand{\versionflag}{V2}

\newcommand{\revise}[2]{#2}

\usepackage{xparse}

\NewDocumentEnvironment{diffversion}{m+b}{%
  \ifthenelse{\equal{\versionflag}{V1}}{%
    #2%
  }{%
    \ifthenelse{\equal{\versionflag}{V2}}{%
      #1%
    }{%
      \begin{minipage}{\linewidth}
        {\color{red}#2\par}
        {\color{blue}#1\par}
      \end{minipage}%
    }%
  }%
}{}

\newcommand{\gitignore}[0]{\texttt{.gitignore}\xspace}
\newcommand{\dotfiles}[0]{\texttt{dotfiles}\xspace}
\newcommand{\agentsmd}[0]{\texttt{AGENTS.md}\xspace}
\newcommand{\claudemd}[0]{\texttt{CLAUDE.md}\xspace}

\newcommand{\codex}[0]{\texttt{Codex}\xspace}
\newcommand{\claude}[0]{\texttt{Claude}\xspace}
\newcommand{\copilot}[0]{\texttt{Copilot}\xspace}
\newcommand{\cursor}[0]{\texttt{Cursor}\xspace}

\newcommand{\initialProjects}[0]{130,621}

\newcommand{\studyenddate}[0]{February 21st, 2026\xspace}

\usepackage{xfp}
\usepackage{xstring} 

\newcommand{\ghtopic}[1]{\texttt{#1}}

\usepackage{tikz}
\usepackage{pgfplots}
\usepackage{pgfplotstable}
\usepackage{xparse}

\pgfplotsset{compat=1.18}

\pgfplotsset{
    sparkline/.style={
        width=4em,
        height=1.2ex,
        axis lines=none,
        ticks=none,
        xlabel={},
        ylabel={},
        every axis plot/.append style={line width=0.4pt},
        clip=false,
        baseline,
        scale only axis,
        xtick=\empty,
        ytick=\empty,
        axis line style={opacity=0},
        enlarge x limits=0.05,
        enlarge y limits=0.1,
    },
    sparkline bar/.style={
        sparkline,
        ybar,
        bar width=0.3em,
        bar shift=0pt,
        enlarge x limits=0.1,
    },
    sparkline line/.style={
        sparkline,
        every axis plot/.append style={mark=none, solid},
    },
       sparkline box/.style={
    width=3em,              
    height=0.8ex,           
    axis lines=none,
    ticks=none,
    xlabel={},
    ylabel={},
    baseline,
    scale only axis,
    xtick=\empty,
    ytick=\empty,
    enlarge x limits=0.05,  
    ymin=-0.5, ymax=0.5,    
}
}

\NewDocumentCommand{\sparklineBarsSimple}{m}{%
    \begin{tikzpicture}[baseline]
        \begin{axis}[sparkline bar]
            \foreach \value [count=\i from 0] in {#1} {
                \pgfmathparse{{\mycolors}[\i]}
                \let\currentcolor\pgfmathresult
                \addplot+[fill=\currentcolor, draw=\currentcolor] coordinates {(\i,\value)};
            }
        \end{axis}
    \end{tikzpicture}%
}

\def\mycolors{{"blue", "red", "green", "orange", "purple", "cyan", "magenta"}}

\NewDocumentCommand{\sparklineBarsCSVbug}{m}{%
    \begin{tikzpicture}[baseline]
        \begin{axis}[sparkline bar]
            \pgfplotstableread[col sep=comma]{#1}\datatable
            \pgfplotstablegetrowsof{\datatable}
            \pgfmathtruncatemacro{\numrows}{\pgfplotsretval-1}
            \foreach \i in {0,...,\numrows} {
                \pgfplotstablegetelem{\i}{value}\of{\datatable}
                \let\myvalue\pgfplotsretval
                \pgfplotstablegetelem{\i}{color}\of{\datatable}
                \let\mycolor\pgfplotsretval
                \addplot+[fill=\mycolor, draw=\mycolor] coordinates {(\i,\myvalue)};
            }
        \end{axis}
    \end{tikzpicture}%
}

\NewDocumentCommand{\sparklineBarsCSV}{m}{%
    \begin{tikzpicture}[baseline]
        \begin{axis}[sparkline bar]
            \pgfplotstableread[col sep=comma]{#1}\datatable
            \pgfplotstablegetrowsof{\datatable}
            \pgfmathtruncatemacro{\numrows}{\pgfplotsretval-1}
            \foreach \i in {0,...,\numrows} {
                \pgfplotstablegetelem{\i}{value}\of{\datatable}
                \let\myvalue\pgfplotsretval
                \pgfplotstablegetelem{\i}{color}\of{\datatable}
                \edef\temp{\noexpand\addplot+[fill=\pgfplotsretval, draw=\pgfplotsretval] coordinates {(\i,\myvalue)};}
                \temp
            }
        \end{axis}
    \end{tikzpicture}%
}

\NewDocumentCommand{\sparklineTikzBars}{O{} m}{%
    \begin{tikzpicture}[baseline, x=0.4em, y=0.6ex]
        \foreach \value [count=\i from 0] in {#2} {
            \pgfmathparse{\value/10}
            \let\scaledvalue\pgfmathresult
            \fill[blue] (\i-0.25, 0) rectangle (\i+0.25, \scaledvalue);
        }
    \end{tikzpicture}%
}

\newcommand\TotalAnalyzedProjects{128,575\xspace}
\newcommand\TotalProjects{128,018\xspace}

\newcommand\ToolUseCount{12,861\xspace}
\newcommand\ToolUsePercent{10.05\%\xspace}
\newcommand\CommitUseInToolDenominator{12,861\xspace}
\newcommand\CommitUseInToolCount{8,301\xspace}
\newcommand\CommitUseInToolPercent{64.54\%\xspace}

\newcommand\IgnoredToolsOnlyCount{2,606\xspace}
\newcommand\IgnoredToolsOnlyPercent{2.04\%\xspace}

\newcommand\IgnoredToolsAllCount{5,020\xspace}
\newcommand\IgnoredToolsAllPercent{3.92\%\xspace}

\newcommand\AllFileBasedUseCount{15,467\xspace}
\newcommand\AllFileBasedUsePercent{12.08\%\xspace}
\newcommand\CommitUseAllFileUsersDenominator{15,467\xspace}
\newcommand\CommitUseAllFileUsersCount{9,440\xspace}
\newcommand\CommitUseAllFileUsersPercent{61.03\%\xspace}
\newcommand\SampledCommitUseDenominator{112,551\xspace}
\newcommand\SampledCommitUseCount{12,951\xspace}
\newcommand\SampledCommitUsePercent{11.51\%\xspace}
\newcommand\ExtrapolationGroupCount{112,551\xspace}
\newcommand\ExtrapolatedCommitUsers{12,951\xspace}

\newcommand\AllTotalUsersEstimate{28,418\xspace}
\newcommand\OverallAdoptionRatePercent{22.20\%\xspace}
\newcommand\HighEstimateCommitUsers{21,220\xspace}
\newcommand\HighEstimateTotalUsersEstimate{36,687\xspace}
\newcommand\HighEstimateOverallAdoptionRatePercent{28.66\%\xspace}

\author{Romain Robbes}
\email{romain.robbes@labri.fr}
\orcid{0000-0003-4569-6868}
\affiliation{%
  \department{INP, LaBRI, UMR 5800}
  \institution{Univ. Bordeaux, CNRS}
  \city{Bordeaux}
  \country{France}
}

\author{Théo Matricon}
\email{theo.matricon@inria.fr}
\orcid{0000-0002-5043-3221}
\affiliation{%
\institution{Univ. Rennes, Inria, CNRS, IRISA}
\city{Rennes}
\country{France}
}

\author{Thomas Degueule}
\email{thomas.degueule@labri.fr}
\orcid{0000-0002-5961-7940}
\affiliation{%
  \department{INP, LaBRI, UMR 5800}
  \institution{Univ. Bordeaux, CNRS}
  \city{Bordeaux}
  \country{France}
}

\author{Andre Hora}
\email{andrehora@dcc.ufmg.br}
\orcid{0000-0003-4900-1330}
\affiliation{%
  \department{Department of Computer Science}
  \institution{UFMG}
  \city{Belo Horizonte}
  \country{Brazil}
}

\author{Stefano Zacchiroli}
\email{stefano.zacchiroli@telecom-paris.fr}
\orcid{0000-0002-4576-136X}
\affiliation{%
  \department{LTCI, Télécom Paris}
  \institution{Institut Polytechnique de Paris}
  \city{Palaiseau}
  \country{France}
}

\begin{document}

\title{Agentic Much? Adoption of Coding Agents on GitHub}

\begin{abstract}
In the first half of 2025, \emph{coding agents} have emerged as a category of development tools that have very quickly transitioned to the practice.
Unlike ``traditional'' code completion LLMs such as Copilot, agents like Cursor, Claude Code, or Codex operate with high degrees of autonomy, up to generating complete pull requests starting from a developer-provided task description.
This new mode of operation is poised to change the landscape in an even larger way than code completion LLMs did, making the need to study their impact critical.
Also, unlike traditional LLMs, coding agents tend to leave more explicit traces in software engineering artifacts, such as co-authoring commits or pull requests.
We leverage these traces to present the first large-scale study (\TotalProjects projects) of the adoption of coding agents on GitHub, finding an estimated adoption rate of \OverallAdoptionRatePercent--\HighEstimateOverallAdoptionRatePercent, which is very high for a technology only a few months old--and increasing.
We carry out an in-depth study of the adopters we identified, finding that adoption is broad: it spans the entire spectrum of project maturity; it includes established organizations; and it concerns diverse programming languages or project topics. At the commit level, we find that commits assisted by coding agents are larger than commits only authored by human developers, and have a large proportion of features and bug fixes. These findings highlight the need for further investigation into the practical use of coding agents.


\end{abstract}

%




\maketitle

\section{Introduction}

In the span of a year, coding agents such as Claude Code, Codex, or \revise{Gemini}{OpenCode} (among many others) have transitioned from promising tools in research papers to practical products seeing practical adoption. While the emergence of LLMs in programming, such as earlier versions of Copilot, has been and is still the subject of extensive study, coding agents, even if they leverage the same underlying technology, are markedly different. What distinguishes coding agents is both the scale of the tasks that they undertake and the degree of autonomy with which they undertake these tasks. Developers can use a traditional LLM to accelerate their programming by completing lines or blocks of code. However, they can (and do--see \Cref{sec:qualitative}) \emph{delegate entire tasks} to coding agents, such as finding and fixing bugs, or implementing new features. Depending on their workflow, coding agents can operate under close supervision (with a developer manually approving each action), up to nearly complete autonomy, \eg, acting on a bug report and submitting a complete pull request for review. 

\revise{The first coding assistant offering ``agentic'' capabilities appeared in 2024. In the spring of 2025, virtually all major AI providers released their take on a coding agent. Since then, the adoption of such programming tools has ballooned (see \Cref{fig:overall-adoption}), in line with their increase in capability. Indeed, studies show that the length of tasks that AI systems can solve at a fixed success rate currently doubles every 7 months \cite{kwa2025measuring}.}{} 
The ability to delegate tasks to a coding agent in such a fashion holds great promise. However, the few human studies on coding agents show contrasting results \cite{becker2025measuring, kumar2025sharp}. We detail the working of coding agents in \Cref{sec:background} and the related work on empirical studies of coding assistants in \Cref{sec:empiricalstudies}.

The paucity of work studying coding agents motivates us to study this from another angle. While completion-based coding assistance embedded in the IDE is hard to track in the wild, since it leaves no explicit traces, coding agents leave abundant traces in software repositories, in the form of configuration files, summaries of knowledge they acquired, or even explicit metadata (e.g., author or co-author information) in commits and pull requests \cite{agentminingpaper}. This paves the way to study the adoption of coding agents in the real world at a very large scale, a point of view that is highly complementary to the existing qualitative studies \cite{stol2018abc}. We detail our methodology, based on the analysis of software artifacts on a very large scale, in \Cref{sec:methodology}.

We find that the adoption of coding agents has been very rapid. \revise{At the end of October 2025}{As of \studyenddate}, in a sample of \TotalProjects projects, \AllFileBasedUsePercent of the projects present traces of adoption at the file level\revise{. At the commit level, the proportion is likely even larger (\SampledCommitUsePercent),}{; a further \SampledCommitUsePercent project show adoption at the commit level} leading us to estimate a total adoption of \OverallAdoptionRatePercent, with a high estimate of up to \HighEstimateOverallAdoptionRatePercent (See \Cref{sec:adoption}).

Beyond quantifying adoption in a binary way, we conduct more detailed analyses to quantify the extent of adoption and its distribution over the project maturity spectrum. We find a bias towards younger projects; beyond that, and somewhat unexpectedly, we find that larger and more mature projects adopt coding agents in comparable proportions to smaller and less mature ones. We also find that a sizeable proportion of projects have adopted coding agents in a pervasive manner, with a small, but non-negligible set of extreme adopters. We detail these findings in \Cref{sec:projects}.

Going further, we study the adoption in more specific contexts; finding adoption in diverse contexts such as established organizations, types of projects as described by their GitHub topics, and programming languages (\Cref{sec:contexts}). We study the evolution of the adoption over time and across tools and organizations, observing a rapid growth since March 2025, and identify that a small number of coding agents account for the majority of adoption (\Cref{sec:evolution}). Finally, we study the contributions authored or co-authored by coding agents at the commit level, finding that they are larger than those authored by humans only (\Cref{sec:contrib-size});  we perform a manual analysis of the content of a sample of commits, to find the types of tasks that are performed with the help of coding agents, finding it is concentrated in \emph{features} and \emph{fixes} (\Cref{sec:qualitative}).

The current extent and the rapid increase of the adoption of coding agents have several implications, both in practice and in research (notably, there are ample avenues for future work). Our study has limitations: in particular, we discuss in which ways our heuristics may lead to overcounts or undercounts of the adoption. We discuss the implications, limitations, and future work related to our study in \Cref{sec:discussion}.





\section{Background}
\label{sec:background}

\subsection{Coding Agents}

\paragraph{The agentic loop}
We use the following definition for coding agents: a \emph{coding agent} is a LLM running in a loop with tool access that aims to complete a given goal. The tools enable the LLMs to have rich interactions with its environment.
The LLM loop, which we call \emph{agent loop}, is simple: the LLM proposes a solution, it gets feedback and then use the feedback to offer another modification, more precisely: 
\begin{enumerate}
\item query the LLM with a sequence of interactions, initially this is just the task description;
\item parse the response given by the LLM for the presence of tool usages and task completion;
\item if tool calls are present: optionally ask the user for permission before using sensitive tools;
\item if task is not completed: add the LLM's response to the sequence of interactions, replacing tool calls with their results.
\end{enumerate}

\begin{figure}
    \centering
    \includegraphics[width=.7\linewidth]{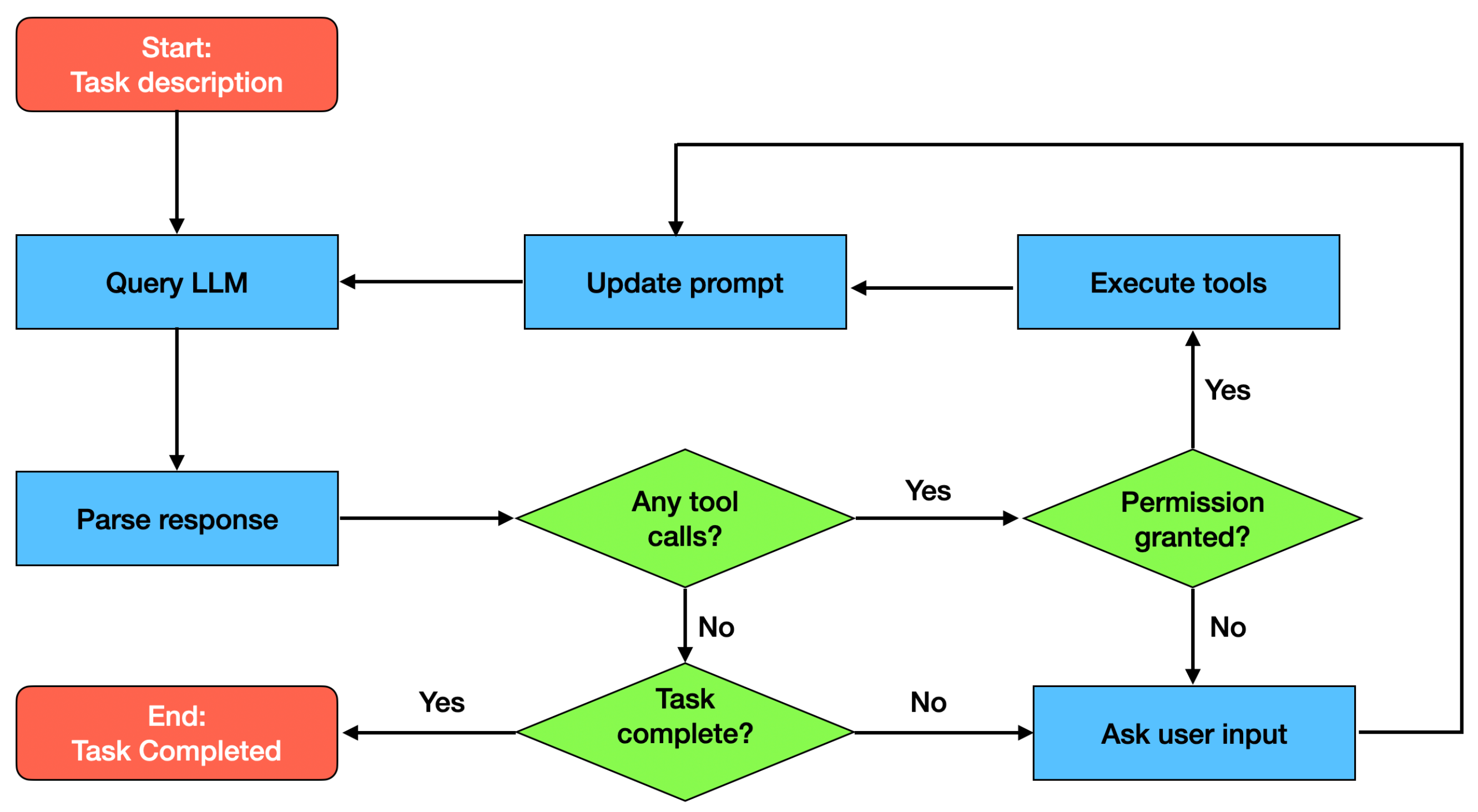}
    \caption{Typical workflow of a coding agent}
    \label{fig:agent-flow}
\end{figure}

\figref{fig:agent-flow} shows a graphical depiction of this process.
The agent notifies the user when it believes it has completed the task, ending the loop.
Some agents go as far as to commit the changes to the repository, or author a pull request.
Depending on its configuration, the agent may insert co-authorship information at this point (e.g.,\texttt{Co-authored-by:copilot-swe-agent}).

The biggest contributor to the performance of coding agents is the LLM. There has been rapid progress of LLMs towards producing structured output or improved tool usage, which are key for agentic use. Furthermore, there has been a recent growth in the reasoning capabilities of agents over long sequence of interactions which is key driving factor for the performance of coding agents.

\paragraph{LLM Harness and tool use} Tools access boosts coding agents with more reliable capabilities, the LLM's harness connects the LLM to these tools. The harness provides the list of available tools to the LLM as part of its prompt. Part of the LLM's training involve tool use, which, from the LLM's point of view, consists in outputting a tool call using a structured format such as JSON.
The harness will query the LLM, and parses the LLM's output in order to detect tool calls. It then invokes the tools on behalf of the LLM, and feeds the output of the tool back to the LLM. Depending on the degree of autonomy desired, the harness may ask the user for confirmation before executing certain tools (\eg, to review code edits, or to approve shell commands).

\paragraph{Tools and context}
Software projects constitute a large context for an LLM, which has a limited context window. For traditional coding assistants, context is often pre-calculated by the IDE, and included to the LLM's prompt. Coding agent allow the same functionality; they also allow developer to manually specify relevant context by mentioning specific files. However, a coding agent can also augment its context: tool calls enable dynamic and interactive queries. This enable coding agents to perform more autonomous workflows, such as:
\begin{itemize}
    \item Adapt its context to the task at hand, by using tool calls to explore the code base dynamically;
    \item Automatic debug of the code thanks to compilation and a potential suite of tests, using the feedback given from these tools the agent can propose solutions.
\end{itemize}

There is a large variety of tools that agents can use. Some coding agents can run arbitrary shell commands in a terminal (this defines a small set of tools, but allows great flexibility), whereas other operates in more controlled environments that restrict the commands they can run.
Some specific tools enable browsing the web to gather information available. Other tools allow coding agents run a project test suite; this is highly valuable for the agent to gain feedback on their work. The specific tools available are user dependent; emerging standards such as the Model Context Protocol (MCP) ease the definition of new tools.

Given a task (e.g., ``implement feature X and make sure the tests pass''), the agent will execute the LLM in a loop until it determines that the task is completed. In the feature implementation example, the agent could start by exploring the code base to find relevant code, then edit the code until it compiles (processing errors from the compiler), and then iterate until tests pass (if they exist; the agent may also be tasked to write tests in the first place).

\paragraph{Autonomy and oversight}

Notably, this process can be lengthy, as the LLM is used multiple times, processing inputs of increasing size (often growing to tens of thousands of tokens). Agents tend to have an asynchronous mode of interaction: they do their work in the background, only stopping when the task is done, or when they require input from the developer. Developers can tend to the agents in between their other tasks. Depending on the degree of autonomy desired, the developer may interrupt the process at any point and provide some guidance to the LLM. Particularly trusting users (or careful developers using sandboxes and extensive test suites) may provide high or full autonomy to the LLM, setting the harness to auto-accept any tool the LLM desires. 

Developers may provide guidance to the agent by providing more detailed instructions (e.g., ``implement feature X in module Y, using library Z''), which will reduce the need for searches, or give feedback on the work (e.g., ``this code does not conform to the project's convention, always do X, not Y"). 
The LLM may also compile this guidance in the form of instructions in text files; with appropriate conventions, the harness will automatically add these files to the LLM's context window in subsequent iterations. The footnote below links to two examples of such files containing guidance about a specific projects, describing their high-level organization, commands to interact with the project, conventions to follow, among others. \footnote{\href{https://github.com/foambubble/foam/blob/main/CLAUDE.md}{Example 1} and \href{https://github.com/shopsys/shopsys/blob/18.0/CLAUDE.md}{Example 2}; the next section has additional examples}
As such knowledge is valuable, such files are often committed to the repository, so that the agent can consult them overtime, and, if necessary, update them. In this study, we look for the presence of these \emph{guidance files}, as markers for the use of LLMs, together with other explicit markers in commits and pull requests. 



\begin{diffversion}{}
\paragraph{Notable coding agents}

Let us present a few of the most popular or notable AI coding agents, at the time of writing:
\begin{itemize}
    \item \textbf{Claude Code}, is Anthropic's take on agentic AI for code. Initially available as a terminal application as a preview in February 2025.  The default guidance file for Claude Code is \claudemd. If allowed, it typically commits on behalf of users. Claude Code can now integrate as a plugin in most code editors, and is since recently available on the web and as GitHub actions; and can recently make pull requests. 
    \item \textbf{Codex} from OpenAI (different from the earlier model with the same name), was released as a command-line application in April 2025, followed by a variant available on the web in May 2025. It can run on a cloud virtual machine which can download repositories before performing tasks on them. The default guidance file for Codex is \agentsmd, which, since August 2025, has become an emerging standard which other agents use. Codex typically create one branch for each task it is assigned, and submits a Pull Request when finishing its work.
    \item \textbf{GitHub Copilot} started as a code completion tool in 2021. Over the years, it added the functionality to query the code base with natural language. It offers an agent mode since May 2025, which can be interacted with from the IDE. It is also directly available within the GitHub user interface, where it can be assigned an issue which it tackles by making a plan, iterating on commits, and submit a draft pull request that it can modify in response to comments. Copilot uses a \texttt{copilot-instructions.md} files for guidance.
    
    
    \item \textbf{Cursor} is based on a fork of Visual Studio Code, with the idea to make the next generation of IDE, an IDE that targets collaborative development between human and AI. Like Copilot, Cursor started as a code completion tool in 2023, before offering a chat mode, and finally an agent in late 2024. The agent is directly available from the different panes of the IDE so that the user can quickly start a conversation or delegate a task to the agent. It integrates with most LLMs. Cursor stores guidance as ``rules'' in a \texttt{.cursorrules} files or a \texttt{.cursor/} directory.
    \item \textbf{Aider} is an open-source framework where the agent lives in your terminal but it can also be plugged into existing IDEs. Aider pioneered the command-line agent format, with development starting as early as 2023. Aider works with any LLM. Aider is also largely developed with the help of the tool itself: for example, 88\% of the code in their last release was authored by their agent. Aider's guidance is often contained in a file called \texttt{CONVENTIONS.md}
    
\end{itemize}

\Cref{fig:agent_timeline} shows a timeline of the release date of a subset of agents, as much as we could determine them. In some cases, we could determine only the products' initial release date (likely too early); nevertheless, the figure shows the drastic increase in the quantity of coding agents over time.
\end{diffversion}

\subsection{Traces and heuristics}
\label{sec:heuristics}

Coding agents often leave visible traces of their use. These traces can be in the form of a specific pattern in commit messages or metadata, a branch name, or the presence of a specific file in a repository.
To detect traces of coding agents activity in this work, we rely on our previous work~\cite{agentminingpaper} in which we produced an extensive list of heuristics about the traces left by more than 50 different coding agents. 
These heuristics were validated with the data found on the GitHub platform. 
Heuristics are, by definition, not perfect, but we ensure that they do not produce many false positives. We provide below a short overview of the traces left by coding agents. Each agent has a specific workflow, and may leave a subset of these traces.

\subsubsection{Files}

Files can be used to configure coding agents and add specific instructions for them.
We distinguish the following file types used to configure agents behavior: \emph{Configuration files}, 
\emph{Rules}, and \emph{Guidance files}.
Configuration files allow to choose between predefined options of the agent in a structured language such as YAML, whereas rules and guidance files enable guiding the agent with custom instructions in natural language.

\emph{Configuration files} indicate that the agent has been set up within the repository.
They contain the most important options for configuring the agent's behavior and thus can offer insight into how the agents work, such as permissions, \ie, the actions the agent can take without waiting for the user's permission.
These settings are highly tool-specific. Some settings influence the presence of absence of other traces, hindering the detection of agents. For instance, Claude Code's \texttt{.claude/settings.json} has an option to disable traces left in commits, while Codex can change the default prefix it uses to create branches.


\emph{Rules and Guidance files} aim to guide the agent through natural language specifications that the agent should follow.
They are generally written in markdown format and are added to the user prompt before sending it to the LLM.
They can have large or specific scopes. They are sometimes generated by an agent.
Their goal is to encode \emph{tacit knowledge} that is necessary to the agent to tackle the tasks; this is why they are often shared among developers and most of the time committed to the repository. Rules are context specific\footnote{\href{https://github.com/adobe/data/blob/main/.cursorrules}{Example cursor rules} with technology stack, programming principles, style issues, and how to test (27 lines).} whereas guidance files are broader.\footnote{\href{https://github.com/Windscribe/Android-App/blob/main/CLAUDE.md}{Example high-level project description} for Claude, with: overview; key features; build, launch, test, and lints commands; project architecture (main modules and classes); development workflow; and typical tasks (239 lines)} Guidance files may contain detailed task descriptions that highlight the steps to solve a specific task \footnote{\href{https://github.com/marcusgoll/robinhood-algo-trading-bot/blob/c05474f63df1aa0a92aee061592ea25dfd13f6d9/specs/order-management/plan.md}{Implementation plan} for a feature, with: context; principal decisions and rationales; architecture of the solution; breakdown of the implementation in phases and steps; description of follow-up tests; risks; and deliverables (188 lines)} or tactics and strategies the LLM should use to tackle problems in general. \footnote{\href{https://github.com/bigcapitalhq/bigcapital/blob/develop/.cursor/commands/speckit.clarify.md}{Example guidance with strategy}, for clarifying specifications that are under-specified (176 lines)}

For example, the guidance file \agentsmd\footnote{\url{https://agents.md}, listing 19 agents using the format on 14/11/25}, originally used by Codex, is a simple open format for guiding coding agents and is adopted by several agents, including, Jules, Cursor, Aider, and Copilot.
Claude uses \claudemd, as stated in its documentation: ``\emph{\claudemd is a special file that Claude automatically pulls into context when starting a conversation. This makes it an ideal place for documenting: common bash commands, core files and utility functions, code style guidelines, testing instructions [...]}''.\footnote{\url{https://www.anthropic.com/engineering/claude-code-best-practices}}

\subsubsection{Commits}

Most agents are given the capability to commit their work automatically or given a tool to do so, for example via the shell.
If allowed to commit, the agent should also write the commit message.
Within commits, agents can advertise their contribution to the commit through different mechanisms:
\begin{itemize}

\item By adding themselves as a \emph{co-author} of the commit, with the agent's user advertised as the main author. In this case, they use the ``\texttt{Co-authored-by:}'' trailer, which is added to the commit. Of note, agents may not adhere to the convention exactly, occasionally using different casings.\footnote{\href{https://github.com/browser-use/browser-use/commit/b2da2fec895577d05f205d1c3049a8e1a02ec089}{Example commit}, co-authored by Claude Code, fixing a bug and adding tests (two files changed, +75/-2 lines).}

\item Less commonly, agents can be recorded as commit authors, similarly to how bots such as dependabot operate. To do so they provide a standard name and email address in commit metadata like Git's "author" and/or "committer" fields.\footnote{\href{https://github.com/microsoft/azuretre/commit/a266b5073353e031a4aa3adcc51d7c37deffa38b}{Example commit}, authored by Copilot, adding a feature and tests (16 files in 13 directories changed, +380/-48 lines)}

\item Some agents also add additional trailer pseudo-headers to the commit message, \eg,\linebreak ``\texttt{Generated~by:~Claude}''.

\end{itemize}
All these traces can be explicitly disabled via configuration files, hiding the participation of agents.

\subsubsection{Users, Issues, and Pull requests}
Specific user accounts can be linked to coding agents, \ie, the activity of those users are in fact agent activities.
This enables users to assign reviews or issues to specific coding agents as users.\footnote{\href{https://github.com/openhands-agent}{Example coding agent account for OpenHands}} Issues can then be used as the descriptions of tasks that are then delegated to coding agents. Some agents, when assigned such an issue or a more general task, will work via Pull Requests (PRs), in which they may author multiple commits depending on the task scope. Developers can interact with the agent via the GitHub interface, for example, delegating additional tasks.\footnote{\href{https://github.com/OpenHands/OpenHands/pull/8420}{Example PR}  with user delegating additional tasks to OpenHands (improving documentation and code consistency)}
A PR provides more information than a commit, with the whole trace of the development and the interaction with GitHub Actions, users, and reviewers.
The complexity of PRs authored by coding agents can range from the quite simple to authoring complex new features from scratch.\footnote{\href{https://github.com/microsoft/vscode/pull/271364}{Example PR} with user interacting with Copilot to implement a feature (3 follow-ups, 8 commits, +350/-15 lines)}

\section{Related work}
\label{sec:empiricalstudies}

\subsection{Studies of LLM-based coding assistants}
\paragraph{Controlled experiments} 
\revise{Early productivity studies showed mixed results. Imai \cite{imai2022github} found higher productivity but lower code quality when using Copilot versus pair programming with humans. Vaithilingam \etal \cite{vaithilingam2022expectation} showed participants failed tasks more often with Copilot than Intellisense due to incorrectly generated code, despite preferring Copilot. However, Peng \etal \cite{peng2023impact} found substantial improvements, with 95 programmers completing HTTP server implementations 55\% faster using Copilot. Security studies yielded contrasting results. Sandoval \etal \cite{sandoval2023lost} found similar overall security bug rates between assisted and control groups, with 36\% of bugs originating from AI suggestions. Perry \etal \cite{perry2023users} found AI-assisted code was less secure across 6 tasks, though participants overestimated their code's security. Asare \etal \cite{asare2023user} found no significant security differences on realistic tasks like sign-in functionality.}{Experiments show mixed results on productivity: Imai \cite{imai2022github} found Copilot improved productivity but reduced code quality versus human pair programming, while Vaithilingam \etal \cite{vaithilingam2022expectation} reported more task failures due to incorrect Copilot suggestions despite user preference. Peng \etal \cite{peng2023impact}, however, found 95 programmers completed HTTP server implementations 55\% faster with Copilot.}

\paragraph{Observation studies} \revise{Barke \etal \cite{barke2023grounded} observed 20 participants using Grounded Theory. They identified two fundamental usage modes: \emph{acceleration}, where developers have clear implementation goals and prefer small, quickly-validated suggestions to maintain flow; and \emph{exploration}, where developers lack clear plans (e.g., working with unfamiliar APIs) and prefer larger suggestions as starting points. This framework explains much of the variation in how developers interact with AI assistants. }{Barke \etal \cite{barke2023grounded} identified two primary usage modes for developers: \emph{acceleration}—where developers with clear goals prefer small, quickly-validated suggestions to maintain flow—and \emph{exploration}—where developers lacking clear plans (e.g., with unfamiliar APIs) favor larger suggestions as starting points. Mozannar \etal \cite{mozannar2022reading} found developers spent 22\% of time evaluating Copilot suggestions versus only 14\% writing new code, indicating that acceptance rates mask the true cognitive effort of AI-assisted programming.}

\paragraph{Surveys and interviews} Liang \etal \cite{liang2024large} surveyed 410 developers, identifying successful use cases (boilerplate, proof of concepts) and barriers (validation time, non-functional requirements). \revise{Mozannar \etal \cite{mozannar2022reading} observed 21 programmers: participants spent 22\% of time evaluating Copilot suggestions versus only 14\% writing new functionality. This suggests that common metrics like acceptance rate significantly underestimate the cognitive overhead and time investment required for AI-assisted programming.}{} Wang \etal \cite{wang2023investigating} studied trust in AI assistants through interviews with developers based on real-world Copilot usage. \revise{Developers emphasized understanding Copilot's limitations and expressed concerns about negative impacts on learning—a key consideration for software engineering practice and education.}{}

\paragraph{Telemetry} \revise{Using telemetry to monitor adoption of coding assistants can be done in controlled environments such as companies.}{} \revise{Murali \etal \cite{murali2023codecompose} reported Meta's deployment of CodeCompose, achieving 22\% acceptance rate and generating 8\% of company-wide code.
Ziegler \etal \cite{ziegler2022productivity} found that among 2,000 Copilot users the best factor to predict perceived productivity was acceptance rate. Izadi \etal \cite{izadi2024language} developed an IDE extension in order to study model failures via the interaction data of 1200 users.}{Murali \etal \cite{murali2023codecompose} reported Meta's CodeCompose achieved a 22\% acceptance rate and generated 8\% of company-wide code. Among 2,000 Copilot users, Ziegler \etal \cite{ziegler2022productivity} found acceptance rate best predicted perceived productivity. }

\paragraph{Usage in software repositories.} 
Tufano \etal \cite{tufano2024unveiling} analyzed ChatGPT usage traces in 467 GitHub instances, developing a taxonomy of 45 software engineering tasks including feature implementation, documentation, software quality, and development processes. \revise{Some usage patterns were unexpected, such as assistance in motivating proposed changes.}{} \revise{Xiao \etal~\cite{xiao2025self_admitted_ai} analyzed more than 1,200 traces of generative AI from ChatGPT and Copilot in a qualitative fashion when usage was explicitly written by developers in their software artifacts. They looked at the targeted tasks of generative AI but also looked the churn before and after adoption of generative AI and found no significant change in churn post-adoption.}{Xiao \etal~\cite{xiao2025self_admitted_ai} analyzed over 1,200 explicit generative AI traces from ChatGPT and Copilot in software artifacts, focusing on targeted tasks and churn changes before and after adoption. They found no significant change in churn post-adoption.} \revise{T}{On the other hand, t}he white papers from GitClear~\cite{hardinggitclear2024, hardinggitclear2025} \revise{paint a different picture. They studied the evolution of churn before and after the adoption of Copilot. They found that churn increased post-adoption as time passes with an also increased incidence of code duplication. However one weakness of this study is the lack in the data of traces of use of Copilot, making it impossible to distinguish between cases where Copilot was used and when it was not, making it a rough approximation that does not exclude other possible factors.}{report increased churn and code duplication after Copilot's release over time. However, their lack of usage traces prevents distinguishing actual Copilot use, limiting the analysis to a rough approximation that may not exclude other confounding factors.}

\subsection{Studies of Coding Agents}

\begin{diffversion}{}
\paragraph{Academic Agents} The field of coding agents is not industry specific, there has been a strong interest in the academic community as goal.
The development of benchmarks such as SWE-Bench~\cite{swebench2024, chowdhury2024swebenchverified} that are now used both by academics and industrials as a standard benchmark to evaluate the performance of their coding agents. SWE-Bench use data from real world Git Hub issues that were successfully merged with enough description and tests in order to assess the ability of the agent to go from the natural language description of the issue to making changes in order to pass the tests.
Coding Agents are also developed by the academic community such as AutoCode Rover~\cite{zhang2024autocoderover} or Agentless~\cite{xia2024agentless}, with a focus on lowering the cost of solving tasks of SWE-Bench.
These papers highlights also flaws in SWE-Bench with tests for some issues that are lacking~\cite{wang2025solved} or the possibility of agents to change the tests or remove features in order to solve tasks.
\end{diffversion}

\paragraph{\revise{Studies of coding agents}{Experiments and observation}} 
\revise{Becker et al.~\cite{becker2025measuring} studied the perceived increase of productivity relative to the effective productivity in a controlled experiments of software developers using Cursor. Software developers believed that Cursor usage increased their productivity by an average of 20\% but actually it decreased their effective productivity by 19\% on average.
The proposed explanation is that developers do not choose efficiently the tasks to delegate to cursor, some are worth it while others are not but many potential factors can have an effect.
Interesting metrics are that only 44\% of code generations are accepted and 9\% of the time is spent waiting on AI to generate code, improving generation speed or quality has clear benefits.}{Becker et al.~\cite{becker2025measuring} found that developers using Cursor overestimated productivity gains by 20\% on average, while effective productivity actually decreased by 19\%. They attribute this to inefficient task delegation, though many factors may contribute. Only 44\% of code generations were accepted, and 9\% of time was spent waiting for AI.}
Importantly, the agent used had limited capabilities \revise{}{(early 2025)}, and was used in interactive rather than autonomous mode.

Kumar et al.~\cite{kumar2025sharp} studied the barriers to success of coding agent use with observations from 19 developers using Cursor. \revise{The developers were split in two groups, one who delegated the whole to the agent and one who split the tasks into several sub tasks with the agent and the agent had to solve the sub tasks.}{The analysis found that developers that delegated the whole task to the agent were less successful than the ones splitting them into subtasks.} \revise{The developers had to provide the agent with contextual information specific to the repository. The main barrier was the lack of tacit knowledge of the agent.}{The main barrier was the lack of project-specific or tacit knowledge of the agent; providing it improved success.}

\revise{Bouzenia and Pradel~\cite{bouzenia2025understanding} studied the interactions traces of agents trying to solve tasks from the SWE-bench benchmark.
These traces are thus purely agent, there is no human interaction except from the initial prompt however at the same time they highlight the diversity of traces between the different less established agents considered. Variability is shown between the length of interaction traces, their type and the differences between successful and failed attempts.}

\paragraph{MSR studies of coding agents}
Concurrent to this work, a few studies using MSR techniques have emerged\revise{(all are currently preprints)}{}. Li \etal present AIDev, a dataset of coding agent pull requests \cite{li2025aidev}, showing a sharp increase in the number of pull requests authored by a subset of agents (Codex, Devin, Copilot, Cursor, and Claude Code). They classify the PRs according to their conventional commit type, with results similar to ours in RQ6.

Mohsenimofidi \etal \cite{mohsenimofidi2025context} performed a preliminary qualitative study of the content found in coding agent guidance files, finding that the content is very diverse, and that the most common categories of content include conventions and best practices, contribution guidelines, project structure, and build and testing instructions, among others.

He \etal study projects adopting an early coding agent, Cursor, to find whether it is associated to increases in productiviy and quality issues using a difference in differences design \cite{he2025speed}. They find that the increase in velocity in the first two months of cursor adoption, while static analysis warnings and code complexity increase over time. 

\revise{}{Bouzenia and Pradel~\cite{bouzenia2025understanding} analyzed autonomous interaction traces from three early academic agents on SWE-bench tasks, revealing significant variability across agents. This variability spans trace length, type, and the differences between successful and failed attempts.}

\revise{}{\paragraph{This work} This work applies MSR techniques to study the specific topic of the \emph{adoption} of coding agents. Compared to experiments and observation studies of users of coding agents, our work has a much larger scale (130,000 repositories vs a few dozens of participants). Our work is primarily quantitative, whereas observation studies can provide qualitative insights. Finally, the existing studies were conducted in early 2025, whereas we include data from 2025 until 2026, offering more recent insights in a fast-changing field.

Compared to the concurrent work studying coding agents via MSR techniques, our study has a different scope (projects, rather than individual artifacts such as files or PRs). Since we study adoption, we try to be as exhaustive as we can in detecting coding agents (we detect dozens of agents), rather than focusing on a small subset of agents. Another aspect of the exhaustivity is the scale of the study, which is comparable to AIDev, but larger than the other MSR studies. Finally, we study adoption via a variety of lenses, such as adoption by programming language, by type of project as denoted by GitHub topics, adoption in specific organizations, or its evolution over time.}
\section{Methodology}
\label{sec:methodology}

\begin{figure}[ht]
    \centering
    \includegraphics[width=\textwidth]{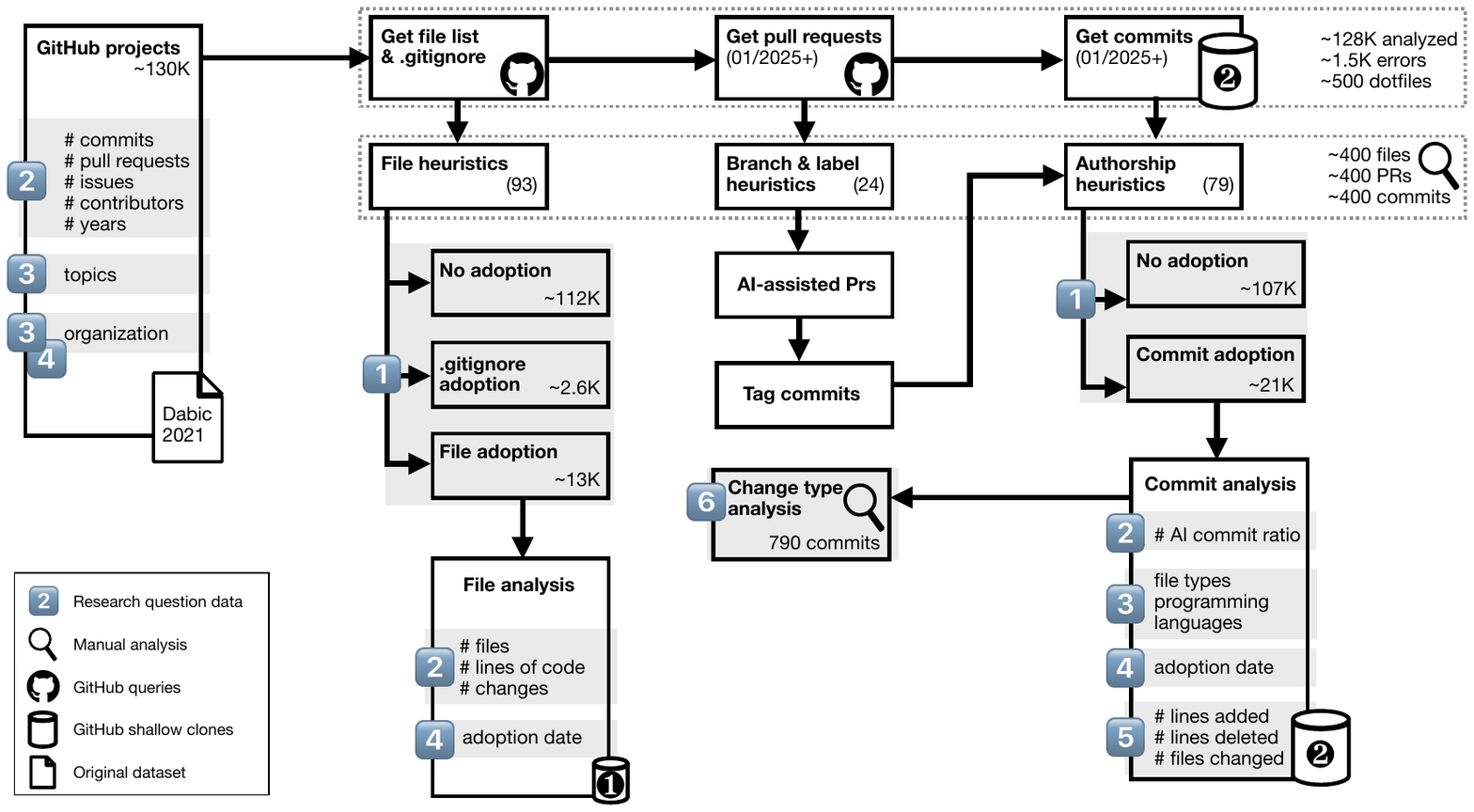}
    \caption{Overall schema of the study, covering the analysis pipeline, the data sources, heuristics, metrics, and research questions.}
    \label{fig:study-schema}
\end{figure}

We study the adoption of coding agents in the real world, at a very large scale, a point of view that is highly complementary to the existing studies, obtained from observations of very few developers~\cite{stol2018abc}.
We decided to target open-source GitHub repositories, so the data comes from Git repositories and is available through the GitHub APIs. \figref{fig:study-schema} presents the study at a glance. \secref{sub:heuristics} details the types of heuristics we use to discover coding agent traces; \secref{sub:pipeline} describe the data gathering pipeline we employ; \secref{sub:metrics} discusses the metrics we gather for the study; finally, \secref{sub:questions} presents the research questions. \revise{}{The code to gather and analyze data from GitHub repositories is available at \url{https://github.com/labri-progress/agent-impact/}, while the data generated by the pipeline  on \studyenddate is available at \url{https://zenodo.org/records/19256968}}.

\subsection{Heuristics for Detecting Coding Agents}
\label{sub:heuristics}

One significant challenge is the diversity of the ways the different agents work, as well as the sheer number of available agents \cite{agentminingpaper}. Some agents are integrated in IDEs, while others operate on the command line; some agents are very autonomous, primarily interacting via pull requests. 

\subsubsection{Inclusion criteria}
We started this study in early 2025. At the time, the capabilities of coding agents were more limited, and they were used with close supervision. Our initial inclusion criteria was for tools that leverage LLMs to automate coding, for which we could find traces in repositories. This includes full-fledged coding agents such as Claude Code, as well as tools that focus on smaller-scale activities, such as fixes automated by tools such as CodeRabbit (as evidenced by commits co-authored by CodeRabbit), whose primary focus is code review.

We note that with time, the increase in capabilities of coding agents means that there is a significant gap in capabilities between full-fledged coding agents, and tools with a more specialized scope. \secref{sec:evolution} investigates adoption of individual tools; the most popular of the specialized scope tools, CodeRabbit, accounts for less than 2\% of the commits we detect, indicating that the adoption is driven by full-fledged coding agents. 

\subsubsection{Heuristic identification}
We employ a multi–faceted approach to identify repositories \revise{that use software‑engineering (SE)}{using coding} agents such as \revise{Copilot,}{} Claude, Codex, and others. The heuristics are grouped into four categories:

\begin{enumerate}
  \item \emph{File\revise{‑based}{s}}: presence of specific configuration files or documentation artifacts that are typical for an agent’s usage (e.g., \revise{presence of }{}a \texttt{.cursorrules} file).
  \item \emph{Author\revise{-based}{s}}: authorship or co-authorship of a commit by a known coding agent (e.g. \texttt{Co-Authored-By: Claude <noreply@anthropic.com>}).
  \item \emph{Branch\revise{‑based}{es}}: creation of branches whose names follow a convention \revise{used by agents}{} (e.g. \texttt{copilot/fix-…}).
  \item \emph{Label\revise{‑based}{s}}: pull requests \revise{or issues }labelled with the corresponding agent name (e.g. \texttt{codex}).
\end{enumerate}


\noindent
\revise{The heuristics were obtained through}{We found the heuristics via} a systematic manual investigation \revise{that involved}{involving} the following steps:

\begin{enumerate}
  \item Mapping agent names to repository identifiers.
  \item Studying documentation to learn how agents are configured and what files they use or generate.
  \item Manually inspecting repositories likely to employ agents and recording relevant artifacts.
  \item Performing targeted GitHub searches to confirm hypotheses, such as:
    \begin{verbatim}
      Path:CLAUDE.md
      Co-authored-by:copilot-swe-agent
      is:pr label:codex
    \end{verbatim}
  \item Cross‑checking results to discover new patterns when a repository uses multiple agents simultaneously.
\end{enumerate}

At the time of writing, we maintain $79$ author-based heuristics, $93$ file‑based heuristics, $20$ branch‑based heuristics, and $4$ label‑based
heuristics for $63$ coding agents.  \textcolor{red}{All the heuristics and tools are documented in our replication package}. In addition, we also maintain a set of general bot
patterns, using authorship heuristics (e.g. \texttt{dependabot}); we use this to filter commits and focus on human-authored and AI-assisted commits. 

It is important to note that the users can hide the traces of AI-assisted commits by removing the metadata or not adding their files to the repository; these commits, therefore, will be identified as human-authored in our study.

\subsubsection{Heuristic validation}
Each heuristic was rigorously validated for false positives. A notable
example concerns the file \texttt{CONVENTIONS.md}. While the Aider
website recommends placing agent‑related information in this file, it is
also commonly used to advertise general project conventions to all
developers.  Because the risk of misclassifying such files outweighs the
potential underestimation of Aider usage, we have excluded
\texttt{CONVENTIONS.md} from our final heuristic set.
Similarly, the \texttt{AGENTS.md} file was originally introduced by Codex; however, it is now used by dozens of coding agents, including Copilot, Jules, Cursor, and Aider.
Consequently, the presence of this file alone is insufficient to reliably identify the coding agent in use.

\newcommand{\TP}[0]{\texttt{TP}\xspace}
\newcommand{\FP}[0]{\texttt{FP}\xspace}
\newcommand{\deleted}[0]{\texttt{404}\xspace}
\newcommand{\unsure}[0]{\texttt{Unsure}\xspace}

\revise{}{To quantify the remaining false positives, we analyzed random samples of commits, files, and pull requests flagged by our heuristics ($\approx$ 400 each). This value exceeds the value needed for a standard confidence level of 95\% with a 5\% error \cite{triola2006elementary}. For each needed sample, we first sampled one project that had both file-level and commit-level adoption, and then sampled among its flagged commits, files, or pull requests, as needed (if the project did not have pull requests, we resampled). Each artefact was checked by two authors to determine whether it was a true positive (\TP) or a false positive (\FP); they could use an \unsure annotation for borderline cases, and a \deleted annotation to indicate that an artefact no longer existed (\eg a file was deleted, a repository was taken private). We then measured the initial disagreement, before discussing to adjudicate differences and refine the annotations: notably, we differentiate between legitimate false positives and borderline cases.}

\revise{}{For files, there were 28 disagreements out of 400 files (7\%). Twelve disagreements were between an annotator finding the file, and the other annotating \deleted (several files moved, or the default github branch changed, between when the study was carried out and when we checked the heuristics). Ten disagrements were between \TP and \unsure, four between \TP and \FP, and two between \FP and \unsure. After adjucating disagreements, we had 15 \deleted files, 370 \TP, and 15 borderline cases (4\%, excluding deleted files), rather than complete false positives. The borderline cases were 2 empty files, and 13 files that were more concerned with agentic code review rather than coding per se; in 8 of those 15 cases, the repository contained other files confirming the use of agents.}

\revise{}{For commits, there were 40 disagreements out of 399 commits (10\%). There were 16 disagreements \FP --- \TP; 16 \TP --- \unsure; and 8 \FP --- \unsure. An important portion of the disagreements were due to an annotator finding more information than the other; in particular, for some commits the information is found in a pull request, rather than the commit itself. After adjudicating disagreements and refining the labels, we found: 2 \deleted, 370 clear \TP, 8 \FP (2\%, excluding \deleted), and 19 borderline cases (5\%, excluding \deleted). Fourteen of the borderline cases are commits corresponding to smaller fixes (\eg code review fixes); the five others correspond to scenarios where the verdict is unclear for the commit itself, but neighbouring commits show clear evidence. Of note, a review of the 8 \FP by another annotator found information that would have classified three of them as \TP (two with Codex, one with Aider).}

\revise{}{For pull requests, there were 16 disagreements out of 399 pull requests (4\%). One pull request no longer existed. All disagreements stemmed from one annotator finding evidence that the other did not; they were found to be true positives. After adjucating disagreements, there was a single false positive out of the 398 remaining pull requests (0.25\%).}

\revise{}{Overall, our evaluation of heuristics found that very few cases are legitimate false positives (0.5 to 1\%), with a higher proportion of borderline cases (3\%). }


%


\subsection{Analysis Pipeline}
\label{sub:pipeline}

\newcommand{\studydate}[0]{February 21st, 2026}
\newcommand{\studydateshort}[0]{21/02/2026}
\newcommand{\studystartdate}[0]{01/01/2025}

Since the adoption of AI agents is very recent and evolving rapidly, recent data is needed. This is why we put a particular focus on automation in this work. We ran the last iteration of our analysis pipeline on \studydate \ and the following \revise{day, to get data on entire months of activity}{two days}. Our study period starts on \studystartdate \ and ends on \studydateshort \footnote{Note to reviewers: when revising this paper, we plan to extend the study period with updated data.}. Early runs of the analysis included 2024: we found that adoption was extremely limited then; as the complexity of our analysis grew, we decided to focus on the period with the most significant adoption. 

We first filter repositories to find relevant active repositories, then we detect agent use and collect statistics.
We perform different analyses based on the overall data that we have gathered this way.

\subsubsection{Data Selection}

We start with a sample using the sampling tool by Dabic et al.~\cite{dabic2021sampling}.  
The original dataset contains repositories with at least 10 stars.
From this dataset, we select projects that satisfy the following characteristics:

\begin{itemize}
    \item not forks,
    \item a minimum of $5\,000$ lines of code (non‑blank, non‑comment),
    \item at least $100$ commits,
    \item active in the last three months (at least one commit),
\end{itemize}

These criteria ensure that we have projects with a reasonable level of maturity, as we enforce minimum thresholds for activity and code size. In particular, we do not want forks to dominate our sample. In most cases, the bulk of development occurs in the “original” project.  In some cases, this may cause us to miss interesting projects--\eg, when an original repository is now inactive but a fork continues development—but we consider this a reasonable compromise.

We accessed the sampling tool to download the dataset used in the remainder of the study on 29/08/2025. After applying the filters above, we are left with a set of \initialProjects \ projects. Considering that GitHub hosts hundreds of millions of projects, we were initially surprised at this relatively low number, given that the sampling tool by Dabic \etal monitors all public projects on GitHub.  However, the minimum threshold of 10 stars excludes the vast majority of repositories, and the exclusion of forks substantially reduces the pool of candidates. The minimum activity and size thresholds reduce it further. 

We initially considered using Software Heritage \cite{DiCosmo2017}, as it is the most comprehensive dataset of public source code. However, Software Heritage lags behind GitHub by a few months, which is problematic in our case, as we want data to be as fresh as possible to study such a rapidly changing phenomenon. Beyond the project names and URLs, this dataset includes an important number of metrics (e.g., number of contributors, number of commits, number of forks, \etc), some of which we use in the remainder of this study.

\revise{At the end of the analysis process described next, we have data on \TotalProjects projects, the difference with the original number (\initialProjects) is due to $\approx900$ failures of the pipeline, and $\approx500$ ``dotfiles'' repositories that we filter out from the analysis as they are not representative of software development. In the vast majority of cases, failures occur when a repository cannot be found on GitHub (typically because it was made private between the time we downloaded the list of repositories and the time we fetched the file list).}{At the end of the analysis process described next, we have data on \TotalAnalyzedProjects projects, the difference with the original number (\initialProjects) is due to $\approx1,500$ failures of the pipeline. In the vast majority of cases, failures occur when a repository cannot be found on GitHub (typically because it was made private between the time we downloaded the list of repositories and the time we fetched the file list). We further filter out $\approx500$ ``dotfiles'' repositories from the analysis as they are not representative of software development; we carry out the analysis on \TotalProjects projects.}

\subsubsection{Identification of Agent Adoption via Tools}
Next, we use our file-based heuristics to identify agent adopters in our sample. For each repository selected in step 1, we query the GitHub REST API to download the file list of the main branch when we execute the study (on \studydate). For very large repositories, the API fails because the file list is too large; in that case, we perform a shallow clone of the repository to obtain the file list. For every repository, we store a single text file containing one file path per line. %

We then search each file list for names that match our compiled set of file‑based heuristics; any repository containing at least one file identified by a heuristic is classified as an adopter at the file level. Nevertheless, we always match all heuristics: by matching all heuristics, we can identify every tool that a repository might use, since a repository may employ more than one agent, and every file that matches heuristics. 

\subsubsection{Identification of Agent Adoption via \gitignore}

Although knowledge in coding agent files is valuable, and thus \emph{should} be shared in the repository, this is not always the case. To have a better coverage of agent adoption, we also look for the presence of agent configuration files in the \gitignore file, which would prevent them from being committed to the repository despite being present. In this way, we identify potential adopters who do not wish to commit certain (or all) agent configuration files. To do so, we query the GitHub API for the main \gitignore file, and we scan it for heuristics in the same way we do for the file list.

\subsubsection{Collecting Agent File Statistics}

For each of the repositories that we have classified as an adopter, we collect fine‑grained
information about the agent configuration files that signal agent usage.

\begin{enumerate}
  \item \textbf{Shallow clone.}  
        We clone each repository with \texttt{--depth 1} and \texttt{--no-checkout}, and populate the index with the information of the files in the latest commit, so that the working tree contains only the minimal information to carry out the next steps, to keep storage needs low (even then, the whole analysis still requires \revise{more than half a terabyte}{more than one terabyte} in total). 

  \item \textbf{Checkout matched files.}  
        After cloning, we check out all agent configuration files--those that matched our heuristics
        (e.g., \claudemd, \texttt{.cursorrules}). This step lets us confirm the correctness of the heuristics by inspecting the actual file contents, if needed.

  \item \textbf{Gather per‑file statistics.}  
        For each agent configuration file, we query its commit history, computing the following descriptive statistics:
        \begin{itemize}
          \item the date on which it was first added,
          \item the total number of commits that modified the file
                (including the initial addition),
          \item the size, in lines, of the most recent version.
        \end{itemize}
\begin{diffversion}{}
  \item \textbf{Aggregate metrics.}  
        From the per‑file data we compute repository‑wide summaries:
        \begin{itemize}
          \item total number of agent configuration files,
          \item cumulative size of all agent configuration file,
          \item overall count of commits that touched any agent configuration file,
          \item \emph{adoption date} – the earliest commit that added an agent
                configuration file in the repository, and
          \item per‑tool adoption dates – the earliest commit of an agent configuration file that is specific for a given agent (e.g., the first commit containing \claudemd).
        \end{itemize}
\end{diffversion}
\end{enumerate}

These metrics provide a quantitative view of how and when agents are introduced into open‑source projects, and they serve as the basis for subsequent analyses, such as maturity classification and usage trends. 



\subsubsection{Commit and Pull Request Level Adoption}

After having identified the repositories that contain agent configuration files,
we perform a more granular analysis on their commit history and pull requests.
\revise{In addition, we also examine a random sample of}{In addition, we carry the same analysis on \emph{all}} non‑adopter repositories
to uncover “silent adopters” (repositories that use agents but have no
explicit agent configuration file).

\begin{diffversion}{}
We estimate the sample size needed for  estimating a proportion of adopters in our population \cite{cochran1977sampling}. For an infinite population, the sample size for a proportion is given by:

\begin{equation}
n_0 = \frac{Z^2 \, p(1 - p)}{e^2}
\end{equation}

where \( Z \) is the critical value corresponding to the desired confidence level,
\( p \) is the estimated proportion, and \( e \) is the margin of error. For a finite population of size \( N \), the corrected sample size is:

\begin{equation}
n = \frac{N \, n_0}{N + n_0 - 1}
\end{equation}

\noindent
This correction accounts for the finite population effect and prevents oversampling when \( N \) is not large. Based on an earlier run of the analysis, we set $N$ = 120,000, assuming the worst-case $p = 0.5$, a 99\% confidence interval and a 1\% margin error yields a sample size of 14,575. We increase the sample size to 16,000 as, after the first analysis run, we expect some projects to become file-level adopters, and others to become private.
\end{diffversion}

\paragraph{Pull Request extraction.}  
For each repository we use the GitHub GraphQL API to download up to 10,000 pull requests (PR) that were created on or after \studystartdate \ and their metadata, including their commits (only \revise{19}{45} projects exceed that number). 
This includes the associated branch of the pull request, as well as any labels associated with it. These two data sources are the two sources of information for our PR identification heuristics. If a pull request matches one of our branch or PR heuristics, it is tagged as \emph{AI‑assisted}.  Inside the PR such an AI-assisted PR, we tag all the commits by the PR author as AI-assisted. This allows us to detect commit activity for agents that do not leave markers in commits, but do so in PRs, such as Codex. For each PR, we also determine its merge status, and merge type (merge, rebase, squash), in order to locate the commits that appear in the main branch.


\paragraph{Commit extraction}
For each repository we study, we perform the following steps:
\begin{enumerate}
  \item \textbf{Repository clone (historical snapshot).}  
Given the storage requirements, we perform a \revise{}{\emph{second}} shallow clone of each project up, \revise{to the start of the study period \studystartdate}{with all the commits from \studystartdate to \studydateshort}. At \revise{that time}{of \studystartdate}, coding agents were still very confidential (as confirmed by an earlier run of the analysis that included 2024): established tools like Copilot and Cursor allowed conversations inside the IDE, but started to propose agent-like functionality in 2025. 

  \item \textbf{Commit log parsing.}  
        For scalability reasons, we do not checkout files or compute diffs. From each cloned repository, we run ``git log'' to obtain:
        \begin{inparaenum}
          \item the commit message;
          \item authorship information, including any ``Co‑authored‑by'' trailers (extracted from the commit message);
          \item the list of files (and file paths) affected by the commit, as well as their status;
          \item file‑change statistics (added/deleted lines) for each file.
        \end{inparaenum}

  \item \textbf{Commit‑level heuristics.}  
        We match each commit against our list of commit‑level heuristics
        (e.g., specific message patterns or authorship attributions).  If any
        heuristic matches, the commit is classified as \emph{AI‑assisted}.

  \item \textbf{Linking commits to pull requests.}  
        For each main branch commit, we check whether it appears in at least one
        AI‑assisted PR, and was labelled as AI-assisted there. We take in account the merge type (commits merge with rebase or squash will have distinct commit hashes \cite{bird2009promises}). If so, we label the commit as
        AI‑assisted.  This choice is motivated by the fact that a commit’s provenance is often clearer when it belongs to an AI‑assisted PR (such as for Codex). For ``squash merged'' PR, this adds some imprecision, since the PR is condensed in a single commit which may be only partially AI-assisted. 
        An alternative would be to treat only
        commits that match our commit‑level heuristics as AI‑assisted,
        regardless of PR association.  We adopt the former strategy while
        documenting the trade‑off.

  \item \textbf{Bot exclusions.}  
        Commits authored by known non‑AI bots (e.g., CI or release bots) 
are identified via a manually curated list and labelled. They are excluded from the metrics computed in this study, as we wish to get insights on the adoption of AI coding agents
in relation to human development activities.
The remaining commits are
classified as human-authored.

  \item \textbf{Filtering out noise.}  
        Merge commits, revert commits, bump‑version releases, and other
        housekeeping changes are removed to focus on substantive work.
\end{enumerate}
The result of this pipeline is a per‑repository catalog of AI‑assisted
commits, both at the PR level (PR‑AI‑assisted) and at the individual commit
level.  These data form the basis for the subsequent metrics detailed next.

\subsection{Metrics}
\label{sub:metrics}
For each adopter repository, we use the following metrics. We indicate in which research questions they are used for convenience; the research questions are described next.

\begin{diffversion}{\paragraph{Overall adoption metrics (RQ1)}
We define overall adoption metrics as the proportion of projects exhibiting various indicators of adoption over all projects. We study adoption at the file level (including explicit files and \gitignore files), at the commit level, and overall.}
\end{diffversion}

\begin{diffversion}{\paragraph{Basic project metrics (RQ2, RQ3)}
From our initial dataset, we extract the following metrics for each project: age in years; number of commits; number of pull requests; number of issues; number of contributors (used in RQ2). We also extract the list of GitHub topics that were assigned to the repository, and to whom it belongs (user or organization); we use these metrics in RQ3 and RQ4.
}
\end{diffversion}

\begin{diffversion}{\paragraph{File adoption statistics (RQ2)}
We aggregate the file-level adoption data into three metrics: the total number of agent configuration files we detected, the cumulative size (in lines of code) of all such files, and the total number of commits that touched any such files.
}
\end{diffversion}

\paragraph{Commit‑level statistics \revise{}{(RQ5)}}  
For every commit that has been labeled as AI‑assisted,
or human-authored, we extract the following raw numbers from ``git log'':
the count of lines added and deleted, and the number of files that were
added, modified, or removed. These figures provide a granular view of how much work is performed in each commit.

\paragraph{\revise{Unified adoption}{Adoption} date \revise{}{(RQ4)}}
We \revise{refine}{define} the \emph{adoption} date as the earliest commit that either adds an agent configuration file \revise{(the tool adoption date)}{to the repository} or that is identified as AI-assisted by our heuristics. \revise{We refine the per-tool adoption date in the same manner.}{We define an adoption date for each agent the project uses, and an overall adoption date as the earliest such date.}

\paragraph{File‑type and language classification \revise{}{(RQ3)}}  
Using GitHub Linguist’s file‑extension database, we classify each file in a
commit as either \emph{source code}, \emph{documentation}, 
\emph{configuration}, or \emph{other} (data, images, etc).  
The same lookup is then used to assign a programming language.  Because we only rely on the extension, we avoid the prohibitive cost of parsing every file’s contents at scale.

Sometimes an extension maps to multiple languages (e.g., ``.rs'' could be the Rust programming language, or the deprecated RenderScript language). In important cases, we resolve the ambiguity by choosing the most popular language for that extension in our dataset. The resulting per‑commit tables contain, for each file type and programming language, the same line‑add/delete statistics as above, enabling downstream analyses such as agent activity by language.

\paragraph{\revise{Temporal aggregation}{Commit ratio (RQ2, RQ3)}}  
\revise{For each repository we aggregate commits over our period of interest, which is the activity from the project's \emph{adoption date}, to the end of the study \studyenddate. We then compute the following metrics}{We compute an \emph{AI-assisted commit ratio} by selecting all human-authored and AI-assisted commits from a project's adoption date until the end of the study. We then define the commit ratio as number of AI-assisted commits divided by all AI-assisted and Human-authored commits. We compute an overall commit ratio, as well as a commit ratio for each type of files (\eg source code, documentation). }


\begin{diffversion}{}
\begin{enumerate}
    \item The ratio $\displaystyle
          \frac{\#\text{AI commits}}{\#\text{all commits}}$, 
        a key indicator of how much work is performed by agents.
    \item Analogous ratios for subsets of commits that touch source code,
          documentation, configuration files, or specific programming
          languages.  
    \item Counts of file additions and modifications per period, broken
          down by file type and language, which reveal the magnitude of
          agent‑driven changes over time.
    \item We also compute the ratio of the unique number of files touched by agents over all files touched.
    \item Finally, we compute the fraction of AI-assisted human authors, that is, the fraction of humans that use AI agents as co-authors for at least one commit during the period.
    
\end{enumerate}
\end{diffversion}

\begin{diffversion}{
In addition we use the overall commit ratio to categorize projects in terms of their \emph{commit-level adoption}. For any chosen time window we classify a repository into one of five commit‑level adoption levels:
}
\paragraph{Commit-level adoption} 
Our first use of these metrics is to categorize projects in terms of their \emph{commit-level adoption}. For any chosen time window we classify a repository into one of five commit‑level adoption levels.
\end{diffversion}

\begin{itemize}
  \item \textbf{No Commits}: The period contains no AI-assisted commits.
  \item \textbf{Experimental}: An AI commit ratio below 1 \%.
  \item \textbf{Limited Use} A ratio between 1 \% and 5 \%, but still below 5 \%. 
  \item \textbf{Consistent Use} AI commits represent 5–20 \% of all activity in the period. 
  \item \textbf{Pervasive} More than 20 \% of commits are AI-assisted, indicating widespread use. 
\end{itemize}

\revise{}{Note that when analyzing categories of projects in RQ2 and RQ3, we filter out projects with a low number of post-adoption commits (less than 10), to avoid very high or low commit ratios due to limited data.}

\subsection{Research questions}
\label{sub:questions}

To carry out our study, we formulate the following research questions:

\begin{itemize}
\item \textbf{RQ1. What is the estimated adoption of coding agents on GitHub?} We answer this question in \Cref{sec:adoption} by analyzing file and commit-level adoption overall.
\item \textbf{RQ2. What are the characteristics of projects adopting coding agents?} We answer this question in \Cref{sec:projects} by analyzing the distribution of commit adoption metrics, and its relationship with basic project metrics, such as size and age.
\item \textbf{RQ3. In which context are coding agents used?} We answer this in \Cref{sec:contexts} by looking at adoption for specific organizations, topics, and programming languages.
\item \textbf{RQ4. How has adoption evolved, and which agents are driving it?} We look at the evolution of adoption over time, and at specific coding agents, in \Cref{sec:evolution}.
\item \textbf{RQ5. How large are AI-assisted contributions?} We look at the size of commits (co)-authored by agents, and compare it to commits authored by humans and bots in \Cref{sec:contrib-size}.
\item \textbf{RQ6. What types of contributions are AI-assisted?} We classify a sample of commits in various types, following the \emph{conventional commit} classification, in \Cref{sec:qualitative}.
\end{itemize}

\section{RQ1: estimation of overall adoption}
\label{sec:adoption}

\tabref{tab:overall_adoption} presents our overall statistics on the datasets of projects we study. We study the adoption along several dimensions: (1) at the file level, on all the projects (with and without the \gitignore file); (2) at the commit level, based on a sample of projects; (3) based on the earlier, we extrapolate for overall adoption rates according to two scenarios.

 
\newcommand{\cmark}{\ding{51}}   
\newcommand{\xmark}{\ding{55}}   

\newcommand{\bothcells}[1]{\multicolumn{2}{c}{\text{#1}}}

\begin{table}[ht]
\centering
\caption{Overall adoption statistics. For each statistic, the middle rows show the subset of the data the adoption is measured in.}
\label{tab:overall_adoption}
\footnotesize
\begin{tabular}{l|ccc|rrr}
\toprule
\textbf{Metric} & \textbf{File} & \textbf{Ignored} & \textbf{Commit} & \textbf{Count} & \textbf{Out of} & \textbf{Percent} \\
\midrule
\midrule
\multicolumn{7}{l}{\textbf{File-level tool usage}} \\
File-level use & \cmark &  -- & -- & \ToolUseCount & \TotalProjects & \ToolUsePercent \\
Ignored files (all) & -- & \cmark & -- & \IgnoredToolsAllCount & \TotalProjects & \IgnoredToolsAllPercent \\
Ignored files (only) & \xmark & \cmark & -- & \IgnoredToolsOnlyCount & \TotalProjects & \IgnoredToolsOnlyPercent \\
All file-level use & \bothcells{\cmark (either)} & -- & \AllFileBasedUseCount & \TotalProjects & \AllFileBasedUsePercent \\
\midrule
\multicolumn{7}{l}{\textbf{Commit-level tool usage}} \\
Commit use (file level) & \cmark & -- & \cmark & \CommitUseInToolCount & \CommitUseInToolDenominator & \CommitUseInToolPercent \\
Commit use (all files) & \bothcells{\cmark (either)} & \cmark & \CommitUseAllFileUsersCount & \CommitUseAllFileUsersDenominator & \CommitUseAllFileUsersPercent \\
Commit use, no files & \xmark & \xmark & \cmark & \SampledCommitUseCount & \SampledCommitUseDenominator & \SampledCommitUsePercent \\
High Estimate Commit Users & \xmark & \xmark  & \cmark & $\approx$ \HighEstimateCommitUsers & \ExtrapolationGroupCount & $\frac{\SampledCommitUsePercent}{\CommitUseAllFileUsersPercent}$ \\ 

\midrule
\multicolumn{7}{l}{\textbf{Adoption estimates}} \\
Conservative Adoption Estimate & -- & -- & -- & \AllTotalUsersEstimate & \TotalProjects & \OverallAdoptionRatePercent \\
High Estimate Overall Adoption Rate & -- & -- & -- & $\approx$ \HighEstimateTotalUsersEstimate & \TotalProjects & \HighEstimateOverallAdoptionRatePercent \\

\bottomrule
\end{tabular}
\end{table}

\subsection{File-based adoption}

The first section of \tabref{tab:overall_adoption} shows adoption of coding agents based on file-level heuristics. Using our file-based heuristics, we find that out of our dataset of \TotalProjects projects, \ToolUseCount have files indicating that they use coding agents in their repositories (\ToolUsePercent). In addition, \IgnoredToolsAllCount projects instruct their version control \emph{not to commit} files corresponding to our heuristics via their \gitignore file; of these \IgnoredToolsOnlyCount projects have \emph{no visible coding agent files} in their repository, save their mention in the \gitignore. Summing both together, we find that \AllFileBasedUseCount (\ToolUseCount + \IgnoredToolsOnlyCount) projects reference coding agent files in their repositories, or \AllFileBasedUsePercent.

\subsection{Commit-level adoption}

The second section of \tabref{tab:overall_adoption} presents commit-level statistics. \revise{We investigate commit-level adoption in various ways: (1) we analyze a sample of projects for which we have \emph{no sign of file-level adoption} for commit-level adoption, and (2) we analyze all the projects that \emph{have} signs of file-level adoption for commit-level adoption to get insights on the overlap between these heuristics. }{}

\begin{diffversion}{}
\paragraph{Sample adoption} As mentioned earlier, we selected a large sample of non-adopters at the file level for this analysis, in order to have a 99\% confidence interval and a small 1\% margin of error. The sample size set between analysis runs was 16,000 projects, out of which \SampledCommitUseDenominator could be analyzed by our pipeline on October 31st, and were still non-adopters (out of a final population of non-adopters of \ExtrapolationGroupCount projects--our base population, minus the \AllFileBasedUseCount file-level adopters). Out of our sample, we find that \SampledCommitUseCount, or \SampledCommitUsePercent are commit-level adopters. Surprisingly, this is a higher percentage than the percentage of projects presenting file-level adoption.
\end{diffversion}

\revise{}{\paragraph{Commit adoption only} Out of \ExtrapolationGroupCount projects that do not have traces of coding agents at the file level (our base population, minus the \AllFileBasedUseCount file-level adopters), we find that \SampledCommitUseCount, or \SampledCommitUsePercent, are commit-level adopters. Surprisingly, this is a similar percentage to the percentage of projects presenting file-level adoption.}

\paragraph{Overlap between file-level and commit-level adoption} File-level adoption and commit-level adoption have some overlap, but it is far from complete. As we have just seen, an important proportion of projects have commit-level adoption, but no file-level adoption. The reverse is also true: an important proportion of projects that have file-level adoption would not be detected by commit-level heuristics. Out of visible file-level users, \revise{a little bit more than half}{almost two thirds} (\CommitUseInToolPercent) are detected by commit-level heuristics; adding projects that are visible only by their \gitignore file, the proportion drops to \CommitUseAllFileUsersPercent.

\subsection{Adoption estimates}

The third \revise{and fourth sections}{section} of \tabref{tab:overall_adoption} extrapolates the adoption rate in the entire population according to two scenarios. 

\revise{}{The \emph{conservative scenario} simply sums up the projects that have adoption at the file level (\CommitUseAllFileUsersDenominator) with the ones that have adoption at the commit level (\ExtrapolatedCommitUsers). This brings us to a total of $\approx$ \AllTotalUsersEstimate, or an adoption rate of \OverallAdoptionRatePercent.}

\begin{diffversion}{}
\paragraph{Conservative scenario} For this scenario, we assume the sample estimate (\SampledCommitUsePercent) is accurate, and simply apply it to the entire population of projects that do not present file-level adoption (\ExtrapolationGroupCount projects). This brings us to an estimate of $approx$ \ExtrapolatedCommitUsers (plus or minus 1\%, with a 99\% confidence interval). If we add the projects identified at the file level (\CommitUseAllFileUsersDenominator), we arrive at a total of $\approx$ \AllTotalUsersEstimate, or an adoption rate of \OverallAdoptionRatePercent.
\end{diffversion}

\begin{diffversion}{}
\paragraph{High estimate} For this scenario, we use the sample estimate, but also consider that a large proportion of projects (almost half) that are visible at the file levels, are not visible at the commit level. 
\end{diffversion}

\revise{}{The \emph{high estimate} scenario is based on the observation that a large proportion of projects (almost half) that are visible at the file levels, are not visible at the commit level. }
Therefore, it is possible that projects have adopted coding agents, without having any visible files in their repository, nor visible commits. An example of this would be a user of Claude Code that stores their configuration files in a separate repository such as a \dotfiles repository, and either instructs Claude Code not to sign the commits, or prefers to commit manually (\Cref{sec:all_together} mentions concrete examples). If we assume that the proportion of such projects is comparable to the ones that have file markers, then only $approx$ \CommitUseAllFileUsersPercent of the projects are detected by the commit heuristics. Following this assumption leads to an estimated number of projects adopting coding agents among those that do not present file-level adoption of $\ExtrapolationGroupCount \times \frac{\SampledCommitUsePercent}{\CommitUseAllFileUsersPercent}=\HighEstimateCommitUsers$. This leads to an estimated percentage of \HighEstimateOverallAdoptionRatePercent in the whole population when we also take into account the file-based users.

Needless to say, given that these projects are very difficult to observe, we can not know their true proportion, making this estimate a very rough approximation. Overall, as of \revise{November 2025}{\studyenddate}, we estimate the true adoption to be somewhere in between our conservative estimate and our high estimate. Regardless of its definitive value, the fact that coding agents, a tool category that saw early products in 2024, and more established products in the Spring of 2025, have already reached 22 to 28\% of adoption among GitHub projects \revise{in the Fall of 2025}{barely a year later} points to an extremely rapid adoption.

\section{RQ2: Adoption and project characteristics}
\label{sec:projects}

To dive into the characteristics of adoption, we first examine the distribution of file and commit-level adoption. We then analyze how adoption varies with high-level project metrics, such as size, age, or activity. 

\subsection{Extant of file and commit level adoption}

\newcommand{\binfiletotalfilesSparkline}{\sparklineTikzBars{103.91, 63.70, 16.02, 8.48, 7.90}}
\newcommand{\binfiletotalfileszeroValue}{52.0\%}
\newcommand{\binfiletotalfilesoneValue}{31.8\%}
\newcommand{\binfiletotalfilestwoValue}{8.0\%}
\newcommand{\binfiletotalfilesthreeValue}{4.2\%}
\newcommand{\binfiletotalfilesfourValue}{3.9\%}
\newcommand{\binfiletotallinesSparkline}{\sparklineTikzBars{10.26, 27.63, 28.78, 58.07, 31.96, 18.58, 24.71}}
\newcommand{\binfiletotallineszeroValue}{5.1\%}
\newcommand{\binfiletotallinesoneValue}{13.8\%}
\newcommand{\binfiletotallinestwoValue}{14.4\%}
\newcommand{\binfiletotallinesthreeValue}{29.0\%}
\newcommand{\binfiletotallinesfourValue}{16.0\%}
\newcommand{\binfiletotallinesfiveValue}{9.3\%}
\newcommand{\binfiletotallinessixValue}{12.4\%}
\newcommand{\binfiletotalchangesSparkline}{\sparklineTikzBars{43.54, 72.77, 32.53, 23.59, 27.58}}
\newcommand{\binfiletotalchangeszeroValue}{21.8\%}
\newcommand{\binfiletotalchangesoneValue}{36.4\%}
\newcommand{\binfiletotalchangestwoValue}{16.3\%}
\newcommand{\binfiletotalchangesthreeValue}{11.8\%}
\newcommand{\binfiletotalchangesfourValue}{13.8\%}
\newcommand{\binfileadoptionratioSparkline}{\sparklineTikzBars{69.28, 14.82, 36.88, 41.45, 37.57}}
\newcommand{\binfileadoptionratiozeroValue}{34.6\%}
\newcommand{\binfileadoptionratiooneValue}{7.4\%}
\newcommand{\binfileadoptionratiotwoValue}{18.4\%}
\newcommand{\binfileadoptionratiothreeValue}{20.7\%}
\newcommand{\binfileadoptionratiofourValue}{18.8\%}

\begin{table}[ht]
\scriptsize
\centering
\caption{Binned distributions of file and commit level adoption metrics}
\label{tab:binned_adoption_metrics}
\begin{tabular}{l|rrrrrrr}
\hline
\toprule
\textbf{Metric} & \textbf{Bin 1} & \textbf{Bin 2} & \textbf{Bin 3} & \textbf{Bin 4} & \textbf{Bin 5} & \textbf{Bin 6} & \textbf{Bin 7} \\
\midrule
Total Files & 1 & 2-5 & 6-10 & 11-20 & 21+ & & \\
\quad \binfiletotalfilesSparkline & \binfiletotalfileszeroValue & \binfiletotalfilesoneValue & \binfiletotalfilestwoValue & \binfiletotalfilesthreeValue & \binfiletotalfilesfourValue & & \\
\midrule
Total Lines & 0-10 & 11-50 & 51-100 & 101-250 & 251-500 & 501-1000 & 1001+ \\
\quad \binfiletotallinesSparkline & \binfiletotallineszeroValue & \binfiletotallinesoneValue & \binfiletotallinestwoValue & \binfiletotallinesthreeValue & \binfiletotallinesfourValue & \binfiletotallinesfiveValue & \binfiletotallinessixValue \\
\midrule
Total Changes & 1 & 2-5 & 6-10 & 11-20 & 21+ & & \\
\quad \binfiletotalchangesSparkline & \binfiletotalchangeszeroValue & \binfiletotalchangesoneValue & \binfiletotalchangestwoValue & \binfiletotalchangesthreeValue & \binfiletotalchangesfourValue & & \\
\midrule
Adoption Ratio & None & Experimental & Limited & Consistent & Pervasive & & \\
\quad \binfileadoptionratioSparkline & \binfileadoptionratiozeroValue & \binfileadoptionratiooneValue & \binfileadoptionratiotwoValue & \binfileadoptionratiothreeValue & \binfileadoptionratiofourValue & & \\
\bottomrule
\end{tabular}
\end{table}

\tabref{tab:binned_adoption_metrics} shows statistics on the distribution of file-level adoption in terms of number of files, lines in these files, and number of changes, and of commit-level adoption. This analysis excludes projects which we detect solely via \gitignore, as we cannot measure these metrics for them. We grouped the metrics in bins to ease the interpretation of the data by providing reference points.

\paragraph{Number of files} We see that more than half of projects (\binfiletotalfileszeroValue) have a single rule or guidance file, and more than 85\% of projects have 5 files or less. On the other hand, a small amount of projects have an extensive amount of guidance files (more than 10: \binfiletotalfilesthreeValue; more than 20: \binfiletotalfilesfourValue), indicating that some projects use either sophisticated guidance for their coding agents, or multiple agents.

\paragraph{Number of lines} In terms of the amount of content in the rules and guidance files, we see that a small number of projects have very short guidance (10 lines or less: \binfiletotallineszeroValue). Such a small amount indicates projects having default or very basic configurations. The bulk of the projects have between \revise{11 and 100 lines (\binfiletotallinesoneValue \ and \binfiletotallinestwoValue, $\approx$ 30\%), and}{} 101 and 250 lines (\binfiletotallinesthreeValue). At the extreme end, \revise{close to}{more than} 10\% of projects have more than 1,000 lines of guidance (\binfiletotallinessixValue), once again indicating sophisticated guidance.

\paragraph{Number of changes} In terms of changes, we see that almost a quarter of projects added their guidance file and never changed it (\binfiletotalchangeszeroValue), while most projects occasionally change or add new guidance files (\binfiletotalchangesoneValue). On the other hand, some projects update their guidance regularly (more than 10 additions or changes: \binfiletotalchangesthreeValue; more than 20: \binfiletotalchangesfourValue).

\paragraph{Commit ratio} In terms of commit ratio (shown on the last row), we see that, as shown in \tabref{tab:overall_adoption}, \binfileadoptionratiozeroValue\xspace of file-using projects have no commit-level activity. Of the remainder, we see that a majority of the projects have significant of commit level adoption: \emph{Consistent} users (5 to 20\% of AI-assisted commits) are at \binfileadoptionratiothreeValue; \emph{Pervasive} users are at \revise{an even higher}{a comparable} \binfileadoptionratiofourValue. Of note, some of these ratios may be high for projects having low activity (\eg, a project with a single post-adoption commit, that happens to be AI-assisted, will have a commit ratio of 100\%). For this reason, in the remainder of the analysis, we only consider projects with at least 10 post-adoption commits. 

\begin{figure}[ht]
    \centering
    \includegraphics[width=0.8\textwidth]{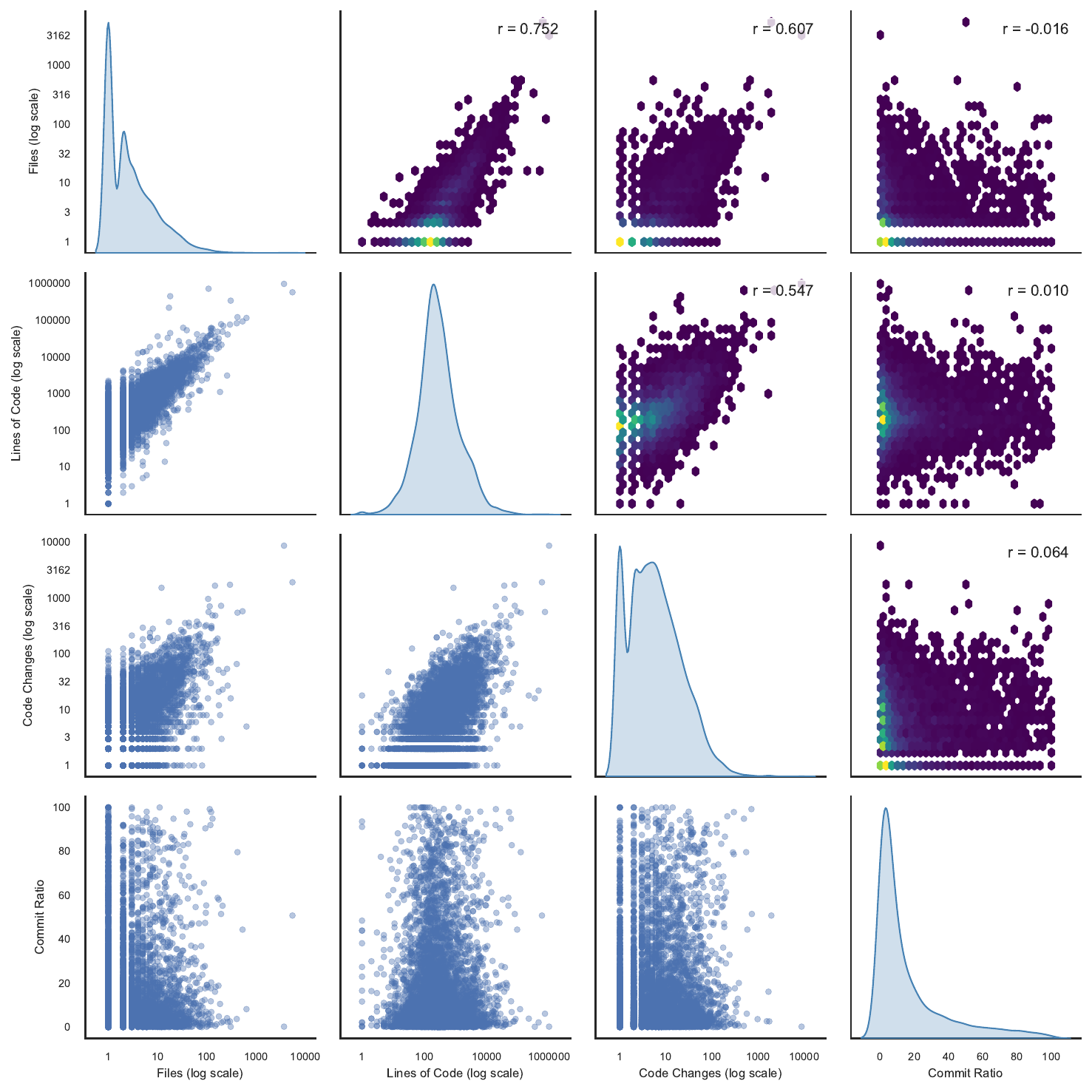}
    \caption{Pairwise relations of file and commit level adoption metrics}
    \label{fig:files-vs-commits}
\end{figure}

\paragraph{Relationship between the metrics} \figref{fig:files-vs-commits} shows the distribution of the four adoption metrics, and how each pair of metrics relate to each other via scatterplots and hexbin plots (showing density); we also compute the correlation. Since many projects have an adoption ratio of 0, we focus only on the ones with a positive commit ratio; further, we include only projects with at least 10 post-adoption commits. The figure shows that there is a high degree of correlation between the file-level adoption metrics: projects that have larger or more numerous files also tend to change them more often. The correlation is strongest between number of files and total content, and somewhat weaker between lines of code and changes. On the other hand, even when filtering for absence of commit ratio and low activity, the correlation between the commit ratio and the file-level activity, if positive, is very low (at best, 0.1 with changes). Clearly, there are more factors at play to explain the relationship between the two categories of metrics. 

Overall, this points to a large proportion of projects that adopt coding agents, but with a modest amount of guidance. A sizeable minority uses a large amount of actively-maintained guidance for coding agents. Whether and how this translates to a large amount of \emph{visible} coding agent activity is another matter: there is a large minority of projects that have a high visible agent usage; however they are not necessarily the ones with the largest amount of guidance.

\subsection{Project characteristics and file-level adoption}

We investigate the characteristics of the projects adopting agents. We first focus on the adoption rate at the file level. As earlier, we count projects that show file-level adoption by usages of \gitignore as adopters. To look into project-level metrics, we split each metric of interest in our dataset into deciles, and we compute the file-level adoption ratio for each decile. This allows us to see whether projects that have larger/smaller metrics tend to have a larger/smaller adoption ratio.

\newcommand{\codelinestooluseSparkline}{\sparklineTikzBars{24.30, 29.10, 32.53, 34.54, 41.06, 45.72, 50.45, 53.26, 56.24, 55.69}}
\newcommand{\contributorstooluseSparkline}{\sparklineTikzBars{33.98, 40.25, 40.69, 38.44, 37.68, 40.35, 40.89, 43.47, 45.73, 62.57}}
\newcommand{\commitstooluseSparkline}{\sparklineTikzBars{35.25, 34.82, 34.45, 35.52, 37.21, 38.62, 40.20, 45.96, 54.25, 66.60}}
\newcommand{\totalissuestooluseSparkline}{\sparklineTikzBars{33.70, 36.35, 38.30, 36.08, 38.14, 40.95, 44.64, 49.71, 71.56}}
\newcommand{\totalpullrequeststooluseSparkline}{\sparklineTikzBars{23.00, 27.40, 30.10, 36.39, 37.28, 39.82, 41.71, 45.88, 53.51, 88.03}}
\newcommand{\ageyearstooluseSparkline}{\sparklineTikzBars{92.31, 50.54, 45.51, 37.74, 36.18, 35.72, 34.77, 30.90, 29.85, 27.86}}
\newcommand{\codelinestooluseDecilezeroBoundary}{5.0K}
\newcommand{\codelinestooluseDecileoneBoundary}{7.5K}
\newcommand{\codelinestooluseDeciletwoBoundary}{11K}
\newcommand{\codelinestooluseDecilethreeBoundary}{15K}
\newcommand{\codelinestooluseDecilefourBoundary}{21K}
\newcommand{\codelinestooluseDecilefiveBoundary}{30K}
\newcommand{\codelinestooluseDecilesixBoundary}{44K}
\newcommand{\codelinestooluseDecilesevenBoundary}{71K}
\newcommand{\codelinestooluseDecileeightBoundary}{129K}
\newcommand{\codelinestooluseDecilenineBoundary}{322K}
\newcommand{\codelinestooluseDecilezero}{6.94\%}
\newcommand{\codelinestooluseDecileone}{8.31\%}
\newcommand{\codelinestooluseDeciletwo}{9.29\%}
\newcommand{\codelinestooluseDecilethree}{9.87\%}
\newcommand{\codelinestooluseDecilefour}{11.73\%}
\newcommand{\codelinestooluseDecilefive}{13.06\%}
\newcommand{\codelinestooluseDecilesix}{14.41\%}
\newcommand{\codelinestooluseDecileseven}{15.22\%}
\newcommand{\codelinestooluseDecileeight}{16.07\%}
\newcommand{\codelinestooluseDecilenine}{15.91\%}
\newcommand{\contributorstooluseDecilezeroBoundary}{0.0}
\newcommand{\contributorstooluseDecileoneBoundary}{1.0}
\newcommand{\contributorstooluseDeciletwoBoundary}{3.0}
\newcommand{\contributorstooluseDecilethreeBoundary}{4.0}
\newcommand{\contributorstooluseDecilefourBoundary}{6.0}
\newcommand{\contributorstooluseDecilefiveBoundary}{10}
\newcommand{\contributorstooluseDecilesixBoundary}{14}
\newcommand{\contributorstooluseDecilesevenBoundary}{21}
\newcommand{\contributorstooluseDecileeightBoundary}{34}
\newcommand{\contributorstooluseDecilenineBoundary}{68}
\newcommand{\contributorstooluseDecilezero}{9.71\%}
\newcommand{\contributorstooluseDecileone}{11.50\%}
\newcommand{\contributorstooluseDeciletwo}{11.63\%}
\newcommand{\contributorstooluseDecilethree}{10.98\%}
\newcommand{\contributorstooluseDecilefour}{10.76\%}
\newcommand{\contributorstooluseDecilefive}{11.53\%}
\newcommand{\contributorstooluseDecilesix}{11.68\%}
\newcommand{\contributorstooluseDecileseven}{12.42\%}
\newcommand{\contributorstooluseDecileeight}{13.07\%}
\newcommand{\contributorstooluseDecilenine}{17.88\%}
\newcommand{\commitstooluseDecilezeroBoundary}{100}
\newcommand{\commitstooluseDecileoneBoundary}{159}
\newcommand{\commitstooluseDeciletwoBoundary}{233}
\newcommand{\commitstooluseDecilethreeBoundary}{324}
\newcommand{\commitstooluseDecilefourBoundary}{446}
\newcommand{\commitstooluseDecilefiveBoundary}{610}
\newcommand{\commitstooluseDecilesixBoundary}{849}
\newcommand{\commitstooluseDecilesevenBoundary}{1.2K}
\newcommand{\commitstooluseDecileeightBoundary}{1.9K}
\newcommand{\commitstooluseDecilenineBoundary}{3.9K}
\newcommand{\commitstooluseDecilezero}{10.07\%}
\newcommand{\commitstooluseDecileone}{9.95\%}
\newcommand{\commitstooluseDeciletwo}{9.84\%}
\newcommand{\commitstooluseDecilethree}{10.15\%}
\newcommand{\commitstooluseDecilefour}{10.63\%}
\newcommand{\commitstooluseDecilefive}{11.04\%}
\newcommand{\commitstooluseDecilesix}{11.49\%}
\newcommand{\commitstooluseDecileseven}{13.13\%}
\newcommand{\commitstooluseDecileeight}{15.50\%}
\newcommand{\commitstooluseDecilenine}{19.03\%}
\newcommand{\totalissuestooluseDecilezeroBoundary}{0.0}
\newcommand{\totalissuestooluseDecileoneBoundary}{3.0}
\newcommand{\totalissuestooluseDeciletwoBoundary}{9.0}
\newcommand{\totalissuestooluseDecilethreeBoundary}{19}
\newcommand{\totalissuestooluseDecilefourBoundary}{35}
\newcommand{\totalissuestooluseDecilefiveBoundary}{60}
\newcommand{\totalissuestooluseDecilesixBoundary}{105}
\newcommand{\totalissuestooluseDecilesevenBoundary}{195}
\newcommand{\totalissuestooluseDecileeightBoundary}{459}
\newcommand{\totalissuestooluseDecilezero}{9.63\%}
\newcommand{\totalissuestooluseDecileone}{10.39\%}
\newcommand{\totalissuestooluseDeciletwo}{10.94\%}
\newcommand{\totalissuestooluseDecilethree}{10.31\%}
\newcommand{\totalissuestooluseDecilefour}{10.90\%}
\newcommand{\totalissuestooluseDecilefive}{11.70\%}
\newcommand{\totalissuestooluseDecilesix}{12.75\%}
\newcommand{\totalissuestooluseDecileseven}{14.20\%}
\newcommand{\totalissuestooluseDecileeight}{20.45\%}
\newcommand{\totalpullrequeststooluseDecilezeroBoundary}{0.0}
\newcommand{\totalpullrequeststooluseDecileoneBoundary}{1.0}
\newcommand{\totalpullrequeststooluseDeciletwoBoundary}{8.0}
\newcommand{\totalpullrequeststooluseDecilethreeBoundary}{22}
\newcommand{\totalpullrequeststooluseDecilefourBoundary}{46}
\newcommand{\totalpullrequeststooluseDecilefiveBoundary}{85}
\newcommand{\totalpullrequeststooluseDecilesixBoundary}{145}
\newcommand{\totalpullrequeststooluseDecilesevenBoundary}{244}
\newcommand{\totalpullrequeststooluseDecileeightBoundary}{435}
\newcommand{\totalpullrequeststooluseDecilenineBoundary}{932}
\newcommand{\totalpullrequeststooluseDecilezero}{6.57\%}
\newcommand{\totalpullrequeststooluseDecileone}{7.83\%}
\newcommand{\totalpullrequeststooluseDeciletwo}{8.60\%}
\newcommand{\totalpullrequeststooluseDecilethree}{10.40\%}
\newcommand{\totalpullrequeststooluseDecilefour}{10.65\%}
\newcommand{\totalpullrequeststooluseDecilefive}{11.38\%}
\newcommand{\totalpullrequeststooluseDecilesix}{11.92\%}
\newcommand{\totalpullrequeststooluseDecileseven}{13.11\%}
\newcommand{\totalpullrequeststooluseDecileeight}{15.29\%}
\newcommand{\totalpullrequeststooluseDecilenine}{25.15\%}
\newcommand{\ageyearstooluseDecilezeroBoundary}{0.0}
\newcommand{\ageyearstooluseDecileoneBoundary}{1.1}
\newcommand{\ageyearstooluseDeciletwoBoundary}{2.0}
\newcommand{\ageyearstooluseDecilethreeBoundary}{2.9}
\newcommand{\ageyearstooluseDecilefourBoundary}{3.8}
\newcommand{\ageyearstooluseDecilefiveBoundary}{4.8}
\newcommand{\ageyearstooluseDecilesixBoundary}{5.8}
\newcommand{\ageyearstooluseDecilesevenBoundary}{7.1}
\newcommand{\ageyearstooluseDecileeightBoundary}{8.7}
\newcommand{\ageyearstooluseDecilenineBoundary}{11}
\newcommand{\ageyearstooluseDecilezero}{26.37\%}
\newcommand{\ageyearstooluseDecileone}{14.44\%}
\newcommand{\ageyearstooluseDeciletwo}{13.00\%}
\newcommand{\ageyearstooluseDecilethree}{10.78\%}
\newcommand{\ageyearstooluseDecilefour}{10.34\%}
\newcommand{\ageyearstooluseDecilefive}{10.21\%}
\newcommand{\ageyearstooluseDecilesix}{9.93\%}
\newcommand{\ageyearstooluseDecileseven}{8.83\%}
\newcommand{\ageyearstooluseDecileeight}{8.53\%}
\newcommand{\ageyearstooluseDecilenine}{7.96\%}


\begin{table}[ht]
\centering
\caption{File adoption statistics versus project-level metrics, by deciles}
\label{tab:file_adoption}
\scriptsize
\begin{tabular}{l|l|l|cccccccccc}
\hline
\toprule\textbf{Metric} & \textbf{Aspect} &  \textbf{Sparkline} & \textbf{D1} & \textbf{D2} & \textbf{D3} & \textbf{D4} & \textbf{D5} & \textbf{D6} & \textbf{D7} & \textbf{D8} & \textbf{D9} & \textbf{D10} \\
\midrule
\midrule
\multirow{2}{*}{LOC} & Deciles &  \multirow{2}{*}{} & \codelinestooluseDecilezeroBoundary & \codelinestooluseDecileoneBoundary & \codelinestooluseDeciletwoBoundary & \codelinestooluseDecilethreeBoundary & \codelinestooluseDecilefourBoundary & \codelinestooluseDecilefiveBoundary & \codelinestooluseDecilesixBoundary & \codelinestooluseDecilesevenBoundary & \codelinestooluseDecileeightBoundary & \codelinestooluseDecilenineBoundary \\
  & File &  \codelinestooluseSparkline & \codelinestooluseDecilezero & \codelinestooluseDecileone & \codelinestooluseDeciletwo & \codelinestooluseDecilethree & \codelinestooluseDecilefour & \codelinestooluseDecilefive & \codelinestooluseDecilesix & \codelinestooluseDecileseven & \codelinestooluseDecileeight & \codelinestooluseDecilenine \\
 \midrule
\multirow{2}{*}{Contributors} & Deciles &  \multirow{2}{*}{} & \contributorstooluseDecilezeroBoundary & \contributorstooluseDecileoneBoundary & \contributorstooluseDeciletwoBoundary & \contributorstooluseDecilethreeBoundary & \contributorstooluseDecilefourBoundary & \contributorstooluseDecilefiveBoundary & \contributorstooluseDecilesixBoundary & \contributorstooluseDecilesevenBoundary & \contributorstooluseDecileeightBoundary & \contributorstooluseDecilenineBoundary \\
  & File &  \contributorstooluseSparkline & \contributorstooluseDecilezero & \contributorstooluseDecileone & \contributorstooluseDeciletwo & \contributorstooluseDecilethree & \contributorstooluseDecilefour & \contributorstooluseDecilefive & \contributorstooluseDecilesix & \contributorstooluseDecileseven & \contributorstooluseDecileeight & \contributorstooluseDecilenine \\
 \midrule
\multirow{2}{*}{Commits} & Deciles &  \multirow{2}{*}{} & \commitstooluseDecilezeroBoundary & \commitstooluseDecileoneBoundary & \commitstooluseDeciletwoBoundary & \commitstooluseDecilethreeBoundary & \commitstooluseDecilefourBoundary & \commitstooluseDecilefiveBoundary & \commitstooluseDecilesixBoundary & \commitstooluseDecilesevenBoundary & \commitstooluseDecileeightBoundary & \commitstooluseDecilenineBoundary \\
  & File &  \commitstooluseSparkline & \commitstooluseDecilezero & \commitstooluseDecileone & \commitstooluseDeciletwo & \commitstooluseDecilethree & \commitstooluseDecilefour & \commitstooluseDecilefive & \commitstooluseDecilesix & \commitstooluseDecileseven & \commitstooluseDecileeight & \commitstooluseDecilenine \\
 \midrule
\multirow{2}{*}{Issues} & Deciles &  \multirow{2}{*}{} & \totalissuestooluseDecilezeroBoundary & \totalissuestooluseDecileoneBoundary & \totalissuestooluseDeciletwoBoundary & \totalissuestooluseDecilethreeBoundary & \totalissuestooluseDecilefourBoundary & \totalissuestooluseDecilefiveBoundary & \totalissuestooluseDecilesixBoundary & \totalissuestooluseDecilesevenBoundary & \totalissuestooluseDecileeightBoundary &  \\
  & File &  \totalissuestooluseSparkline & \totalissuestooluseDecilezero & \totalissuestooluseDecileone & \totalissuestooluseDeciletwo & \totalissuestooluseDecilethree & \totalissuestooluseDecilefour & \totalissuestooluseDecilefive & \totalissuestooluseDecilesix & \totalissuestooluseDecileseven & \totalissuestooluseDecileeight &  \\
 \midrule
\multirow{2}{*}{Pull Request} & Deciles &  \multirow{2}{*}{} & \totalpullrequeststooluseDecilezeroBoundary & \totalpullrequeststooluseDecileoneBoundary & \totalpullrequeststooluseDeciletwoBoundary & \totalpullrequeststooluseDecilethreeBoundary & \totalpullrequeststooluseDecilefourBoundary & \totalpullrequeststooluseDecilefiveBoundary & \totalpullrequeststooluseDecilesixBoundary & \totalpullrequeststooluseDecilesevenBoundary & \totalpullrequeststooluseDecileeightBoundary & \totalpullrequeststooluseDecilenineBoundary \\
  & File &  \totalpullrequeststooluseSparkline & \totalpullrequeststooluseDecilezero & \totalpullrequeststooluseDecileone & \totalpullrequeststooluseDeciletwo & \totalpullrequeststooluseDecilethree & \totalpullrequeststooluseDecilefour & \totalpullrequeststooluseDecilefive & \totalpullrequeststooluseDecilesix & \totalpullrequeststooluseDecileseven & \totalpullrequeststooluseDecileeight & \totalpullrequeststooluseDecilenine \\
 \midrule
\multirow{2}{*}{Age (years)} & Deciles &  \multirow{2}{*}{} & \ageyearstooluseDecilezeroBoundary & \ageyearstooluseDecileoneBoundary & \ageyearstooluseDeciletwoBoundary & \ageyearstooluseDecilethreeBoundary & \ageyearstooluseDecilefourBoundary & \ageyearstooluseDecilefiveBoundary & \ageyearstooluseDecilesixBoundary & \ageyearstooluseDecilesevenBoundary & \ageyearstooluseDecileeightBoundary & \ageyearstooluseDecilenineBoundary \\
  & File &  \ageyearstooluseSparkline & \ageyearstooluseDecilezero & \ageyearstooluseDecileone & \ageyearstooluseDeciletwo & \ageyearstooluseDecilethree & \ageyearstooluseDecilefour & \ageyearstooluseDecilefive & \ageyearstooluseDecilesix & \ageyearstooluseDecileseven & \ageyearstooluseDecileeight & \ageyearstooluseDecilenine \\
 \bottomrule
\end{tabular}
\end{table}

\tabref{tab:file_adoption} presents the deciles and adoption ratio for 6 metrics: size in (non-blank, non-comment) lines of code, age in years, number of contributors, commits, issues, and pull requests. For better readability, we also depict the trends using Sparklines \cite{Tufte2004_SparklineTheoryAndPractice}--word-size graphical representation of trends or distributions of data. While there is a common conception that coding agents work best with ``greenfield'' projects \cite{Osmani2025_AI_Assisted_SE_Productivity}, as it is much easier to write "new" code without needing to integrate it in a large codebase, our results are more nuanced. We describe each metric in turn.

\begin{itemize}

\item \textbf{Lines of Code:} we first check whether larger projects have more or less adoption than smaller projects. We find that, somewhat contrary to expectations, adoption is \emph{higher} for larger projects. Projects in the smallest decile have an adoption rate \revise{hovering around 5\%}{of \codelinestooluseDecilezero}, while the largest project categories (from \codelinestooluseDecilesixBoundary\xspace lines of code and up) more than double the adoption rate \revise{at around 10\%}{(\codelinestooluseDecileseven--\codelinestooluseDecilenine)}. This trend maintains itself for the very largest decile (\codelinestooluseDecilenineBoundary\xspace lines of code and up).

\item \textbf{Contributors:} in terms of contributors, \revise{adoption hovers between 7 and 8\% for all deciles, except the largest decile (69 contributors and up), in which the adoption reaches 11\%}{adoption grows smoothly between \contributorstooluseDecilezero\xspace and \contributorstooluseDecileeight, but grows sharply for the largest decile (\contributorstooluseDecilenineBoundary\xspace and up, \contributorstooluseDecilenine)}. Note that the contributor count (extracted from GitHub) includes all contributors (human, bots, agents). However, this does not impact the trend that the very largest projects have more adoption.

\item \textbf{Commits:} \revise{for commits, both the smallest decile (7.7\%) and the largest deciles (1000+) show larger adoption, with the very largest decile (4,000+ commits) having the largest adoption rate at close to 12\%. This confirms the impression that agents are adopted in more active projects, although they do contribute to this activity themselves. For the smallest decile, this might reflect the fact that new projects starting from scratch (``greenfield'' projects) are indeed more likely to start using agents.}{for commits, adoption hovers at around 10\% for the smallest deciles, before growing smoothly for the largest half of projects, reaching \commitstooluseDecilenine\xspace for the largest decile (\commitstooluseDecilenineBoundary\xspace commits and up)}.

\item \textbf{Issues:} as with other metrics, we find that projects with more issues tend to have a higher adoption rate. This is especially the case of the top decile (\totalissuestooluseDecileeightBoundary+ issues), where the adoption rate is \totalissuestooluseDecileeight. 

\item \textbf{Pull Requests:} the tendency that projects in larger deciles have higher adoption is further confirmed by pull requests, where the difference is even more stark, with the highest decile \revise{jumping to close to 16\% of adoption}{adoption jumping from \totalpullrequeststooluseDecileeight\xspace to \totalpullrequeststooluseDecilenine\xspace}. Note that, depending on their workflows, agents can, and do, contribute pull requests to projects. Thus, it is likely that the stark difference observed is at least partly self-fulfilling, with projects having a heavy usage of coding agents seeing a fast growth in pull requests.

\item \textbf{Age:} While larger metric values are associated with higher usage for the other metrics, age goes strongly in the other direction. Younger projects have a far higher adoption rate than older projects, with a factor of \revise{4}{3} difference between the smallest decile (1 year or less, \ageyearstooluseDecilezero), and the largest (more than a decade, \ageyearstooluseDecilenine). On its own, this does comfort the narrative that coding agents are best used in greenfield projects; however, we see that even old projects see a non-negligible adoption rate\revise{ (close to 5\%)}{almost 8\%}.

\end{itemize}

Thus, we see two trends at play here: clearly, younger projects have a much higher adoption than older projects. On the other hand, there is also a trend towards larger projects with more contributors, commits, issues, and pull requests having a higher adoption rate. This trend is partially self-fulfilling, since agents are contributors, and can contribute commits and pull requests to a project. 

We applied a Chi-square goodness-of-fit test to assess whether the observed values are evenly distributed across categories or tend to be concentrated; in our case, the categories correspond to deciles.
We also computed the effect size associated with the Chi-square test.
The results show statistically significant differences for all metrics (\emph{p-value} $<$ 0.01), indicating that the distributions are not uniform across bins. The effect sizes are \emph{small} for lines of code, contributors, commits, and issues, and \emph{medium} for pull requests and age.

\newcommand{\onepluscodelinestooluseSparkline}{\sparklineTikzBars{8.59, 10.42, 11.87, 12.33, 14.47, 16.22, 18.50, 20.22, 22.64, 23.32}}
\newcommand{\onepluscontributorstooluseSparkline}{\sparklineTikzBars{11.07, 13.38, 13.53, 13.30, 14.03, 15.58, 16.36, 17.67, 19.06, 26.82}}
\newcommand{\onepluscommitstooluseSparkline}{\sparklineTikzBars{8.64, 9.90, 11.25, 12.56, 13.68, 15.20, 16.18, 19.13, 23.35, 28.73}}
\newcommand{\oneplustotalissuestooluseSparkline}{\sparklineTikzBars{10.20, 11.76, 12.13, 12.78, 13.07, 14.03, 15.85, 17.55, 20.94, 30.98}}
\newcommand{\oneplustotalpullrequeststooluseSparkline}{\sparklineTikzBars{6.61, 8.93, 9.66, 11.66, 12.42, 14.20, 16.04, 18.54, 22.31, 38.57}}
\newcommand{\oneplusageyearstooluseSparkline}{\sparklineTikzBars{22.28, 19.76, 17.24, 15.91, 15.33, 15.02, 14.76, 12.95, 12.44, 12.14}}


\newcommand{\twopluscodelinestooluseSparkline}{\sparklineTikzBars{8.10, 9.70, 11.00, 11.34, 13.57, 14.75, 17.16, 19.07, 21.75, 23.12}}
\newcommand{\twopluscontributorstooluseSparkline}{\sparklineTikzBars{10.09, 12.06, 12.25, 12.13, 13.01, 14.05, 15.32, 16.89, 18.52, 26.66}}
\newcommand{\twopluscommitstooluseSparkline}{\sparklineTikzBars{7.46, 9.00, 10.16, 11.41, 12.22, 13.70, 15.75, 18.46, 22.33, 29.12}}
\newcommand{\twoplustotalissuestooluseSparkline}{\sparklineTikzBars{9.28, 11.01, 11.02, 10.90, 12.14, 13.22, 14.93, 16.66, 19.77, 31.22}}
\newcommand{\twoplustotalpullrequeststooluseSparkline}{\sparklineTikzBars{5.52, 7.72, 9.17, 10.59, 11.59, 12.99, 14.67, 17.81, 21.20, 38.45}}
\newcommand{\twoplusageyearstooluseSparkline}{\sparklineTikzBars{19.50, 16.97, 16.50, 15.06, 15.19, 14.75, 14.29, 12.77, 12.24, 12.04}}
To tease apart these two effects, we re-ran the same analysis, excluding projects younger than one year. While the values of all the deciles and their adoption ratios changed slightly, we see the same trends, with higher adoption for higher metric values, and decrease of adoption with age (\eg, lines of code: \onepluscodelinestooluseSparkline; contributors: \onepluscontributorstooluseSparkline\revise{}{;commits: \onepluscommitstooluseSparkline}). The main changes were that, as expected, the very high peak for the first decile in age shrank (\oneplusageyearstooluseSparkline)\revise{; in addition, the adoption in commit deciles rose steadily (\onepluscommitstooluseSparkline), instead of having a higher first decile and a flat progression. This is in line with our hypothesis that this may come from new projects.}{.}

We also carried out the same analysis, excluding projects younger than two years, finding again similar results (\eg: lines of code: \twopluscodelinestooluseSparkline ; issues: \twoplustotalissuestooluseSparkline ; pull requests: \twoplustotalpullrequeststooluseSparkline ). \revise{This comforts the observation that there are two distinct effects: a large effect of age for younger projects; however when removing young projects, there is still a tendency towards higher adoption of coding agents in projects with higher numbers of lines of code, contributors, commits, issues, or pull requests.}{These two analyses comfort the observation that there are two distinct effects: a large effect of age for younger projects; however, even when removing young projects, the trend towards higher adoption of coding agents in projects with higher numbers of lines of code, contributors, commits, issues, or pull requests is still clearly noticeable.}

\subsection{Project characteristics and commit-level adoption}

\tabref{tab:overall_commit_ratios} presents our analysis of the AI-assisted commit ratio of the projects in our dataset, since their adoption of coding agents. For this analysis, we only consider projects that have at least 10 commits since adoption; this avoids cases of projects having a very high commit ratio but little activity (\eg, a project that has a 100\% AI-assisted commit ratio, but a single commit). We carry out this analysis for several categories of projects, and the per-decile analysis we applied for the file-level usage. We use the commit ratio categories defined in \secref{sec:methodology} (Experimental: $< 1\%$; Limited: $< 5\%$; Consistent: $< 20\%$; Pervasive: $>= 20\%$).

\newcommand{\AllFilescodelinesNoneSparkline}{\sparklineTikzBars{32.92, 31.01, 31.53, 28.93, 28.43, 24.04, 24.52, 21.05, 20.78, 24.11}}
\newcommand{\AllFilescodelinesExperimentalSparkline}{\sparklineTikzBars{4.66, 3.91, 5.49, 7.07, 7.81, 9.15, 11.55, 11.81, 14.96, 15.21}}
\newcommand{\AllFilescodelinesLimitedSparkline}{\sparklineTikzBars{15.30, 16.46, 17.72, 20.53, 21.78, 24.71, 22.94, 26.87, 29.43, 29.76}}
\newcommand{\AllFilescodelinesConsistentSparkline}{\sparklineTikzBars{22.03, 24.85, 21.63, 25.44, 24.52, 24.88, 26.10, 27.20, 22.19, 21.11}}
\newcommand{\AllFilescodelinesPervasiveSparkline}{\sparklineTikzBars{25.10, 23.77, 23.63, 18.04, 17.46, 17.22, 14.88, 13.06, 12.64, 9.81}}
\newcommand{\AllFilescontributorsNoneSparkline}{\sparklineTikzBars{42.01, 35.82, 30.82, 28.44, 26.93, 25.60, 22.64, 21.00, 16.90, 12.35}}
\newcommand{\AllFilescontributorsExperimentalSparkline}{\sparklineTikzBars{5.96, 6.92, 7.82, 7.78, 8.59, 9.13, 10.93, 9.44, 9.73, 16.22}}
\newcommand{\AllFilescontributorsLimitedSparkline}{\sparklineTikzBars{12.36, 17.04, 16.79, 19.14, 23.14, 23.38, 23.51, 27.30, 29.84, 35.97}}
\newcommand{\AllFilescontributorsConsistentSparkline}{\sparklineTikzBars{17.56, 16.78, 22.02, 25.31, 23.22, 24.91, 25.44, 27.64, 30.42, 27.56}}
\newcommand{\AllFilescontributorsPervasiveSparkline}{\sparklineTikzBars{22.11, 23.44, 22.56, 19.32, 18.11, 16.98, 17.48, 14.63, 13.11, 7.90}}
\newcommand{\AllFilesageyearsNoneSparkline}{\sparklineTikzBars{22.18, 25.65, 27.31, 27.15, 27.99, 27.47, 27.87, 27.97, 25.76, 28.21}}
\newcommand{\AllFilesageyearsExperimentalSparkline}{\sparklineTikzBars{9.62, 10.48, 10.43, 10.27, 10.18, 7.07, 8.45, 7.55, 8.59, 9.07}}
\newcommand{\AllFilesageyearsLimitedSparkline}{\sparklineTikzBars{21.70, 24.64, 24.44, 24.52, 21.02, 20.08, 22.18, 24.77, 22.55, 19.31}}
\newcommand{\AllFilesageyearsConsistentSparkline}{\sparklineTikzBars{21.94, 24.90, 23.65, 21.02, 24.98, 27.47, 23.51, 25.43, 25.24, 22.08}}
\newcommand{\AllFilesageyearsPervasiveSparkline}{\sparklineTikzBars{24.56, 14.33, 14.17, 17.04, 15.83, 17.91, 17.99, 14.27, 17.87, 21.33}}
\newcommand{\AllFileswithcommitscodelinesNoneSparkline}{\sparklineTikzBars{0.00, 0.00, 0.00, 0.00, 0.00, 0.00, 0.00, 0.00, 0.00, 0.00}}
\newcommand{\AllFileswithcommitscodelinesExperimentalSparkline}{\sparklineTikzBars{6.69, 6.24, 8.17, 10.90, 11.24, 13.05, 14.30, 15.10, 19.07, 20.29}}
\newcommand{\AllFileswithcommitscodelinesLimitedSparkline}{\sparklineTikzBars{22.79, 23.95, 27.47, 27.58, 31.67, 32.01, 31.10, 34.85, 37.57, 38.78}}
\newcommand{\AllFileswithcommitscodelinesConsistentSparkline}{\sparklineTikzBars{33.33, 36.21, 30.19, 37.00, 33.37, 33.03, 34.85, 34.05, 27.47, 28.00}}
\newcommand{\AllFileswithcommitscodelinesPervasiveSparkline}{\sparklineTikzBars{37.19, 33.60, 34.17, 24.52, 23.72, 21.91, 19.75, 16.00, 15.89, 12.93}}
\newcommand{\AllFileswithcommitscontributorsNoneSparkline}{\sparklineTikzBars{0.00, 0.00, 0.00, 0.00, 0.00, 0.00, 0.00, 0.00, 0.00, 0.00}}
\newcommand{\AllFileswithcommitscontributorsExperimentalSparkline}{\sparklineTikzBars{10.28, 11.17, 10.93, 11.45, 12.24, 13.14, 12.73, 11.53, 12.46, 19.50}}
\newcommand{\AllFileswithcommitscontributorsLimitedSparkline}{\sparklineTikzBars{21.31, 26.62, 23.63, 32.83, 29.77, 31.10, 31.19, 36.19, 36.91, 41.27}}
\newcommand{\AllFileswithcommitscontributorsConsistentSparkline}{\sparklineTikzBars{30.28, 27.60, 36.12, 29.97, 34.44, 32.22, 34.40, 36.07, 35.29, 30.95}}
\newcommand{\AllFileswithcommitscontributorsPervasiveSparkline}{\sparklineTikzBars{38.13, 34.61, 29.32, 25.75, 23.55, 23.54, 21.67, 16.21, 15.34, 8.28}}
\newcommand{\AllFileswithcommitsageyearsNoneSparkline}{\sparklineTikzBars{0.00, 0.00, 0.00, 0.00, 0.00, 0.00, 0.00, 0.00, 0.00, 0.00}}
\newcommand{\AllFileswithcommitsageyearsExperimentalSparkline}{\sparklineTikzBars{12.36, 14.09, 14.04, 14.54, 14.00, 9.75, 11.72, 10.48, 11.57, 12.63}}
\newcommand{\AllFileswithcommitsageyearsLimitedSparkline}{\sparklineTikzBars{27.89, 33.15, 32.83, 35.35, 28.78, 27.69, 30.74, 34.40, 30.37, 26.90}}
\newcommand{\AllFileswithcommitsageyearsConsistentSparkline}{\sparklineTikzBars{28.19, 33.48, 32.71, 28.63, 34.42, 37.87, 32.60, 35.31, 34.00, 30.76}}
\newcommand{\AllFileswithcommitsageyearsPervasiveSparkline}{\sparklineTikzBars{31.56, 19.28, 20.43, 21.48, 22.80, 24.70, 24.94, 19.82, 24.07, 29.71}}
\newcommand{\FilelevelcodelinesNoneSparkline}{\sparklineTikzBars{0.00, 0.00, 0.00, 0.00, 0.00, 0.00, 0.00, 0.00, 0.00, 0.00}}
\newcommand{\FilelevelcodelinesExperimentalSparkline}{\sparklineTikzBars{6.15, 6.28, 8.08, 9.74, 10.64, 13.85, 12.82, 14.87, 18.97, 19.85}}
\newcommand{\FilelevelcodelinesLimitedSparkline}{\sparklineTikzBars{22.41, 24.36, 25.64, 26.92, 31.15, 30.90, 31.54, 33.97, 36.15, 38.67}}
\newcommand{\FilelevelcodelinesConsistentSparkline}{\sparklineTikzBars{32.78, 35.26, 28.97, 37.95, 33.08, 32.18, 36.03, 34.74, 28.33, 27.78}}
\newcommand{\FilelevelcodelinesPervasiveSparkline}{\sparklineTikzBars{38.67, 34.10, 37.31, 25.38, 25.13, 23.08, 19.62, 16.41, 16.54, 13.70}}
\newcommand{\FilelevelcontributorsNoneSparkline}{\sparklineTikzBars{0.00, 0.00, 0.00, 0.00, 0.00, 0.00, 0.00, 0.00, 0.00, 0.00}}
\newcommand{\FilelevelcontributorsExperimentalSparkline}{\sparklineTikzBars{9.97, 11.73, 10.18, 11.17, 11.74, 12.10, 12.32, 11.45, 12.28, 18.61}}
\newcommand{\FilelevelcontributorsLimitedSparkline}{\sparklineTikzBars{20.47, 24.81, 23.99, 32.32, 29.65, 31.22, 29.75, 35.91, 35.17, 41.34}}
\newcommand{\FilelevelcontributorsConsistentSparkline}{\sparklineTikzBars{30.01, 26.91, 35.43, 29.78, 33.95, 32.21, 35.12, 35.91, 36.70, 30.94}}
\newcommand{\FilelevelcontributorsPervasiveSparkline}{\sparklineTikzBars{39.55, 36.54, 30.40, 26.73, 24.65, 24.47, 22.80, 16.73, 15.86, 9.11}}
\newcommand{\FilelevelageyearsNoneSparkline}{\sparklineTikzBars{0.00, 0.00, 0.00, 0.00, 0.00, 0.00, 0.00, 0.00, 0.00, 0.00}}
\newcommand{\FilelevelageyearsExperimentalSparkline}{\sparklineTikzBars{12.10, 13.72, 14.38, 13.86, 12.55, 8.84, 10.61, 10.77, 11.54, 12.80}}
\newcommand{\FilelevelageyearsLimitedSparkline}{\sparklineTikzBars{27.31, 32.85, 31.78, 33.83, 29.02, 27.02, 31.03, 34.44, 28.79, 25.99}}
\newcommand{\FilelevelageyearsConsistentSparkline}{\sparklineTikzBars{27.66, 33.11, 33.17, 28.46, 35.16, 37.37, 32.76, 34.97, 34.89, 30.09}}
\newcommand{\FilelevelageyearsPervasiveSparkline}{\sparklineTikzBars{32.93, 20.32, 20.68, 23.85, 23.27, 26.77, 25.60, 19.81, 24.77, 31.11}}
\newcommand{\FilesignoredonlycodelinesNoneSparkline}{\sparklineTikzBars{0.00, 0.00, 0.00, 0.00, 0.00, 0.00, 0.00, 0.00, 0.00, 0.00}}
\newcommand{\FilesignoredonlycodelinesExperimentalSparkline}{\sparklineTikzBars{10.89, 6.93, 9.90, 12.87, 15.84, 17.82, 16.83, 18.81, 20.79, 23.76}}
\newcommand{\FilesignoredonlycodelinesLimitedSparkline}{\sparklineTikzBars{25.74, 22.77, 36.63, 30.69, 39.60, 35.64, 29.70, 40.59, 52.48, 40.59}}
\newcommand{\FilesignoredonlycodelinesConsistentSparkline}{\sparklineTikzBars{36.63, 42.57, 40.59, 37.62, 25.74, 35.64, 39.60, 26.73, 15.84, 29.70}}
\newcommand{\FilesignoredonlycodelinesPervasiveSparkline}{\sparklineTikzBars{26.73, 27.72, 12.87, 18.81, 18.81, 10.89, 13.86, 13.86, 10.89, 5.94}}
\newcommand{\FilesignoredonlycontributorsNoneSparkline}{\sparklineTikzBars{0.00, 0.00, 0.00, 0.00, 0.00, 0.00, 0.00, 0.00, 0.00, 0.00}}
\newcommand{\FilesignoredonlycontributorsExperimentalSparkline}{\sparklineTikzBars{12.41, 6.80, 15.38, 15.62, 16.35, 20.88, 14.29, 14.42, 14.29, 25.74}}
\newcommand{\FilesignoredonlycontributorsLimitedSparkline}{\sparklineTikzBars{27.01, 40.78, 20.51, 33.33, 30.77, 31.87, 38.78, 37.50, 53.06, 40.59}}
\newcommand{\FilesignoredonlycontributorsConsistentSparkline}{\sparklineTikzBars{32.12, 33.01, 43.59, 32.29, 38.46, 30.77, 32.65, 34.62, 23.47, 31.68}}
\newcommand{\FilesignoredonlycontributorsPervasiveSparkline}{\sparklineTikzBars{28.47, 19.42, 20.51, 18.75, 14.42, 16.48, 14.29, 13.46, 9.18, 1.98}}
\newcommand{\FilesignoredonlyageyearsNoneSparkline}{\sparklineTikzBars{0.00, 0.00, 0.00, 0.00, 0.00, 0.00, 0.00, 0.00, 0.00, 0.00}}
\newcommand{\FilesignoredonlyageyearsExperimentalSparkline}{\sparklineTikzBars{11.88, 19.80, 12.50, 13.98, 17.00, 25.45, 16.67, 13.73, 10.78, 12.12}}
\newcommand{\FilesignoredonlyageyearsLimitedSparkline}{\sparklineTikzBars{25.74, 35.64, 45.54, 36.56, 35.00, 32.73, 26.67, 32.35, 42.16, 40.40}}
\newcommand{\FilesignoredonlyageyearsConsistentSparkline}{\sparklineTikzBars{37.62, 26.73, 31.25, 34.41, 29.00, 30.91, 42.22, 37.25, 28.43, 34.34}}
\newcommand{\FilesignoredonlyageyearsPervasiveSparkline}{\sparklineTikzBars{24.75, 17.82, 10.71, 15.05, 19.00, 10.91, 14.44, 16.67, 18.63, 13.13}}
\newcommand{\CommitsonlycodelinesNoneSparkline}{\sparklineTikzBars{0.00, 0.00, 0.00, 0.00, 0.00, 0.00, 0.00, 0.00, 0.00, 0.00}}
\newcommand{\CommitsonlycodelinesExperimentalSparkline}{\sparklineTikzBars{4.88, 8.40, 8.40, 8.61, 9.75, 12.66, 15.46, 17.63, 21.37, 27.70}}
\newcommand{\CommitsonlycodelinesLimitedSparkline}{\sparklineTikzBars{33.09, 30.71, 34.44, 35.37, 38.80, 39.21, 39.21, 40.98, 39.42, 38.28}}
\newcommand{\CommitsonlycodelinesConsistentSparkline}{\sparklineTikzBars{41.49, 40.46, 38.69, 37.76, 35.37, 34.44, 32.16, 29.98, 30.71, 25.10}}
\newcommand{\CommitsonlycodelinesPervasiveSparkline}{\sparklineTikzBars{20.54, 20.44, 18.46, 18.26, 16.08, 13.69, 13.17, 11.41, 8.51, 8.92}}
\newcommand{\CommitsonlycontributorsNoneSparkline}{\sparklineTikzBars{0.00, 0.00, 0.00, 0.00, 0.00, 0.00, 0.00, 0.00, 0.00, 0.00}}
\newcommand{\CommitsonlycontributorsExperimentalSparkline}{\sparklineTikzBars{8.94, 10.86, 10.35, 12.56, 12.12, 12.99, 12.98, 12.64, 15.86, 26.20}}
\newcommand{\CommitsonlycontributorsLimitedSparkline}{\sparklineTikzBars{32.12, 33.20, 34.63, 35.24, 36.65, 38.76, 36.81, 39.45, 40.80, 42.62}}
\newcommand{\CommitsonlycontributorsConsistentSparkline}{\sparklineTikzBars{31.93, 36.68, 37.09, 37.07, 35.98, 36.05, 37.55, 35.88, 32.24, 26.40}}
\newcommand{\CommitsonlycontributorsPervasiveSparkline}{\sparklineTikzBars{27.01, 19.26, 17.93, 15.12, 15.25, 12.20, 12.66, 12.03, 11.10, 4.78}}
\newcommand{\CommitsonlyageyearsNoneSparkline}{\sparklineTikzBars{0.00, 0.00, 0.00, 0.00, 0.00, 0.00, 0.00, 0.00, 0.00, 0.00}}
\newcommand{\CommitsonlyageyearsExperimentalSparkline}{\sparklineTikzBars{16.29, 15.21, 13.89, 15.38, 12.45, 11.28, 10.40, 10.05, 12.46, 17.28}}
\newcommand{\CommitsonlyageyearsLimitedSparkline}{\sparklineTikzBars{39.90, 38.99, 36.64, 35.18, 35.15, 37.20, 37.20, 36.75, 37.38, 34.76}}
\newcommand{\CommitsonlyageyearsConsistentSparkline}{\sparklineTikzBars{29.69, 32.17, 35.36, 34.97, 37.12, 36.38, 36.40, 37.49, 32.18, 34.76}}
\newcommand{\CommitsonlyageyearsPervasiveSparkline}{\sparklineTikzBars{14.12, 13.64, 14.12, 14.47, 15.28, 15.14, 16.00, 15.71, 17.98, 13.19}}
\newcommand{\AllFilesNoneValue}{26.73\%}
\newcommand{\AllFilesExperimentalValue}{9.16\%}
\newcommand{\AllFilesLimitedValue}{22.55\%}
\newcommand{\AllFilesConsistentValue}{24.00\%}
\newcommand{\AllFilesPervasiveValue}{17.56\%}
\newcommand{\AllFileswithcommitsNoneValue}{0.00\%}
\newcommand{\AllFileswithcommitsExperimentalValue}{12.51\%}
\newcommand{\AllFileswithcommitsLimitedValue}{30.78\%}
\newcommand{\AllFileswithcommitsConsistentValue}{32.75\%}
\newcommand{\AllFileswithcommitsPervasiveValue}{23.97\%}
\newcommand{\FilelevelNoneValue}{0.00\%}
\newcommand{\FilelevelExperimentalValue}{12.13\%}
\newcommand{\FilelevelLimitedValue}{30.17\%}
\newcommand{\FilelevelConsistentValue}{32.71\%}
\newcommand{\FilelevelPervasiveValue}{24.99\%}
\newcommand{\FilesignoredonlyNoneValue}{0.00\%}
\newcommand{\FilesignoredonlyExperimentalValue}{15.45\%}
\newcommand{\FilesignoredonlyLimitedValue}{35.45\%}
\newcommand{\FilesignoredonlyConsistentValue}{33.07\%}
\newcommand{\FilesignoredonlyPervasiveValue}{16.04\%}
\newcommand{\CommitsonlyNoneValue}{0.00\%}
\newcommand{\CommitsonlyExperimentalValue}{13.49\%}
\newcommand{\CommitsonlyLimitedValue}{36.95\%}
\newcommand{\CommitsonlyConsistentValue}{34.62\%}
\newcommand{\CommitsonlyPervasiveValue}{14.95\%}

\begin{table}[ht]
\scriptsize
\centering
\caption{Commit adoption across different project categories}
\label{tab:overall_commit_ratios}
\begin{tabular}{l|r|rrrrr}
\hline
\toprule
\textbf{Category} & \textbf{N} & \textbf{None} & \textbf{Experimental} & \textbf{Limited} & \textbf{Consistent} & \textbf{Pervasive} \\
\midrule
All Files & 12,027 & \AllFilesNoneValue & \AllFilesExperimentalValue & \AllFilesLimitedValue & \AllFilesConsistentValue & \AllFilesPervasiveValue \\
\quad by LOC & & \AllFilescodelinesNoneSparkline & \AllFilescodelinesExperimentalSparkline & \AllFilescodelinesLimitedSparkline & \AllFilescodelinesConsistentSparkline & \AllFilescodelinesPervasiveSparkline \\
\quad by Contributors & & \AllFilescontributorsNoneSparkline & \AllFilescontributorsExperimentalSparkline & \AllFilescontributorsLimitedSparkline & \AllFilescontributorsConsistentSparkline & \AllFilescontributorsPervasiveSparkline \\
\quad by Age & & \AllFilesageyearsNoneSparkline & \AllFilesageyearsExperimentalSparkline & \AllFilesageyearsLimitedSparkline & \AllFilesageyearsConsistentSparkline & \AllFilesageyearsPervasiveSparkline \\
\midrule
All Files with commits & 8,812 & \AllFileswithcommitsNoneValue & \AllFileswithcommitsExperimentalValue & \AllFileswithcommitsLimitedValue & \AllFileswithcommitsConsistentValue & \AllFileswithcommitsPervasiveValue \\
\quad by LOC & & \AllFileswithcommitscodelinesNoneSparkline & \AllFileswithcommitscodelinesExperimentalSparkline & \AllFileswithcommitscodelinesLimitedSparkline & \AllFileswithcommitscodelinesConsistentSparkline & \AllFileswithcommitscodelinesPervasiveSparkline \\
\quad by Contributors & & \AllFileswithcommitscontributorsNoneSparkline & \AllFileswithcommitscontributorsExperimentalSparkline & \AllFileswithcommitscontributorsLimitedSparkline & \AllFileswithcommitscontributorsConsistentSparkline & \AllFileswithcommitscontributorsPervasiveSparkline \\
\quad by Age & & \AllFileswithcommitsageyearsNoneSparkline & \AllFileswithcommitsageyearsExperimentalSparkline & \AllFileswithcommitsageyearsLimitedSparkline & \AllFileswithcommitsageyearsConsistentSparkline & \AllFileswithcommitsageyearsPervasiveSparkline \\
\midrule
File level & 7,802 & \FilelevelNoneValue & \FilelevelExperimentalValue & \FilelevelLimitedValue & \FilelevelConsistentValue & \FilelevelPervasiveValue \\
\quad by LOC & & \FilelevelcodelinesNoneSparkline & \FilelevelcodelinesExperimentalSparkline & \FilelevelcodelinesLimitedSparkline & \FilelevelcodelinesConsistentSparkline & \FilelevelcodelinesPervasiveSparkline \\
\quad by Contributors & & \FilelevelcontributorsNoneSparkline & \FilelevelcontributorsExperimentalSparkline & \FilelevelcontributorsLimitedSparkline & \FilelevelcontributorsConsistentSparkline & \FilelevelcontributorsPervasiveSparkline \\
\quad by Age & & \FilelevelageyearsNoneSparkline & \FilelevelageyearsExperimentalSparkline & \FilelevelageyearsLimitedSparkline & \FilelevelageyearsConsistentSparkline & \FilelevelageyearsPervasiveSparkline \\
\midrule
Files ignored only & 1,010 & \FilesignoredonlyNoneValue & \FilesignoredonlyExperimentalValue & \FilesignoredonlyLimitedValue & \FilesignoredonlyConsistentValue & \FilesignoredonlyPervasiveValue \\
\quad by LOC & & \FilesignoredonlycodelinesNoneSparkline & \FilesignoredonlycodelinesExperimentalSparkline & \FilesignoredonlycodelinesLimitedSparkline & \FilesignoredonlycodelinesConsistentSparkline & \FilesignoredonlycodelinesPervasiveSparkline \\
\quad by Contributors & & \FilesignoredonlycontributorsNoneSparkline & \FilesignoredonlycontributorsExperimentalSparkline & \FilesignoredonlycontributorsLimitedSparkline & \FilesignoredonlycontributorsConsistentSparkline & \FilesignoredonlycontributorsPervasiveSparkline \\
\quad by Age & & \FilesignoredonlyageyearsNoneSparkline & \FilesignoredonlyageyearsExperimentalSparkline & \FilesignoredonlyageyearsLimitedSparkline & \FilesignoredonlyageyearsConsistentSparkline & \FilesignoredonlyageyearsPervasiveSparkline \\
\midrule
Commits only & 9,640 & \CommitsonlyNoneValue & \CommitsonlyExperimentalValue & \CommitsonlyLimitedValue & \CommitsonlyConsistentValue & \CommitsonlyPervasiveValue \\
\quad by LOC & & \CommitsonlycodelinesNoneSparkline & \CommitsonlycodelinesExperimentalSparkline & \CommitsonlycodelinesLimitedSparkline & \CommitsonlycodelinesConsistentSparkline & \CommitsonlycodelinesPervasiveSparkline \\
\quad by Contributors & & \CommitsonlycontributorsNoneSparkline & \CommitsonlycontributorsExperimentalSparkline & \CommitsonlycontributorsLimitedSparkline & \CommitsonlycontributorsConsistentSparkline & \CommitsonlycontributorsPervasiveSparkline \\
\quad by Age & & \CommitsonlyageyearsNoneSparkline & \CommitsonlyageyearsExperimentalSparkline & \CommitsonlyageyearsLimitedSparkline & \CommitsonlyageyearsConsistentSparkline & \CommitsonlyageyearsPervasiveSparkline \\
\bottomrule
\end{tabular}
\end{table}

\paragraph{Caveats} Note that this analysis comes with important caveats. In particular, there are good reasons to think the AI-assisted commit ratio is an \emph{under estimate} of the actual one, since: 1) not all agents sign their commits or tag their pull requests; 2) even if they do, this behaviour can often be turned off; and 3) some developers may use agents, but prefer to commit manually. This can cause projects to have a commit ratio of 0 if all the activity is unobserved, or just to reduce the commit ratio if, \eg a developer disables commit signing at a given moment. We expand on this in \secref{sec:discussion}. 

\paragraph{Overall adoption} The first section of \tabref{tab:overall_commit_ratios} shows the distribution of the commit ratio for all the projects that have file adoption markers. Overall, \AllFilesNoneValue \xspace of projects have a ratio of AI-assisted commit of 0 (this is smaller than the value among all adopters, $\approx$46\%, as we filter projects with few commits for this analysis). Out of the projects that have AI-assisted commits, the smallest category is the ``Experimental'' category (\AllFilesExperimentalValue\xspace). The second smallest is the ``Pervasive'' category, the one with the largest degree of AI assistance (\AllFilesPervasiveValue\xspace). The bulk of the remaining projects are roughly equally distributed between the ``Limited'' and ``Consistent'' category.

\newcommand{\AllFileswithcommitsMajorityValue}{5.25\%}
\newcommand{\AllFileswithcommitsExtremeValue}{3.09\%}

Since we can not know for certain whether the projects with a ratio of 0\% have absolutely no AI-assisted commits or not, we filter them out in the subsequent analysis. Of the remaining projects, we see that a significant proportion of projects have a very high usage of coding agents, with almost a quarter (\AllFileswithcommitsPervasiveValue) of such projects having more than one in five commits that are AI-assisted. We find this number to be very high, especially considering that it is likely to be an undercount. We note that some projects have very high ratios of AI assistance at the commit level. In particular, \AllFileswithcommitsMajorityValue \ of projects with non-zero commit adoption have the \emph{majority} of their commits that are AI-assisted; \AllFileswithcommitsExtremeValue \ have three quarters or more of their commits that are AI-assisted.

As in the prior section, we also applied a Chi-square goodness-of-fit test to assess whether the observed values are evenly distributed across the commit ratio categories.
We also computed the effect size associated with the Chi-square test.
The results show statistically significant differences for all metrics (\emph{p-value} $<$ 0.01), confirming that the distributions are not uniform across categories. The effect sizes are \revise{\emph{medium} for all metrics}{\emph{medium} for all files, and \emph{large} for the other metrics}.



\paragraph{Adoption across categories} The bottom three sections of \tabref{tab:overall_commit_ratios} focus on specific categories of projects, and can be directly compared with the second section, since they all filter projects with no AI-assisted commits (the projects detected by the commit-level heuristics have this by definition). Notably, we see that projects that do not have visible files in their repositories (either detected through \gitignore or through commit-level heuristics only) tend to have lower levels of adoption than the ones detected by file-level heuristics. In particular, they have approximately \revise{half}{two thirds} as much projects in the ``Pervasive'' category , while having more projects in the remaining categories. In both cases, the relative growth is larger in the ``Experimental'' and ``Limited'' categories ($\approx$ \revise{15 to 30}{10 to 25}\%), while it is considerably smaller for the ``Consistent'' categories ($\approx$ \revise{5 to 10}{2 to 5}\%).

\paragraph{Variation with project metrics} Looking at the variation of the distribution of commit ratio categories across project metrics in deciles paints a contrasting picture with the file-level adoption. We focus on lines of code and contributors, but we see similar trends in the other metrics (commits, issues, and pull requests). In all cases, larger or more established projects (\eg with more contributors) tend to have a \emph{smaller} ratio of AI-assisted commits: across deciles, the proportion of ``Experimental'' projects rises steadily, while the proportion of ``Pervasive'' projects decreases. The pattern is less clear for the ``Limited'' and ``Consistent'' categories: for file-level adopters, the ``Limited'' category rises with deciles, and the ``Consistent'' category is stable; for projects detected via \gitignore or commits, the ``Limited'' category seems to rise with deciles, while the ``Consistent'' category seems to decrease, although the trends are noisier due to the smaller size of the sample. 

For Age, the trends are less clear: the first decile tends to have a higher proportion of ``Pervasive'' projects (except for commit-only projects). Beyond the first decile, the trend appears to be relatively stable across all the categories.

If we compare to the file-level adoption, we see that if larger or more established projects have more file-level adoption, this does not translate to a higher level of commit adoption. This echoes the earlier finding that the correlation of the metrics is very low, but gives us additional insights. Adopters that are larger or more established seem to use coding agents in more limited settings than smaller projects. In terms of age, we still see a trend that younger projects are more extensive adopters, similar to the file-level adoption. Beyond that, the age of a project in itself does not appear to be correlated to the intensity of use of coding agents. While larger, more established projects have less adoption at the commit level, it is worth noting that a small but significant proportion of these projects have high levels of adoption: for instance, the proportion of ``Pervasive'' file-level adopters in the largest LOC decile \revise{is close to}{is more than} 5\%. 

A final remark is that the proportion of projects that have \emph{no AI-assisted commits} is decreasing with deciles. The effect is particularly visible for the number of contributors. Given that each contributor can have their own setting and workflow for a coding agent, it is possible that projects with more contributors would have a higher probability that at least some of them leave visible traces, leading to this observed behavior. 

\section{RQ3: contexts of use of coding agents}
\label{sec:contexts}

We refine our analysis of adoption by looking at specific contexts: we start with the adoption in specific organizations, before looking at projects that have specific GitHub topics, before analysing adoption by programming languages.


\subsection{Adoption in specific organizations}

\newcommand{\TopTwentyMacroToolUse}{21.13\%}
\newcommand{\TopTwentyMicroToolUse}{17.21\%}

\newcommand{\MicrosoftToolUse}{25.80\%}
\newcommand{\MicrosoftRepoCount}{938}
\newcommand{\GoogleToolUse}{9.86\%}
\newcommand{\GoogleRepoCount}{629}
\newcommand{\ApacheToolUse}{8.99\%}
\newcommand{\ApacheRepoCount}{534}
\newcommand{\AmazonToolUse}{12.74\%}
\newcommand{\AmazonRepoCount}{416}
\newcommand{\HashicorpToolUse}{2.27\%}
\newcommand{\HashicorpRepoCount}{132}
\newcommand{\MetaToolUse}{14.81\%}
\newcommand{\MetaRepoCount}{108}
\newcommand{\NvidiaToolUse}{15.89\%}
\newcommand{\NvidiaRepoCount}{107}
\newcommand{\GrafanaToolUse}{24.47\%}
\newcommand{\GrafanaRepoCount}{94}
\newcommand{\JetbrainsToolUse}{17.24\%}
\newcommand{\JetbrainsRepoCount}{87}
\newcommand{\ElasticToolUse}{13.41\%}
\newcommand{\ElasticRepoCount}{82}
\newcommand{\MozillaToolUse}{10.96\%}
\newcommand{\MozillaRepoCount}{73}
\newcommand{\AutomatticToolUse}{35.38\%}
\newcommand{\AutomatticRepoCount}{65}
\newcommand{\ShopifyToolUse}{14.52\%}
\newcommand{\ShopifyRepoCount}{62}
\newcommand{\AlibabaToolUse}{22.81\%}
\newcommand{\AlibabaRepoCount}{57}
\newcommand{\RustToolUse}{3.85\%}
\newcommand{\RustRepoCount}{52}
\newcommand{\TencentToolUse}{7.84\%}
\newcommand{\TencentRepoCount}{51}
\newcommand{\KubernetesToolUse}{4.00\%}
\newcommand{\KubernetesRepoCount}{50}
\newcommand{\HuggingfaceToolUse}{16.33\%}
\newcommand{\HuggingfaceRepoCount}{49}
\newcommand{\SymfonyToolUse}{4.08\%}
\newcommand{\SymfonyRepoCount}{49}
\newcommand{\IobrokerToolUse}{100.00\%}
\newcommand{\IobrokerRepoCount}{48}


\begin{table}[ht]
\scriptsize
\caption{File-based adoption for top 20 organizations}
\label{tab:organization_adoption}
\begin{tabular}{l|rr@{\hspace{0.5cm}}|@{\hspace{0.5cm}}l|rr}
\toprule
\textbf{Organization} & \textbf{Adoption pct} & \textbf{Repositories} & \textbf{Organization} & \textbf{Adoption pct} & \textbf{Repositories} \\
\midrule
Microsoft & \MicrosoftToolUse & \MicrosoftRepoCount & Mozilla & \MozillaToolUse & \MozillaRepoCount \\
Google & \GoogleToolUse & \GoogleRepoCount & Automattic & \AutomatticToolUse & \AutomatticRepoCount \\
Apache & \ApacheToolUse & \ApacheRepoCount & Shopify & \ShopifyToolUse & \ShopifyRepoCount \\
Amazon & \AmazonToolUse & \AmazonRepoCount & Alibaba & \AlibabaToolUse & \AlibabaRepoCount \\
Hashicorp & \HashicorpToolUse & \HashicorpRepoCount & Rust & \RustToolUse & \RustRepoCount \\
Meta & \MetaToolUse & \MetaRepoCount & Tencent & \TencentToolUse & \TencentRepoCount \\
Nvidia & \NvidiaToolUse & \NvidiaRepoCount & Kubernetes & \KubernetesToolUse & \KubernetesRepoCount \\
Grafana & \GrafanaToolUse & \GrafanaRepoCount & Huggingface & \HuggingfaceToolUse & \HuggingfaceRepoCount \\
Jetbrains & \JetbrainsToolUse & \JetbrainsRepoCount & Symfony & \SymfonyToolUse & \SymfonyRepoCount \\
Elastic & \ElasticToolUse & \ElasticRepoCount & Iobroker & \IobrokerToolUse & \IobrokerRepoCount \\
\bottomrule
\end{tabular}
\end{table}

\newcommand{\industryincrease}[0]{$\approx$ 42 \%\xspace}

Since our dataset contains projects from large organizations, we were interested in seeing the variation in adoption across these. This is especially relevant since projects belonging to large organizations are generally more mature than other GitHub projects. To do so, we started with a list of the top 100 organizations according to gitstar-ranking \footnote{\url{https://gitstar-ranking.com/organizations}}. We then manually grouped some organizations into groups representing companies (e.g., Microsoft, GitHub, Azure, DotNet are all parts of Microsoft).  We then simply compute the file-level adoption ratio for the top twenty of these organizations (for each one, and overall). Over all the repositories in the top twenty organizations, file-level adoption is at \TopTwentyMicroToolUse. For reference, the file-level adoption over all the repositories is \AllFileBasedUsePercent. This means that among open-source repositories of top organizations, there is a higher adoption of coding agents (a relative increase of \industryincrease). 

Looking at specific organizations, \tabref{tab:organization_adoption} shows the top 20 organizations and their file-level adoption rate. We can see that there are large variations, with some organizations well below the average. \revise{Meta is the lowest (\MetaToolUse); however, we know that they have deployed code completion tools internally \cite{murali2023codecompose}. It is possible that the usage in internal repositories and external repositories is different.}{HashiCorp is the lowest (\HashicorpToolUse), followed by Rust, Kubernetes, and Symphony (\RustToolUse, \KubernetesToolUse, and \SymfonyToolUse, respectively)}. Other organizations, on the other hand, are well above the average, notably Microsoft, which, with a few other organizations, has close to double the mean adoption (\MicrosoftToolUse); given that Microsoft, through GitHub Copilot, is itself a provider of a leading coding agent, there is certainly a strong interest in using coding agents in practice. \revise{}{Iobroker is of particular interest, having transitioned all of their projects to use agents; the few repositories we checked adopted Copilot}.

\subsection{Adoption by topic}

\subsubsection{Overall adoption by topic}

\begin{figure}
    \centering
    \includegraphics[width=\linewidth]{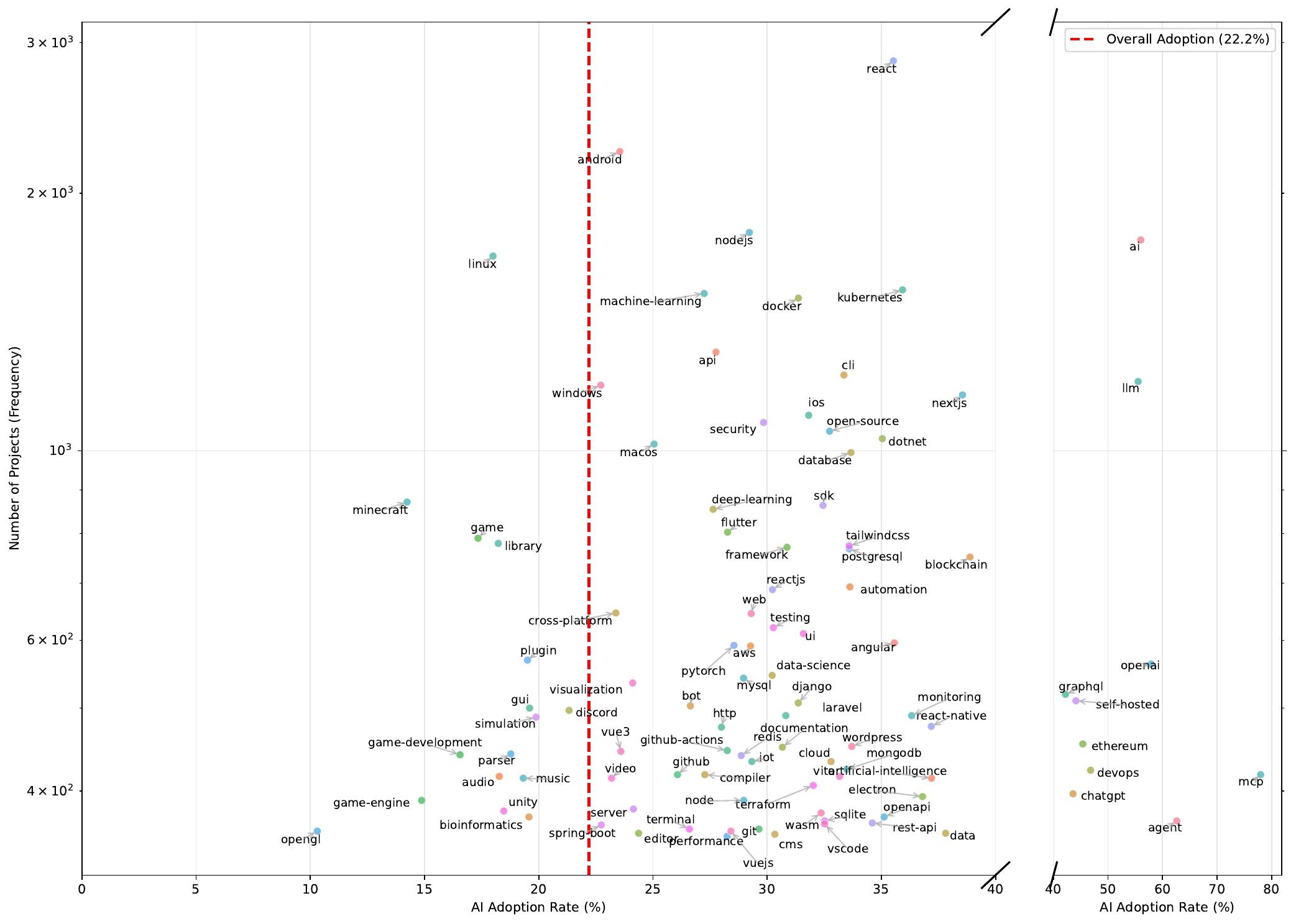}
    \caption{Topics of GitHub projects showing \revise{file-based }{}evidence of the adoption of coding agents, for topics associated to at least 350 projects\revise{ (with file-based evidence of use)}{}. Each topic is plotted with adoption rate, \ie, ratio of projects using agents among projects with the same topic, on the X axis; and the number of projects using agents on the Y axis (log scale).}
    \label{fig:file-adoption-by-topic}
\end{figure}

\Cref{fig:file-adoption-by-topic} shows the GitHub topics of projects with evidence of \revise{file-based}{} adoption of coding agents.
The figure focuses on the top topics in our dataset, \ie, it only shows topics associated to $\geq 350$ projects\revise{ (with file-based evidence of adoption)}; this threshold was chosen to have both a large enough number of project and ensuring the figure is legible. Note that projects can, and often are, associated to multiple topics; hence a single project can contribute to multiple points in this visualization. For each retained topic we show: the adoption rate on the X axis, as a percentage of projects for that topic that use agents; and the number (frequency) of adopting projects on the Y axis, on a log scale. We filter out programming languages (which we see next), and split the figure in two sections to increase legibility.

The average adoption rate in this dataset is around \revise{8}{22}\%, and very scattered around topics (X axis).
Adoption is below average, for topics like \ghtopic{minecraft}, system and network programming topics (\eg, those related to game engines, \revise{\ghtopic{windows}}{\ghtopic{opengl},} \ghtopic{linux}, or network protocols like \ghtopic{http}). Even topics related to AI but on the more foundational side, like \ghtopic{machine-learning} and \ghtopic{deep-learning} are near the average. While some topics are below average, we stress that very few topics have extremely low adoption. \revise{}{To check whether topics below the 350 threshold could have lower adoption, we generated alternative versions of the figure with lower thresholds (down to 100, but far less legible). We found a handful of topics with  between 7\% and 10\% of adoption; it is possible that more niche topics would have lower adoption.}

In between the average and around \revise{17}{35}\% (which is still far lower than the upper-end of the scale, above \revise{40}{70}\%), we find several topics related to Web development (\eg, React, \ghtopic{rest-api}, \ghtopic{wasm}, \ghtopic{electron}), others related to cloud and automation (\ghtopic{cloud}, \ghtopic{docker}, \ghtopic{terraform}, \ghtopic{devops}), as well as many specific database management systems.
We also see a few cross-cutting themes, and technologies, like \ghtopic{open-source}, \ghtopic{dotnet}, and \ghtopic{vscode} (the latter presumably related to coding agents integrated with the popular IDE).

Among the topics with highest adoption ratio (\revise{20\%--60\%}{35--80\%}) some entries can be interpreted as ``dog fooding'', \ie, projects that either develop agents themselves (and use agents for agent development!), \eg: \ghtopic{agent}.
Other high-adoption topics are for neighboring technologies, like \ghtopic{mcp}, but also \ghtopic{ai}, \ghtopic{llm}, \ghtopic{chatgpt}, and \ghtopic{openai}.
Other high-adopter topics are more surprising, like \ghtopic{angular}, \ghtopic{nextjs}, and \ghtopic{ethereum}; the communities behind blockchains and some frontend Web development frameworks appear to be more likely to use coding agents than other technology communities.
Investigating why this is the case is beyond the scope of this paper, and left as interesting future work to pursue.

\subsubsection{Commit-level adoption by topic}

\begin{figure}
    \centering
    \includegraphics[width=0.85\linewidth]{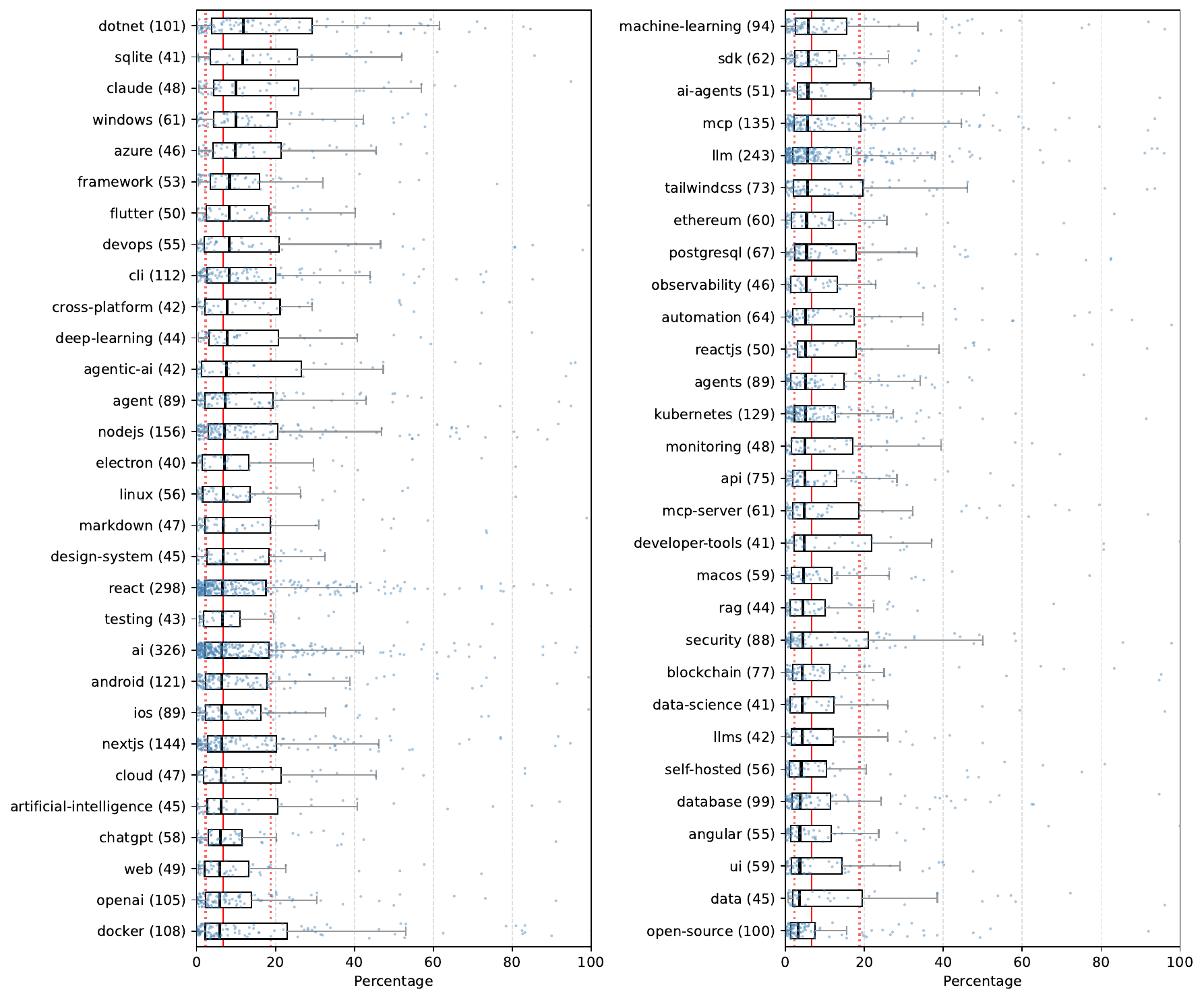}
    \caption{Commit Ratio distribution by topic of commit-level adopters projects. The red lines shows the overall median commit ratio, while the dotted red lines show quartiles.}
    \label{fig:commit-adoption-by-topic}
\end{figure}

\newcommand\medianOverallCommitRatio{6.7\%\xspace}
\newcommand\lowQOverallCommitRatio{2.3\%\xspace}
\newcommand\highQOverallCommitRatio{18.6\%\xspace}

\newcommand\medianCodeCommitRatio{7.8\%\xspace}
\newcommand\lowQCodeCommitRatio{2.6\%\xspace}
\newcommand\highQCodeCommitRatio{22.7\%\xspace}

\Cref{fig:commit-adoption-by-topic} shows the distribution of commit ratios among commit-level adopters categorized by topic. \revise{In black, on the overall data, that is over all topics, we show in dotted lines the first and third quartiles and in solid line the median.
In red, we plot the same lines but for the specific topic.}{A boxplot shows the distribution for a given topic, while a stripplot gives insights on the distribution, such as outliers. The figure also shows the median commit ratio over all commit adopters (\medianOverallCommitRatio) as a red line, and the quartiles (\lowQOverallCommitRatio and \highQOverallCommitRatio) as red dotted lines.} As earlier, we only consider projects with at least 10 post-adoption commits\revise{, and select topics with at least 40 commit-level adopters as a tradeoff between completeness and legibility}.

Unlike \Cref{fig:file-adoption-by-topic} which shows a high discrepancy of file-level adoption depending on the topic, \Cref{fig:commit-adoption-by-topic} shows a more homogeneous commit ratio across topics.
This observation confirms the data from \Cref{fig:files-vs-commits}, where we saw that the correlation between file-level and commit-level activity is very low: similarly, there appears to be little correlation between adoption rate based on topics, and commit ratio based on topics.
We would have thought that topics linked to AI, LLMs or agents had a higher commit ratio than other topics such as \ghtopic{android} or \ghtopic{linux}, but that is not the case\revise{, it is actually the opposite}. 
For instance, the topics \ghtopic{openai} and \ghtopic{llm} have a \revise{30\%}{very high} file-level adoption rate whereas their median commit ratio \revise{at nearly 5\% which is under the}{is at or slightly below the} average median commit ratio among commit-level adopters. One possible reason for that is that developers for these topics have a higher awareness of the workings of coding agents, and are more likely to change their configuration, leading to a decrease in observable traces.

\revise{Instead, \ghtopic{android} which has close to 8\% file-level adoption, has a median commit ratio of nearly 8\%. This is also the case for \ghtopic{linux}, which had below average file-level adoption.} The leaders in commit ratio are \ghtopic{dotnet}, \ghtopic{sqlite}, and \ghtopic{claude} which have a median commit ratio a few percentage higher than the overall median. \revise{There is not a lot of variance by topic except for the \ghtopic{claude} topic which has a significantly higher third quartile compared to other topics.}{Some topics have somewhat lower commit ratios distributions, notably \ghtopic{ui}, \ghtopic{data}, and \ghtopic{open-source}}.

\subsubsection{Takeaway}
The main takeaway of this analysis is that adoption is \emph{broad}: at the file level, it spans a diversity of topics, with topics with very high adoption, but surprisingly few topics with very limited adoption. At the commit level, adoption is more uniform, showing again the diversity of topics in which coding agents are used.

\subsection{Adoption by programming language}

\begin{figure}
    \centering
    \includegraphics[width=0.85\linewidth]{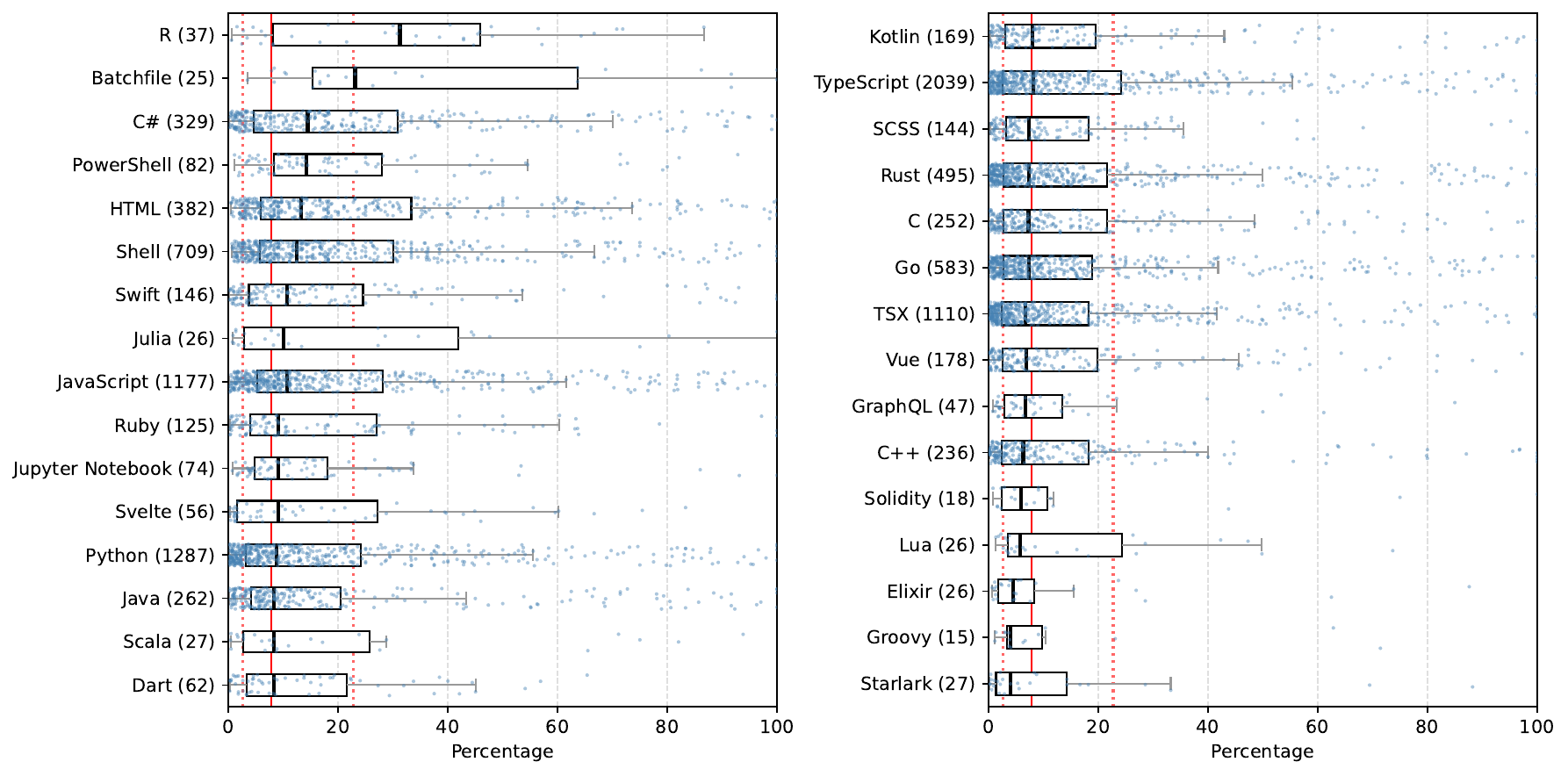}
    \caption{Commit Ratio distribution by programming language of commit-level adopters projects}
    \label{fig:commit-adoption-by-languages}
\end{figure}


For programming languages we only show the commit-level data analysis, as it is much more precise. When computing commit ratio, we count commits that affect files belonging to a given programming language (identified by GitHub linguist). We only consider projects with at least 10 post-adoption commits, but apply this filter for the language of interest.

\Cref{fig:commit-adoption-by-languages} shows the distribution of commit ratios among commit-level adopters categorized by programming language. 
\revise{In black, on the overall data, that is over all programming languages (\ie the commit ratio among commits that change source code), we show in dotted lines the first and third quartiles and in solid line the median. In red, we plot the same lines but for the specific programming language.}{For each language, a boxplot shows the distribution of commit ratios, while a stripplot gives more insights such as on the outliers. The figure also shows the median commit ratio over all commit adopters (\medianCodeCommitRatio) as a red line, and the quartiles (\lowQCodeCommitRatio and \highQCodeCommitRatio) as red dotted lines. Interestingly, the commit ratios on source code are higher than the overall commit ratios, showing that, indeed, coding agents are used for coding, but also for other activities (such as documentation)}.  

The data shows some differences but overall most single programming languages see comparable commit ratios within $\pm$10\%.
Surprisingly, languages such as C, C++ or Rust who were expected to have lower commit ratios, due to their low-level nature, are in the middle of the pack, with close to the same median as all programming languages overall. Widely used languages such as Javascript, Python, Java have also median commit ratios close to the overall median. 

Interestingly, some less popular languages such as Dart, Swift, have commit ratios comparable to the most popular languages\revise{}{, while Kotlin and Julia even have relatively higher commit ratios}. This shows that coding agents have some usefulness beyond the very popular languages with ample training data. However, less established languages such as GraphQL, SCSS, Svelte and Elixir are trailing in median commit ratio compared to the others. Doing this analysis for additional languages would be interesting, but our filters reduce the number of repositories, preventing this analysis.

\revise{}{Some languages are positive outliers, notably R and Batch, although the relatively low number of projects prevents us to make strong conclusions.} \revise{The leaders}{Among more established languages, the leaders are C\#}, Shell, HTML, and PowerShell which have medians up to nearly 8\% above the overall median. Shell languages are good use cases for coding agents: they typically are small-size scripts, in languages that are common, yet are used less frequently day-to-day by developers. 

Notice that there is a high variance in commit ratio for most programming languages. Most commit-level adopters have commit ratios under the 20\% but there are always some projects with very high commit ratio from 40\% up to 100\%. This is reflected in the upper quartile, which tends to be higher than in \Cref{fig:commit-adoption-by-topic}. 

All in all, this analysis also comforts the idea that adoption is broad: it is not limited to a small subset of very popular languages, but extends also to less popular ones.
\section{RQ4: evolution of adoption over time, and tool-specific adoption}
\label{sec:evolution}

In this section, we investigate the evolution of adoption over time, as well as the adoption of specific tools. 

\subsection{Evolution of adoption over time}

\figref{fig:overall-adoption} presents the cumulative adoption of coding accross all the projects for which we could identify it. We record as adoption date the earliest date for which we have evidence of adoption, either through files or through commits. We can see several inflections point on the curve:


\newcommand{\totaladoptingprojects}{26\,883}
\newcommand{\firstadoptiondate}{2025-01-01}
\newcommand{\lastadoptiondate}{2026-02-23}
\newcommand{\adoptiontimespandays}{418}

\newcommand{\totaltooladoptions}{42\,060}
\newcommand{\uniquetools}{54}
\newcommand{\displayedtools}{54}
\newcommand{\toptoolname}{Copilot}
\newcommand{\toptoolcount}{13\,890}
\newcommand{\copilotcount}{13\,890}
\newcommand{\claudecodecount}{12\,053}
\newcommand{\genericcount}{4\,416}
\newcommand{\cursorcount}{2\,965}
\newcommand{\codexcount}{2\,147}
\newcommand{\geminicount}{1\,256}
\newcommand{\coderabbitcount}{1\,249}
\newcommand{\julescount}{873}
\newcommand{\aidercount}{520}
\newcommand{\devincount}{421}
\newcommand{\ampcount}{263}
\newcommand{\clinecount}{242}
\newcommand{\opencodecount}{227}
\newcommand{\sourcerycount}{196}
\newcommand{\windsurfcount}{172}
\newcommand{\juniecount}{158}
\newcommand{\openhandscount}{112}
\newcommand{\serenacount}{98}
\newcommand{\kirocount}{94}
\newcommand{\qwencodercount}{65}
\newcommand{\othertoolscount}{635}

\newcommand{\totalorganizationadoptions}{42\,060}
\newcommand{\uniqueorganizations}{18\,325}
\newcommand{\displayedorganizations}{35}
\newcommand{\toporganizationname}{microsoft}
\newcommand{\toporganizationcount}{619}
\newcommand{\microsoftcount}{619}
\newcommand{\apachecount}{240}
\newcommand{\openshiftcount}{159}
\newcommand{\googlecount}{156}
\newcommand{\amazoncount}{100}
\newcommand{\getsentrycount}{93}
\newcommand{\automatticcount}{92}
\newcommand{\grafanacount}{90}
\newcommand{\datadogcount}{89}
\newcommand{\pulumicount}{87}
\newcommand{\nvidiacount}{79}
\newcommand{\elasticcount}{70}
\newcommand{\micronautcount}{70}
\newcommand{\iobrokercount}{63}
\newcommand{\mattermostcount}{54}
\newcommand{\tencentcount}{50}
\newcommand{\huggingfacecount}{50}
\newcommand{\reowncomcount}{44}
\newcommand{\langchainaicount}{42}
\newcommand{\scimlcount}{40}
\newcommand{\shiguredocount}{37}
\newcommand{\shopifycount}{36}
\newcommand{\mongodbcount}{36}
\newcommand{\capgocount}{36}
\newcommand{\getstreamcount}{36}

\newcommand{\minsubsetsize}{40}
\newcommand{\minfrequency}{250}
\newcommand{\totalprojects}{26,883}
\newcommand{\projectsusingOneTools}{17,774}
\newcommand{\projectsusingTwoTools}{5,570}
\newcommand{\projectsusingThreeTools}{2,072}
\newcommand{\projectsusingFourPlusTools}{1,467}
\newcommand{\projectsusingFourTools}{874}
\newcommand{\projectsusingFiveTools}{340}
\newcommand{\projectsusingSixTools}{153}
\newcommand{\projectsusingSevenTools}{55}
\newcommand{\projectsusingEightTools}{24}
\newcommand{\projectsusingNineTools}{13}
\newcommand{\projectsusingTenTools}{2}
\newcommand{\projectsusingElevenTools}{1}
\newcommand{\projectsusingTwelveTools}{2}
\newcommand{\projectsusingThirteenTools}{3}
\newcommand{\meantoolsperproject}{1.564036751850612}
\newcommand{\mediantoolsperproject}{1.0}
\newcommand{\maxtoolsperproject}{17}
\newcommand{\mintoolsperproject}{1}
\newcommand{\projectsafterfiltering}{26\,703}
\newcommand{\projectsfilteredout}{180}
\newcommand{\toolsanalyzed}{11}
\newcommand{\totalintersections}{703}
\newcommand{\projectsinshownintersections}{24\,632}
\newcommand{\projectsinfilteredintersections}{2\,071}
\newcommand{\projectswithmultipletools}{59}
\newcommand{\largestintersectionsize}{8\,346}
\newcommand{\largestintersectiontools}{Copilot}
\newcommand{\intersectioncopilot}{Copilot (8346)}
\newcommand{\intersectioncopilotsize}{8\,346}
\newcommand{\intersectioncopilottoolcount}{1}
\newcommand{\intersectionclaudecode}{Claude\_Code (5656)}
\newcommand{\intersectionclaudecodesize}{5\,656}
\newcommand{\intersectionclaudecodetoolcount}{1}
\newcommand{\intersectionclaudecodecopilot}{Claude\_Code $\cap$ Copilot (1732)}
\newcommand{\intersectionclaudecodecopilotsize}{1\,732}
\newcommand{\intersectionclaudecodecopilottoolcount}{2}
\newcommand{\intersectiongeneric}{Generic (837)}
\newcommand{\intersectiongenericsize}{837}
\newcommand{\intersectiongenerictoolcount}{1}
\newcommand{\intersectioncursor}{Cursor (833)}
\newcommand{\intersectioncursorsize}{833}
\newcommand{\intersectioncursortoolcount}{1}
\newcommand{\intersectionclaudecodegeneric}{Claude\_Code $\cap$ Generic (731)}
\newcommand{\intersectionclaudecodegenericsize}{731}
\newcommand{\intersectionclaudecodegenerictoolcount}{2}
\newcommand{\intersectioncodex}{Codex (558)}
\newcommand{\intersectioncodexsize}{558}
\newcommand{\intersectioncodextoolcount}{1}
\newcommand{\intersectionclaudecodecopilotgeneric}{Claude\_Code $\cap$ Copilot $\cap$ Generic (440)}
\newcommand{\intersectionclaudecodecopilotgenericsize}{440}
\newcommand{\intersectionclaudecodecopilotgenerictoolcount}{3}
\newcommand{\intersectioncopilotgeneric}{Copilot $\cap$ Generic (417)}
\newcommand{\intersectioncopilotgenericsize}{417}
\newcommand{\intersectioncopilotgenerictoolcount}{2}
\newcommand{\intersectionclaudecodecursor}{Claude\_Code $\cap$ Cursor (401)}
\newcommand{\intersectionclaudecodecursorsize}{401}
\newcommand{\intersectionclaudecodecursortoolcount}{2}
\newcommand{\intersectiongemini}{Gemini (344)}
\newcommand{\intersectiongeminisize}{344}
\newcommand{\intersectiongeminitoolcount}{1}
\newcommand{\intersectioncoderabbit}{Coderabbit (340)}
\newcommand{\intersectioncoderabbitsize}{340}
\newcommand{\intersectioncoderabbittoolcount}{1}
\newcommand{\intersectionjules}{Jules (286)}
\newcommand{\intersectionjulessize}{286}
\newcommand{\intersectionjulestoolcount}{1}
\newcommand{\intersectioncopilotcursor}{Copilot $\cap$ Cursor (228)}
\newcommand{\intersectioncopilotcursorsize}{228}
\newcommand{\intersectioncopilotcursortoolcount}{2}
\newcommand{\intersectionclaudecodecopilotcursor}{Claude\_Code $\cap$ Copilot $\cap$ Cursor (205)}
\newcommand{\intersectionclaudecodecopilotcursorsize}{205}
\newcommand{\intersectionclaudecodecopilotcursortoolcount}{3}
\newcommand{\intersectionaider}{Aider (204)}
\newcommand{\intersectionaidersize}{204}
\newcommand{\intersectionaidertoolcount}{1}
\newcommand{\intersectionclaudecodecodex}{Claude\_Code $\cap$ Codex (179)}
\newcommand{\intersectionclaudecodecodexsize}{179}
\newcommand{\intersectionclaudecodecodextoolcount}{2}
\newcommand{\intersectioncopilotcoderabbit}{Copilot $\cap$ Coderabbit (168)}
\newcommand{\intersectioncopilotcoderabbitsize}{168}
\newcommand{\intersectioncopilotcoderabbittoolcount}{2}
\newcommand{\intersectioncopilotcodex}{Copilot $\cap$ Codex (163)}
\newcommand{\intersectioncopilotcodexsize}{163}
\newcommand{\intersectioncopilotcodextoolcount}{2}
\newcommand{\intersectioncodexgeneric}{Codex $\cap$ Generic (132)}
\newcommand{\intersectioncodexgenericsize}{132}
\newcommand{\intersectioncodexgenerictoolcount}{2}
\newcommand{\intersectioncopilotgemini}{Copilot $\cap$ Gemini (128)}
\newcommand{\intersectioncopilotgeminisize}{128}
\newcommand{\intersectioncopilotgeminitoolcount}{2}
\newcommand{\intersectionclaudecodecursorgeneric}{Claude\_Code $\cap$ Cursor $\cap$ Generic (126)}
\newcommand{\intersectionclaudecodecursorgenericsize}{126}
\newcommand{\intersectionclaudecodecursorgenerictoolcount}{3}
\newcommand{\intersectionclaudecodecopilotcodex}{Claude\_Code $\cap$ Copilot $\cap$ Codex (120)}
\newcommand{\intersectionclaudecodecopilotcodexsize}{120}
\newcommand{\intersectionclaudecodecopilotcodextoolcount}{3}
\newcommand{\intersectionclaudecodecopilotcursorgeneric}{Claude\_Code $\cap$ Copilot $\cap$ Cursor $\cap$ Generic (117)}
\newcommand{\intersectionclaudecodecopilotcursorgenericsize}{117}
\newcommand{\intersectionclaudecodecopilotcursorgenerictoolcount}{4}
\newcommand{\intersectionclaudecodecopilotcodexgeneric}{Claude\_Code $\cap$ Copilot $\cap$ Codex $\cap$ Generic (112)}
\newcommand{\intersectionclaudecodecopilotcodexgenericsize}{112}
\newcommand{\intersectionclaudecodecopilotcodexgenerictoolcount}{4}
\newcommand{\intersectionclaudecodecoderabbit}{Claude\_Code $\cap$ Coderabbit (104)}
\newcommand{\intersectionclaudecodecoderabbitsize}{104}
\newcommand{\intersectionclaudecodecoderabbittoolcount}{2}
\newcommand{\intersectiondevin}{Devin (103)}
\newcommand{\intersectiondevinsize}{103}
\newcommand{\intersectiondevintoolcount}{1}
\newcommand{\intersectionclaudecodecodexgeneric}{Claude\_Code $\cap$ Codex $\cap$ Generic (96)}
\newcommand{\intersectionclaudecodecodexgenericsize}{96}
\newcommand{\intersectionclaudecodecodexgenerictoolcount}{3}
\newcommand{\intersectionclaudecodegemini}{Claude\_Code $\cap$ Gemini (89)}
\newcommand{\intersectionclaudecodegeminisize}{89}
\newcommand{\intersectionclaudecodegeminitoolcount}{2}
\newcommand{\intersectioncopilotjules}{Copilot $\cap$ Jules (83)}
\newcommand{\intersectioncopilotjulessize}{83}
\newcommand{\intersectioncopilotjulestoolcount}{2}
\newcommand{\intersectionclaudecodecopilotcoderabbit}{Claude\_Code $\cap$ Copilot $\cap$ Coderabbit (72)}
\newcommand{\intersectionclaudecodecopilotcoderabbitsize}{72}
\newcommand{\intersectionclaudecodecopilotcoderabbittoolcount}{3}
\newcommand{\intersectioncopilotcodexgeneric}{Copilot $\cap$ Codex $\cap$ Generic (72)}
\newcommand{\intersectioncopilotcodexgenericsize}{72}
\newcommand{\intersectioncopilotcodexgenerictoolcount}{3}
\newcommand{\intersectioncursorgeneric}{Cursor $\cap$ Generic (65)}
\newcommand{\intersectioncursorgenericsize}{65}
\newcommand{\intersectioncursorgenerictoolcount}{2}
\newcommand{\intersectionsourcery}{Sourcery (64)}
\newcommand{\intersectionsourcerysize}{64}
\newcommand{\intersectionsourcerytoolcount}{1}
\newcommand{\intersectionclaudecodeaider}{Claude\_Code $\cap$ Aider (62)}
\newcommand{\intersectionclaudecodeaidersize}{62}
\newcommand{\intersectionclaudecodeaidertoolcount}{2}
\newcommand{\intersectiongeminijules}{Gemini $\cap$ Jules (60)}
\newcommand{\intersectiongeminijulessize}{60}
\newcommand{\intersectiongeminijulestoolcount}{2}
\newcommand{\intersectionclaudecodecopilotgemini}{Claude\_Code $\cap$ Copilot $\cap$ Gemini (60)}
\newcommand{\intersectionclaudecodecopilotgeminisize}{60}
\newcommand{\intersectionclaudecodecopilotgeminitoolcount}{3}
\newcommand{\intersectionclaudecodejules}{Claude\_Code $\cap$ Jules (57)}
\newcommand{\intersectionclaudecodejulessize}{57}
\newcommand{\intersectionclaudecodejulestoolcount}{2}
\newcommand{\intersectionjunie}{Junie (56)}
\newcommand{\intersectionjuniesize}{56}
\newcommand{\intersectionjunietoolcount}{1}
\newcommand{\intersectioncline}{Cline (52)}
\newcommand{\intersectionclinesize}{52}
\newcommand{\intersectionclinetoolcount}{1}
\newcommand{\intersectionclaudecodecopilotcodexcursorgeneric}{Claude\_Code $\cap$ Copilot $\cap$ Codex $\cap$ Cursor $\cap$ Generic (50)}
\newcommand{\intersectionclaudecodecopilotcodexcursorgenericsize}{50}
\newcommand{\intersectionclaudecodecopilotcodexcursorgenerictoolcount}{5}
\newcommand{\intersectionamp}{Amp (48)}
\newcommand{\intersectionampsize}{48}
\newcommand{\intersectionamptoolcount}{1}
\newcommand{\intersectionclaudecodegeminigeneric}{Claude\_Code $\cap$ Gemini $\cap$ Generic (47)}
\newcommand{\intersectionclaudecodegeminigenericsize}{47}
\newcommand{\intersectionclaudecodegeminigenerictoolcount}{3}
\newcommand{\intersectioncopilotcline}{Copilot $\cap$ Cline (46)}
\newcommand{\intersectioncopilotclinesize}{46}
\newcommand{\intersectioncopilotclinetoolcount}{2}
\newcommand{\intersectionclaudecodecopilotgeminigeneric}{Claude\_Code $\cap$ Copilot $\cap$ Gemini $\cap$ Generic (45)}
\newcommand{\intersectionclaudecodecopilotgeminigenericsize}{45}
\newcommand{\intersectionclaudecodecopilotgeminigenerictoolcount}{4}
\newcommand{\intersectionclaudecodedevin}{Claude\_Code $\cap$ Devin (45)}
\newcommand{\intersectionclaudecodedevinsize}{45}
\newcommand{\intersectionclaudecodedevintoolcount}{2}
\newcommand{\intersectioncopilotcursorgeneric}{Copilot $\cap$ Cursor $\cap$ Generic (43)}
\newcommand{\intersectioncopilotcursorgenericsize}{43}
\newcommand{\intersectioncopilotcursorgenerictoolcount}{3}
\newcommand{\intersectionclaudecodeamp}{Claude\_Code $\cap$ Amp (41)}
\newcommand{\intersectionclaudecodeampsize}{41}
\newcommand{\intersectionclaudecodeamptoolcount}{2}
\newcommand{\intersectionclaudecodecopilotjules}{Claude\_Code $\cap$ Copilot $\cap$ Jules (37)}
\newcommand{\intersectionclaudecodecopilotjulessize}{37}
\newcommand{\intersectionclaudecodecopilotjulestoolcount}{3}
\newcommand{\intersectionclaudecodecopilotaider}{Claude\_Code $\cap$ Copilot $\cap$ Aider (37)}
\newcommand{\intersectionclaudecodecopilotaidersize}{37}
\newcommand{\intersectionclaudecodecopilotaidertoolcount}{3}
\newcommand{\intersectioncopilotaider}{Copilot $\cap$ Aider (37)}
\newcommand{\intersectioncopilotaidersize}{37}
\newcommand{\intersectioncopilotaidertoolcount}{2}
\newcommand{\intersectionopenhands}{OpenHands (37)}
\newcommand{\intersectionopenhandssize}{37}
\newcommand{\intersectionopenhandstoolcount}{1}
\newcommand{\intersectioncopilotsourcery}{Copilot $\cap$ Sourcery (34)}
\newcommand{\intersectioncopilotsourcerysize}{34}
\newcommand{\intersectioncopilotsourcerytoolcount}{2}
\newcommand{\intersectioncursorcoderabbit}{Cursor $\cap$ Coderabbit (33)}
\newcommand{\intersectioncursorcoderabbitsize}{33}
\newcommand{\intersectioncursorcoderabbittoolcount}{2}
\newcommand{\intersectionjulesgeneric}{Jules $\cap$ Generic (32)}
\newcommand{\intersectionjulesgenericsize}{32}
\newcommand{\intersectionjulesgenerictoolcount}{2}
\newcommand{\intersectionclaudecodecodexcursorgeneric}{Claude\_Code $\cap$ Codex $\cap$ Cursor $\cap$ Generic (31)}
\newcommand{\intersectionclaudecodecodexcursorgenericsize}{31}
\newcommand{\intersectionclaudecodecodexcursorgenerictoolcount}{4}
\newcommand{\intersectionclaudecodecodexcursor}{Claude\_Code $\cap$ Codex $\cap$ Cursor (31)}
\newcommand{\intersectionclaudecodecodexcursorsize}{31}
\newcommand{\intersectionclaudecodecodexcursortoolcount}{3}
\newcommand{\intersectiongeminigeneric}{Gemini $\cap$ Generic (30)}
\newcommand{\intersectiongeminigenericsize}{30}
\newcommand{\intersectiongeminigenerictoolcount}{2}
\newcommand{\intersectionwindsurf}{Windsurf (30)}
\newcommand{\intersectionwindsurfsize}{30}
\newcommand{\intersectionwindsurftoolcount}{1}
\newcommand{\meanintersectionsize}{417}
\newcommand{\medianintersectionsize}{83}

\begin{figure}[ht]
    \centering
    \includegraphics[width=0.9\textwidth]{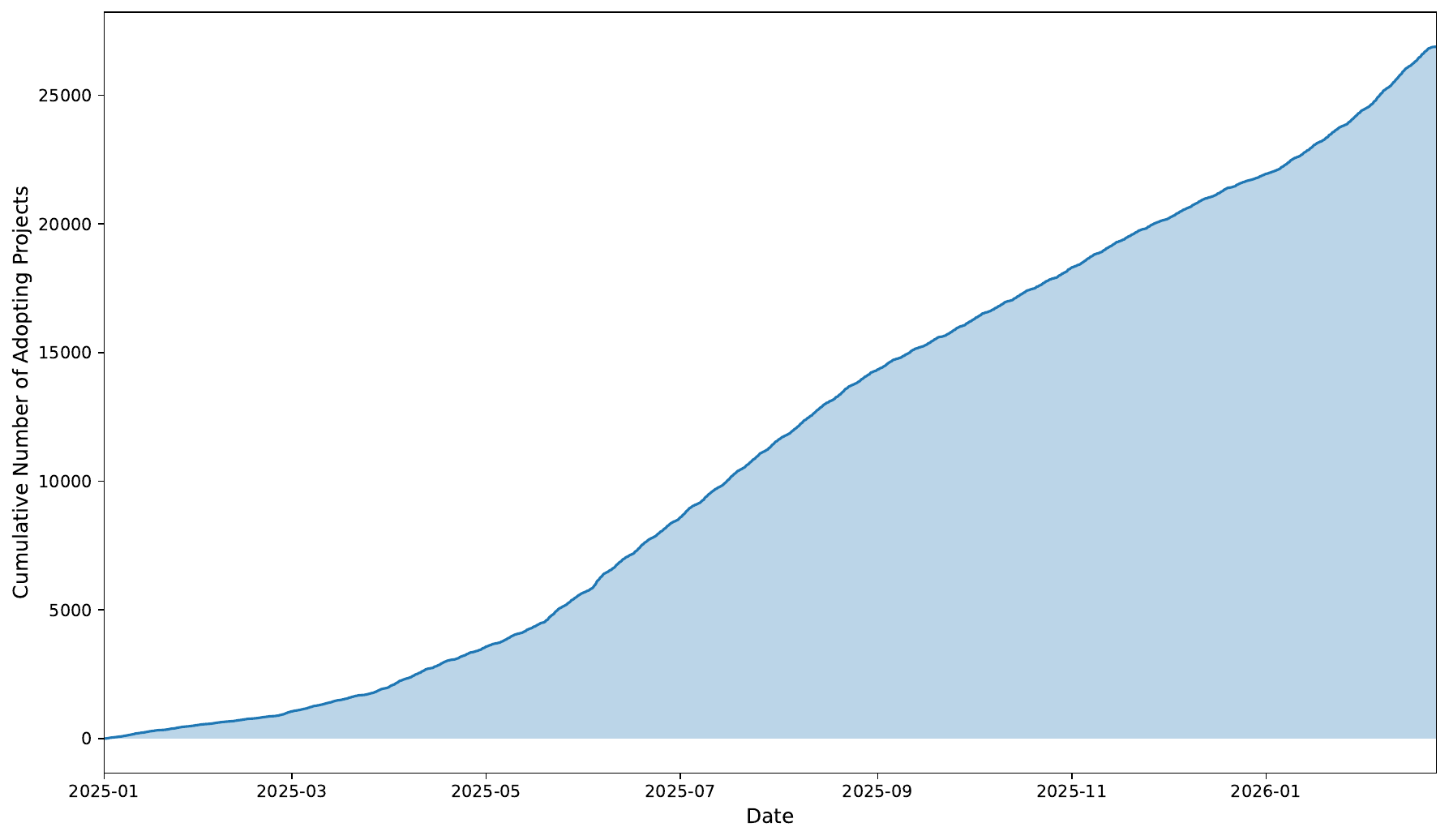}
    \caption{Cumulative adoption of any tool across all \totaladoptingprojects\ projects from \firstadoptiondate\ to \lastadoptiondate.}
    \label{fig:overall-adoption}
\end{figure}

\begin{itemize}
\item At the end of February 2025, the slope increases, indicating a faster take-off; this corresponds to the initial release of Claude Code.
\item The slope increases further around mid-May 2025; from then on the adoption occured at the fastest rate. Around this period, multiple coding agents were released (e.g. Codex, Gemini, Jules), which likely raised the popularity of the whole category.
\item \revise{Finally, adoption seems to slow down slightly at the end of August 2025. However, the current}{Adoption seems to slow down in the fall of 2025, from September to December, but the} slope is still higher than it was before the second inflection point.
\item \revise{}{Finally, there is a fourth inflection point at the start of 2026, where adoption picks up sharply once again.}
\end{itemize}

\revise{Overall, we see a fast increase of adoption, with early signs of slowing down, consistent with an S-curve. It is however early to determine if that is the case; tracking the adoption over the next few months will be useful to confirm if the adoption is indeed slowing down. The current rate of increase of adoption is still in a decidedly onward slope.}{Overall, we see a fast increase of adoption. An earlier version of this study detected hints of a possible slow down in the fall of 2025, but, as of early 2026, the trend has picked up again. This is consistent with the reported increase in capabilities of the models released at the end of 2025 (e.g. Claude Opus 4.5), with training particularly focusing on improving agentic tasks and tool use. Improvement in agentic capabilities has continued, with additional model releases in February and March 2026; however it is too early to foresee their impact on adoption}.

\subsection{Adoption per tool}

\begin{figure}[ht]
    \centering
    \includegraphics[width=0.9\textwidth]{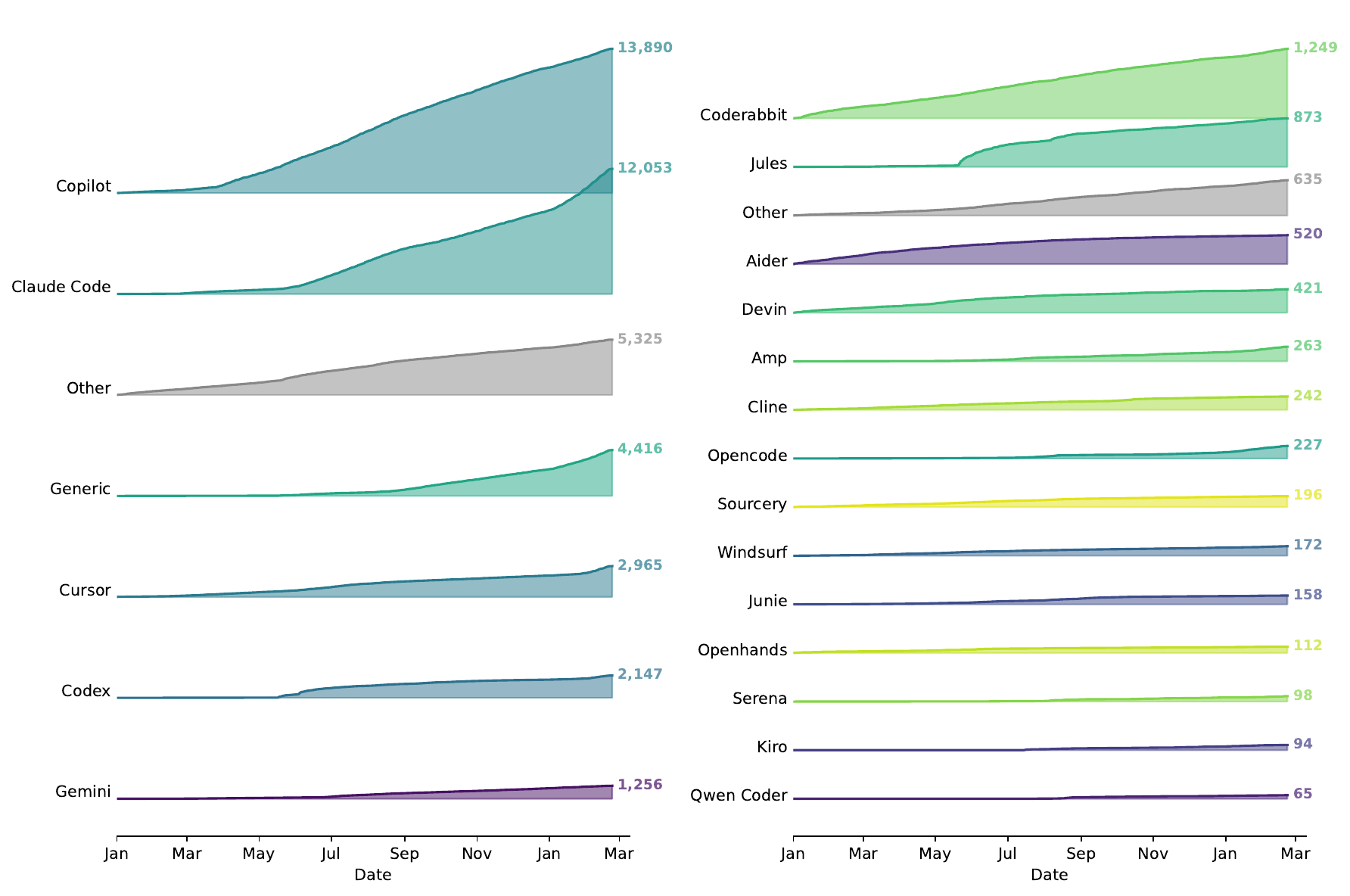}
    \caption{Evolution of the number of projects (Y axis) that adopt each tool over time for the \displayedtools\ tools, with \toptoolname\ being the most adopted (\toptoolcount\ projects).}
    \label{fig:tool-adoption}
\end{figure}

\figref{fig:tool-adoption} shows the evolution of the adoption for the \uniquetools \ specific tools that we track, via a two-panel ridgeline plot. Overall, the \totaladoptingprojects \ projects have adopted \totaltooladoptions \ tools; thus some projects adopt multiple tools (we analyse this next).

We separate the visualization in two parts as there are wide differences in terms of popularity across tools. The left part shows the 6 most popular tools (all with more than 500 adopting projects), as well as an ``other'' category for the remaining tools. The right panel zooms in on this ``other'' category, specifically on the next 15 most popular tools (50 or more adopting projects). The remaining 34 tools that we identified are shown in the rightmost ``other'' category; these tools have a more limited adoption, but make a sizeable minority together.

Of note, the ``Generic'' category is made up of projects that use the \agentsmd guidance file. This was initially the file used for the \codex coding agent; however, a standardization effort has started on this file format as the default guidance file for coding agents \footnote{\url{https://agents.md}}. This is why we can no longer attribute it with certainty to \codex, although we suspect that the majority of its usage stems from \codex. 

\figref{fig:tool-adoption} shows that a minority of tools show a significantly higher adoption than the rest. The top 5 tools (assuming most of Generic tool usage is \codex) account for more than 80\% of the overall adoption. Just \claude and \copilot are responsible for more than half of the adoption. 

If we remove the \intersectioncodexgenericsize \ projects that adopt both (see \figref{fig:tool-coadoption}), and estimate that $\approx$ 90\% of the remainder is likely projects adopting \codex ($\approx$ 3,850) rather than other agents, we arrive at an estimate of $\approx 6,000$ projects adopting \codex, which would place it in a clear third position behind \claude and \copilot. We stress this is a very rough estimate.

In terms of trends, we clearly see all the inflection points on the \claude trendline\revise{ with a sharper inflection upwards at the end}. On the other hand, the trendlines of \copilot and \cursor and more regular\revise{}{, although \cursor picks up at the last inflection point}. The trend for \codex, if we also account for the ``Generic'' trend, \revise{shows an interesting behaviour, since it rises more starting in August/September, unlike the remainder of the coding agents, and the overall trend. If this continues, it is possible that the slowdown we see in the general trendline does not materialize}{also shows a significant increase starting in the second half of 2025, with a sharper turn upwards in 2026}.  

In the remainder of projects, we see different trends: some projects such as Aider, Devin, or Coderabbit have been available for longer (starting in 2024), and seem to rise steadily over time. Others (such as \revise{}{Amp, OpenCode, }Kiro or Serena) are more recent agents, who may increase in popularity in the future.

\subsection{Co-adoption of coding agents}



As mentioned earlier, the \totaladoptingprojects~projects have adopted \totaltooladoptions~tools, representing more than 1.6 adopted agents per adopting project\revise{. As such, it is interesting to }{, which is why we }analyse the projects that have adopted more than one coding agent. \revise{}{Overall, two thirds of projects use a single tool (\projectsusingOneTools\ projects); the remaining third uses multiple tools, confirming significant co-adoption. More precisely, \projectsusingTwoTools\ projects use two tools, \projectsusingThreeTools\ use three, and \projectsusingFourPlusTools\ use four tools of more, with outliers using ten or more.} \revise{We select the coding agents having at least 100 adopting projects, and count the number of co-occurrences of multiple agents. }{}

\begin{figure}[ht]
    \centering
    \includegraphics[width=1.0\textwidth]{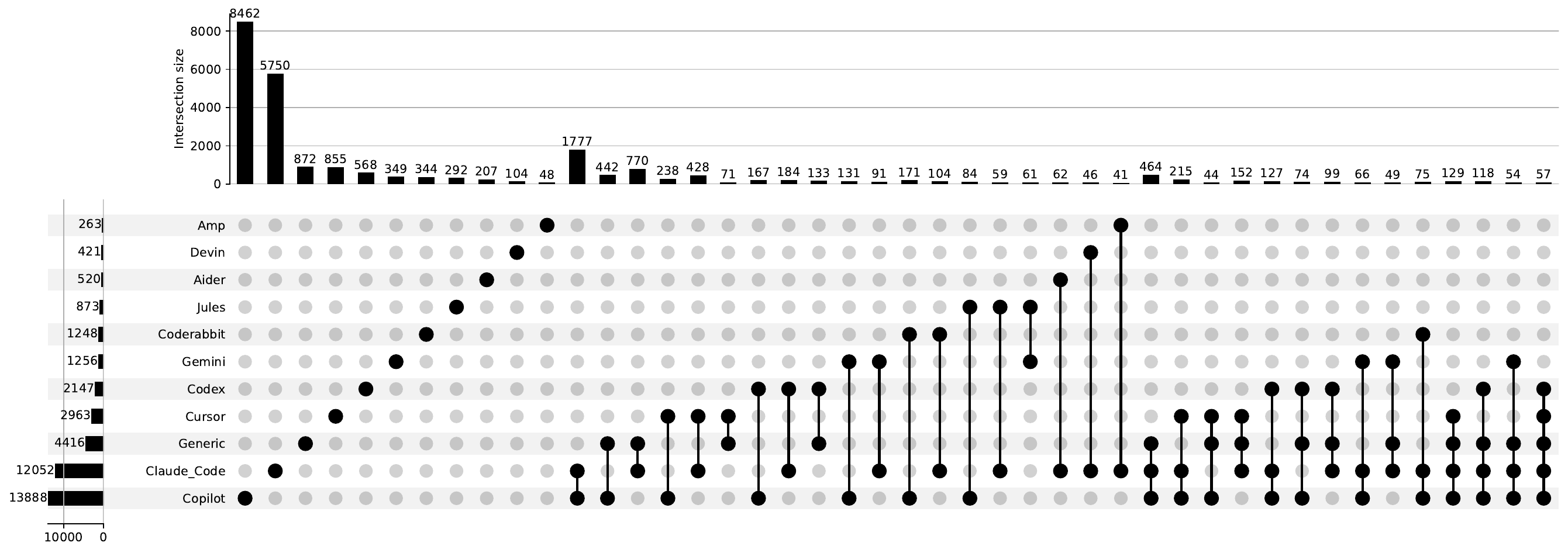}
    \caption{UpSet plot showing tool co-adoption patterns among \toolsanalyzed\ tools with at least \minsubsetsize\ projects per intersection.}
    \label{fig:tool-coadoption}
\end{figure}

\figref{fig:tool-coadoption} presents an UpSet plot of the co-occurrences. An UpSet plot is similar to a Venn diagram, but can scale more gracefully as the number of categories increase \cite{lex2014upset}. 
The figure displays the number of projects that use multiple agents, filtering for cases where at least \revise{30}{40} projects adopt more than one combination of agents\revise{}{,  limiting to agents with at least 250 adopters for legibility}. Each agent is represented by a horizontal line. Each combination (ranging from one to \revise{four}{five} agents in the figure) is represented as a series of points, connected by a line if there are more than one coding agents in the combination. Simpler combination and more frequent combinations are on the left, so the figure starts with single usage of the most popular agents, and ends with the complex combinations. The plot also includes horizontal and vertical bar charts to show: 1) the count of projects using a particular agent \revise{in the horizontal bar chart}{vertically}, and 2) the count of projects using a particular agent combination \revise{in the vertical bar charts}{horizontally}. \revise{The figure shows that most projects (\projectsusingOneTool\xspace in total) use a single agent, but a significant minority has used more than one. Unsurprisingly, the most popular combinations involve the most popular agents (\eg \intersectionclaudecodecopilot). It also shows that, as expected, the intersection between \codex and the ``Generic'' category is large: \intersectioncodexgeneric. And it also gives insights on relatively popular combinations of three or four agents: \intersectioncodexcopilotclaudecode, \intersectioncopilotclaudecodegeneric, and \intersectioncodexcopilotclaudecodegeneric, indicating that the amount of projects that likely use all of \claude, \copilot and \codex together, has around 200 projects. Another popular combination is \intersectioncopilotclaudecodecursor. All in all, \projectsusingFourTools\xspace use four different agents, while \projectsusingFiveToolsOrMore\xspace projects use five or more agents}{The figure gives insights on common agent combinations; \eg, the most popular is \intersectionclaudecodecopilot. }

\subsection{Adoption evolution in organizations}

\begin{figure}[ht]
    \centering
    \includegraphics[width=0.9\textwidth]{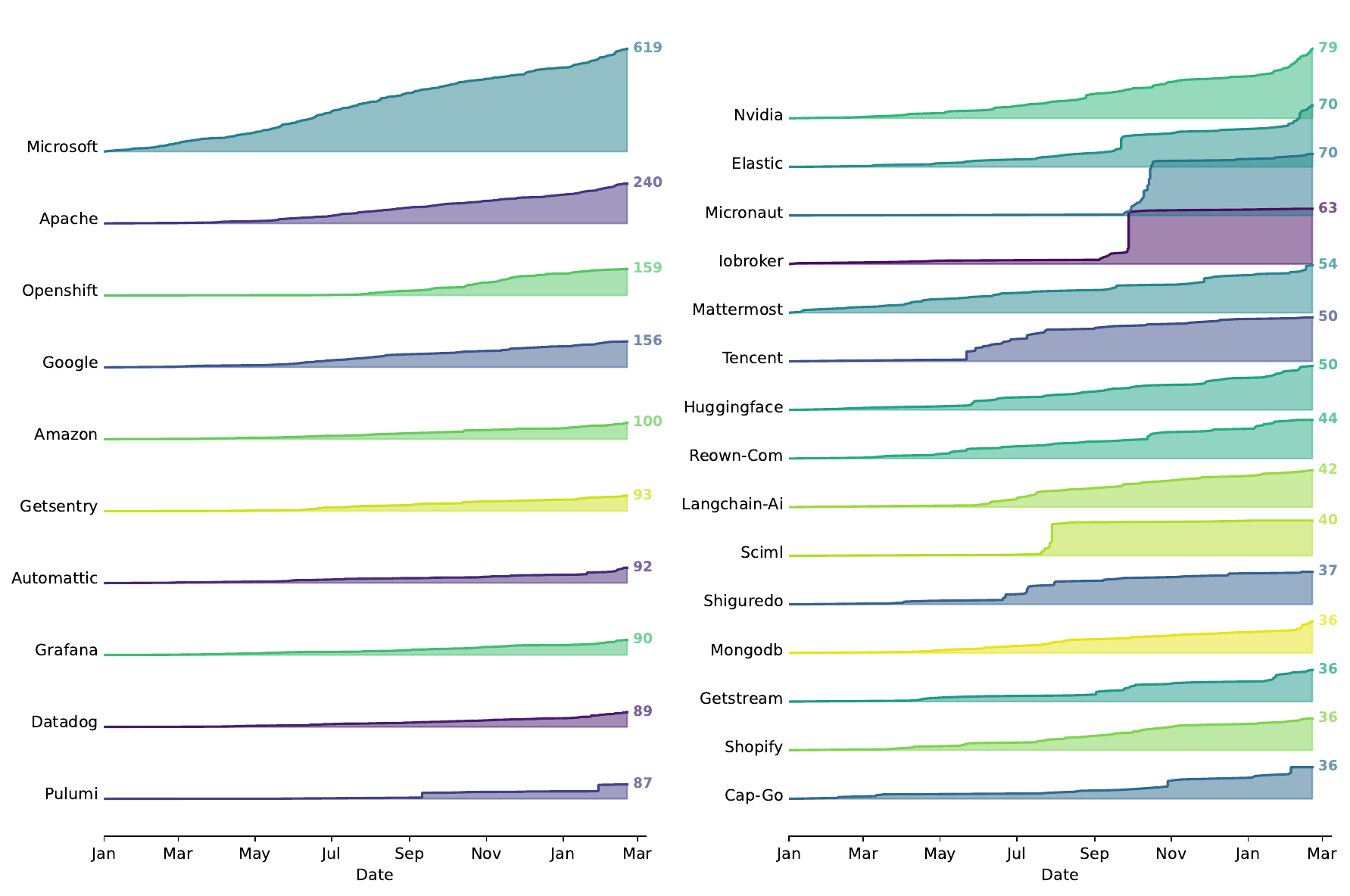}
    \caption{Evolution of the number of projects per organization (Y axis) that adopt coding agents over time across the top \displayedorganizations\ organizations, with \toporganizationname\ leading at \toporganizationcount\ adoptions.}
    \label{fig:organization-adoption}
\end{figure}

\figref{fig:organization-adoption} shows the evolution of adoption across the larger organization in our dataset over time. Note that the list is different from the top 20 organizations we had earlier, as this one focuses on the organizations with the most adoption. We once again use a two-sided ridgeline plot. 

We see that at the organization level, the adoption has a few distinct profiles. In particular some organizations, once they start to adopt coding agents, do so very quickly (\eg, Micronaut), and in some cases, almost overnight (\eg, iobroker), or by waves (Pulumi). Other organizations have a more regular adoption (\eg Apache, Google). 

We also see similar inflection points than in the overall adoption curve (\figref{fig:overall-adoption}. Notably, some organization start to adopt coding agents in February or March (\eg Automattic, Antiwork, Posthog), and there is a significant takeoff in the May to July period (\eg Getsentry, Datadog, Reown). \revise{The last months see slower adoption overall, save a few exceptions (\eg Getsentry, Openshift, Grafana)}{Finally, there is clearly a significant inflection point in late 2025/early 2026, with several organizations seeing a rise at that time (\eg(Nvidia, Elastic, Mongodb, among others).}

Very few top organizations had adoption as soon as January 2025: Microsoft is the exception, being the earliest and the largest overall adopter. Given that they are a provider of coding agents with Copilot, Microsoft certainly has some incentives to adopt it itself and serve as an example.

\section{RQ5: Size of AI contributions}
\label{sec:contrib-size}

\newcommand{\AddedLinesHumanMedian}{11}
\newcommand{\AddedLinesHumanQOne}{3}
\newcommand{\AddedLinesHumanQThree}{41}
\newcommand{\AddedLinesHumanUpperWhisker}{98}

\newcommand{\AddedLinesBotMedian}{4}
\newcommand{\AddedLinesBotQOne}{1}
\newcommand{\AddedLinesBotQThree}{10}
\newcommand{\AddedLinesBotUpperWhisker}{24}

\newcommand{\AddedLinesAIMedian}{31}
\newcommand{\AddedLinesAIQOne}{6}
\newcommand{\AddedLinesAIQThree}{114}
\newcommand{\AddedLinesAIUpperWhisker}{276}

\newcommand{\DeletedLinesHumanMedian}{5}
\newcommand{\DeletedLinesHumanQOne}{2}
\newcommand{\DeletedLinesHumanQThree}{18}
\newcommand{\DeletedLinesHumanUpperWhisker}{42}

\newcommand{\DeletedLinesBotMedian}{3}
\newcommand{\DeletedLinesBotQOne}{1}
\newcommand{\DeletedLinesBotQThree}{7}
\newcommand{\DeletedLinesBotUpperWhisker}{16}

\newcommand{\DeletedLinesAIMedian}{7}
\newcommand{\DeletedLinesAIQOne}{2}
\newcommand{\DeletedLinesAIQThree}{23}
\newcommand{\DeletedLinesAIUpperWhisker}{54}

\newcommand{\TotalFilesHumanMedian}{2}
\newcommand{\TotalFilesHumanQOne}{1}
\newcommand{\TotalFilesHumanQThree}{3}
\newcommand{\TotalFilesHumanUpperWhisker}{6}

\newcommand{\TotalFilesBotMedian}{2}
\newcommand{\TotalFilesBotQOne}{1}
\newcommand{\TotalFilesBotQThree}{2}
\newcommand{\TotalFilesBotUpperWhisker}{3}

\newcommand{\TotalFilesAIMedian}{2}
\newcommand{\TotalFilesAIQOne}{1}
\newcommand{\TotalFilesAIQThree}{4}
\newcommand{\TotalFilesAIUpperWhisker}{8}


\newcommand{\AddedLinesHumanBinOneToFive}{29.20}
\newcommand{\AddedLinesHumanBinSixToTwoZero}{23.30}
\newcommand{\AddedLinesHumanBinTwoOneToFiveZero}{14.52}
\newcommand{\AddedLinesHumanBinFiveOneToOneZeroZero}{9.88}
\newcommand{\AddedLinesHumanBinOneZeroOneToFiveZeroZero}{15.67}
\newcommand{\AddedLinesHumanBinFiveZeroOneToOneZeroZeroZero}{3.37}
\newcommand{\AddedLinesHumanBinOneZeroZeroOnePlus}{4.06}
\newcommand{\AddedLinesBotBinOneToFive}{43.06}
\newcommand{\AddedLinesBotBinSixToTwoZero}{28.79}
\newcommand{\AddedLinesBotBinTwoOneToFiveZero}{9.60}
\newcommand{\AddedLinesBotBinFiveOneToOneZeroZero}{6.26}
\newcommand{\AddedLinesBotBinOneZeroOneToFiveZeroZero}{8.23}
\newcommand{\AddedLinesBotBinFiveZeroOneToOneZeroZeroZero}{1.54}
\newcommand{\AddedLinesBotBinOneZeroZeroOnePlus}{2.52}
\newcommand{\AddedLinesAIBinOneToFive}{18.58}
\newcommand{\AddedLinesAIBinSixToTwoZero}{18.01}
\newcommand{\AddedLinesAIBinTwoOneToFiveZero}{14.43}
\newcommand{\AddedLinesAIBinFiveOneToOneZeroZero}{11.64}
\newcommand{\AddedLinesAIBinOneZeroOneToFiveZeroZero}{23.34}
\newcommand{\AddedLinesAIBinFiveZeroOneToOneZeroZeroZero}{6.28}
\newcommand{\AddedLinesAIBinOneZeroZeroOnePlus}{7.72}

\newcommand{\DeletedLinesHumanBinOneToFive}{40.00}
\newcommand{\DeletedLinesHumanBinSixToTwoZero}{25.48}
\newcommand{\DeletedLinesHumanBinTwoOneToFiveZero}{13.09}
\newcommand{\DeletedLinesHumanBinFiveOneToOneZeroZero}{7.51}
\newcommand{\DeletedLinesHumanBinOneZeroOneToFiveZeroZero}{9.72}
\newcommand{\DeletedLinesHumanBinFiveZeroOneToOneZeroZeroZero}{1.81}
\newcommand{\DeletedLinesHumanBinOneZeroZeroOnePlus}{2.39}
\newcommand{\DeletedLinesBotBinOneToFive}{48.13}
\newcommand{\DeletedLinesBotBinSixToTwoZero}{27.50}
\newcommand{\DeletedLinesBotBinTwoOneToFiveZero}{8.43}
\newcommand{\DeletedLinesBotBinFiveOneToOneZeroZero}{5.63}
\newcommand{\DeletedLinesBotBinOneZeroOneToFiveZeroZero}{7.19}
\newcommand{\DeletedLinesBotBinFiveZeroOneToOneZeroZeroZero}{1.18}
\newcommand{\DeletedLinesBotBinOneZeroZeroOnePlus}{1.93}
\newcommand{\DeletedLinesAIBinOneToFive}{35.25}
\newcommand{\DeletedLinesAIBinSixToTwoZero}{25.76}
\newcommand{\DeletedLinesAIBinTwoOneToFiveZero}{14.27}
\newcommand{\DeletedLinesAIBinFiveOneToOneZeroZero}{8.18}
\newcommand{\DeletedLinesAIBinOneZeroOneToFiveZeroZero}{10.98}
\newcommand{\DeletedLinesAIBinFiveZeroOneToOneZeroZeroZero}{2.30}
\newcommand{\DeletedLinesAIBinOneZeroZeroOnePlus}{3.25}

\newcommand{\TotalFilesHumanBinOne}{42.56}
\newcommand{\TotalFilesHumanBinTwo}{18.19}
\newcommand{\TotalFilesHumanBinThreeToFive}{14.80}
\newcommand{\TotalFilesHumanBinSixToOneZero}{12.29}
\newcommand{\TotalFilesHumanBinOneOneToTwoZero}{6.69}
\newcommand{\TotalFilesHumanBinTwoOneToFiveZero}{3.64}
\newcommand{\TotalFilesHumanBinFiveOnePlus}{1.83}
\newcommand{\TotalFilesBotBinOne}{40.55}
\newcommand{\TotalFilesBotBinTwo}{35.31}
\newcommand{\TotalFilesBotBinThreeToFive}{11.12}
\newcommand{\TotalFilesBotBinSixToOneZero}{6.10}
\newcommand{\TotalFilesBotBinOneOneToTwoZero}{3.09}
\newcommand{\TotalFilesBotBinTwoOneToFiveZero}{2.04}
\newcommand{\TotalFilesBotBinFiveOnePlus}{1.78}
\newcommand{\TotalFilesAIBinOne}{36.51}
\newcommand{\TotalFilesAIBinTwo}{17.52}
\newcommand{\TotalFilesAIBinThreeToFive}{17.04}
\newcommand{\TotalFilesAIBinSixToOneZero}{14.21}
\newcommand{\TotalFilesAIBinOneOneToTwoZero}{7.69}
\newcommand{\TotalFilesAIBinTwoOneToFiveZero}{4.56}
\newcommand{\TotalFilesAIBinFiveOnePlus}{2.46}

Beyond the commit ratio analysis, in this section, we compare the size of AI-assisted contributions with the size of human-written contributions. To do so, we measure the size of commits with three metrics: 1), the number of lines added in the commit; 2), the number of lines deleted in the commit; and 3), the number of files involved in the commit (added, deleted, modified, or renamed). We limit the number of metrics to simplify the presentation. We then use our heuristics to determine if a commit was authored by a human, a software bot, or was assisted by a coding agent. We do this analysis for all commits by file-level adopters that have commit-level adoption. This allows us to minimize the number of AI-assisted commits mislabelled as human contributions, although this mitigation is not perfect (\eg, a contributor may decide to stop their coding agent signing commits, while still using it). 

\begin{figure}[ht]
    \centering
    \includegraphics[width=0.9\textwidth]{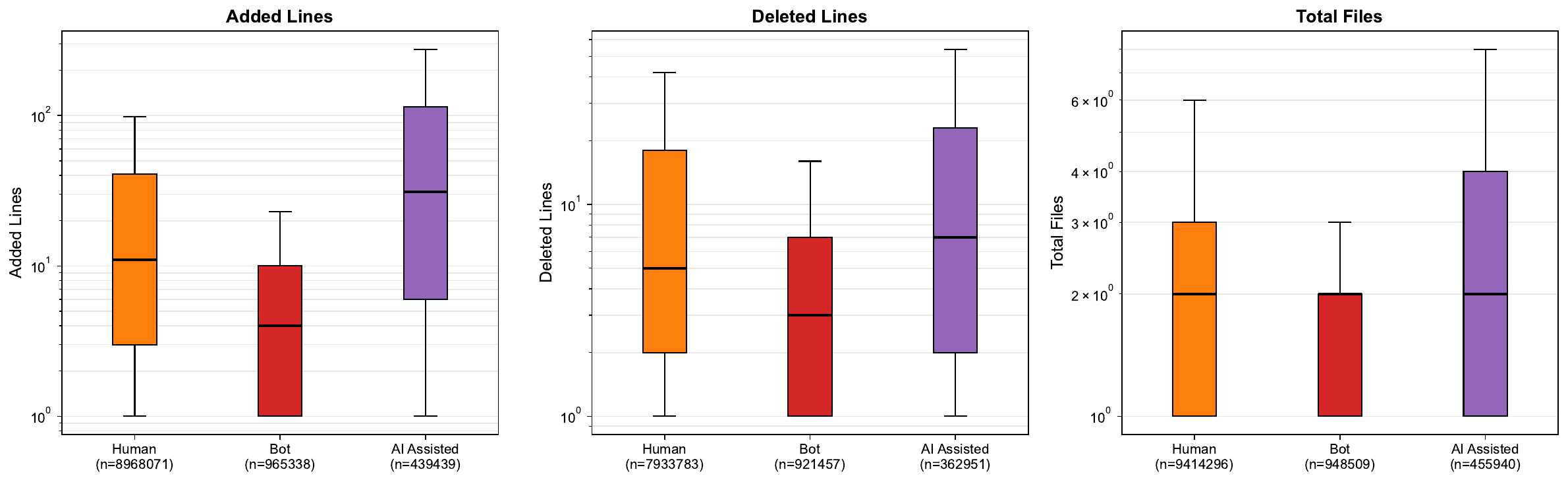}
    \caption{Distribution for commit size in terms of lines and files by type of contribution.}
    \label{fig:commit-boxplots}
\end{figure}

\figref{fig:commit-boxplots} shows the distribution of the commit sizes for the three types of commits. We clearly see that across all three metrics, bots make smaller contributions than humans, whereas coding agents make larger ones. For instance, the median number of added lines for a human contribution is \AddedLinesHumanMedian, while for a bot, it is only \AddedLinesBotMedian; on the other hand, for AI-assisted commits, it rises to \AddedLinesAIMedian, a value triple the median for humans, and closer to the third quartile of human contributions (\AddedLinesHumanQThree). The third quartile of AI-assisted contributions is considerably larger (\AddedLinesAIQThree). 

For deleted lines, the differences are not as stark: we see that the amount of deleted lines per commits for bots is smaller than the other two categories (median: \DeletedLinesBotMedian), while AI-assisted contributions are larger (median: \DeletedLinesAIMedian)  than Human contributions (median: \DeletedLinesHumanMedian). The same is true at the file level. However, the metric being coarser, the differences are smaller: the median bot commit involves a single file, while both AI-assisted and human commits involve two files. Only in the upper part of the distribution do we see a difference (Q3 for humans: \TotalFilesHumanQThree; for AI-assisted commits: \TotalFilesAIQThree).





\begin{figure}[ht]
    \centering
    \includegraphics[width=\textwidth]{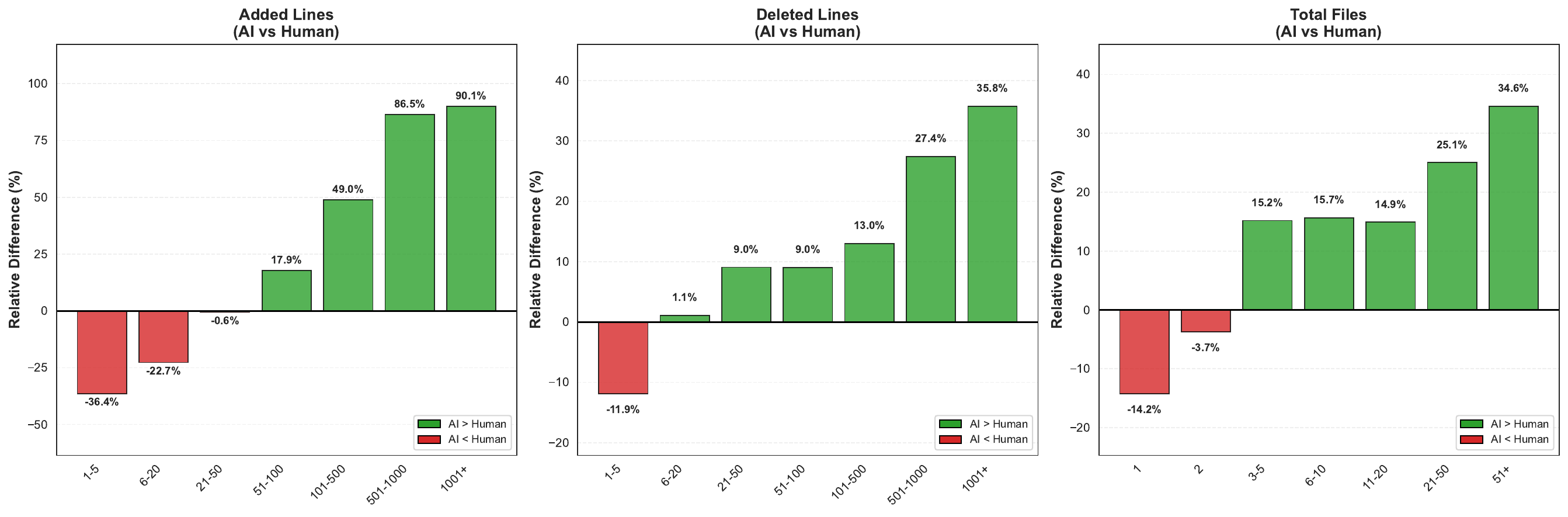}
    \caption{Comparison between the frequency among size categories of human-authored and AI-assisted commits (lines added, lines deleted, and files involved).}
    \label{fig:commit-diffs}
\end{figure}

Another way to look at the data is to categorize it into discrete size categories. \figref{fig:commit-diffs} shows the difference in the proportion of size categories for commits, focusing on human-authored and AI-assisted commits. The figure shows that the trend observed in \figref{fig:commit-boxplots} is even more true at the edges of the distribution. 

For line additions, the smallest category of commits (adding one to 5 lines) is \revise{41\%}{36\%} smaller when they are AI-assisted. On the other hand, the largest commits are almost twice as frequent for AI-assisted commits; this holds true even for commits that add more than 1,000 lines. 

We see the same trend, albeit less pronounced, for line deletions: smaller commits (1 to 5 deleted lines) are \revise{13.5\%}{12\%} less frequent in AI-assisted commits, while the larger commits (thousands of deleted lines) are \revise{close to half as much more frequent}{36\% more frequent}. Files follow the same pattern, with single-file AI-assisted commits being close to \revise{20\%}{15\%} less frequent, while commits with more than 20 files are 30\% more frequent.

In summary, AI-assisted commits tend to add more lines, in more files, than human-authored commits. In addition, and perhaps somewhat surprisingly, they also delete more code than human-authored commits. Taken together, these trends point towards an increased churn for AI-assisted commits, and for commits changing multiple files with increased frequency. This increased churn and non-local changes, in turn, brings questions on the sustainability of coding agents over time: further study is necessary to evaluate their impact on code quality.

\section{RQ6: Types of commits authored by coding agents}
\label{sec:qualitative}



As semi-autonomous development assistants, coding agents can contribute in various ways to code bases and support developers across multiple aspects of software development: fix bugs, write documentation, refactor code, write test cases, add new features, \etc
Starting from the observation that the adoption of coding agents has surged, with this research question we now focus on \emph{which} tasks agents perform:
\emph{What kind of contributions, visible in version control system histories, do developers perform with the help of coding agents?}

To answer this question, we qualitatively examine concrete contributions, at the granularity of individual commits, made by or with agents in git repositories that have adopted them, to better understand the tasks they are used for.

\subsection{Dataset}
Given the sheer amount of agent-authored commits in our adoption data and the cost of qualitative analysis, we resort to studying a random sample of commit-level contributions.
Among the coding agents identified in \Cref{sec:evolution}, Claude Code leads in popularity (see \Cref{fig:tool-adoption}), which makes it a good candidate for qualitative analyzes.
Claude Code also defaults to annotating its commits with the \texttt{Co-authored-by} trailer, making the identification of its contributions at the commit level both practical and robust.\footnote{\url{https://docs.claude.com/en/docs/claude-code/settings}}

From the 90,321 commits identified as Claude-authored in our adoption dataset, we draw a sample of 790 for further qualitative analysis using Cochran’s formula for categorical classification into seven categories with a 5\% margin of error and a 95\% confidence level~\cite{goodman1965simultaneous}.
We then adjust for the finite population size.
As our previous test used a confidence level of 99\% and ran 9 statistical tests with p-values < $10^{-16}$, we apply Bonferroni’s correction~\cite{rupert2012simultaneous} to adjust the interval from 95\% to 96\%, maintaining an overall 95\% probability that all our results are correct.
These calculations yield a required sample size of 759 commits;~we analyze 790 to be conservative.
For each commit, we extract the first line of the commit message, diff statistics (added and deleted lines and files), the kind of files impacted by the changes (source code, documentation, configuration, data), and the programming languages used in impacted files.

\Cref{tab:claude-commits-stats} shows some descriptive statistics of the sample.

\begin{table}
    \scriptsize
    \caption{Descriptive statistics of the random sample of 790 commits authored by Claude Code, qualitatively analyzed to determine their type.}
    \centering
    \begin{minipage}[t]{0.415\textwidth}
    \vspace{0pt}
    \begin{tabular}{l|r|r|r|r}
    \toprule
    \textbf{Metric} & \textbf{Min.} & \textbf{Med.} & \textbf{Avg.} & \textbf{Max.} \\
    \midrule
    \midrule
    Lines added     & 0 & 70 & 3743.0 & 1.6\,M  \\
    Lines deleted   & 0 &  9 &  909.0 & 576\,K  \\
    Files added     & 0 &  1 & 18.2 & 10.7\,K \\
    Files modified  & 0 &  1 &  3.7 & 138     \\
    Files deleted   & 0 &  0 &    3.0 & 1959    \\
    \bottomrule
    \end{tabular}
    \end{minipage}%
    \begin{minipage}[t]{0.25\textwidth}
    \vspace{0pt}
    \begin{tabular}{l|r}
    \toprule
    \textbf{Type of} & \textbf{Commit} \\
    \textbf{changed files} & \textbf{count} \\
    \midrule
    \midrule
    source code	& 573 \\
    documentation & 296 \\
    data & 287 \\
    configuration & 202 \\
    other & 68 \\
    build & 17 \\
    image & 4 \\
    \bottomrule
    \end{tabular}
    \end{minipage}%
    \begin{minipage}[t]{0.25\textwidth}
    \vspace{0pt}
    \begin{tabular}{l|r}
    \toprule
    \textbf{Language of} & \textbf{Commit count} \\
    \textbf{changed files} & \textbf{(top-10)} \\
    \midrule
    \midrule
    Markdown & 254 \\
    TypeScript & 213 \\
    XML & 106 \\
    JSON & 105 \\
    Python & 102 \\
    YAML & 92 \\
    JavaScript & 73 \\
    Rust & 57 \\
    Go & 35 \\
    TOML & 33 \\
    \bottomrule
    \end{tabular}
    \end{minipage}
    \label{tab:claude-commits-stats}
\end{table}

\subsection{Protocol}
As manually inspecting the code changes in a large set of commits is impractical (4,652 modified lines per commit on average in our sample), we classify commits based on the content of their messages.
In particular, we rely on the Conventional Commits specification,\footnote{\url{https://www.conventionalcommits.org/en/v1.0.0/}} a convention that has seen growing adoption over the past years which encodes the nature of the changes made in a commit within the commit message itself (\eg \textit{feat} for new features, \textit{fix} for bug corrections, \textit{docs} for documentation)~\cite{zeng2025first}.
Originally designed to ease semantic versioning, it defines two core categories: \textit{fix} for bug patching (corresponding to \textit{patch} version increments) and \textit{feat} for new features (corresponding to \textit{minor} version increments).
In practice, this set of categories is open-ended and a popular extension is that of semantic commits, which defines seven types of tasks: \texttt{chore}, \texttt{docs}, \texttt{feat}, \texttt{fix}, \texttt{refactor}, \texttt{style}, and \texttt{test}.\footnote{\url{https://gist.github.com/joshbuchea/6f47e86d2510bce28f8e7f42ae84c716}}
A commit following that convention has a message prefixed with the name of one of these categories, followed by a colon (\eg \texttt{feat: Enable filtering clients per organization}).
This convention enables a structured and scalable way to infer the intent and type of work represented by each commit without examining the underlying code.
By leveraging this convention, we can analyze contribution patterns systematically and with reasonable confidence in the accuracy of the classification.

To our surprise, the majority of Claude commits in our sample already follows this convention:~513 of the 790 commits (65\%) can automatically be classified into one of the seven categories by matching their prefix.
Indeed, coding agents may imitate the style of commit messages made by developers (in the current project or in their training data), or be explicitly instructed to follow that specific convention.
When a commit can be classified automatically, we retain the classification made by the agent or developer who made the commit.
Then, two authors were tasked to manually and individually classify the remaining 277 commits into one of the seven categories by inspecting their commit message and diff, with an initial agreement on 250 of the 277 commits (90\%).
The remaining 27 commits were finally classified after discussion.

\subsection{Results}

In a recent study, \citeauthor{zeng2025first} analyze 88,704 \emph{human-authored} commit messages in 116 repositories from diverse domains following the Conventional Commit specification.
They find that \textit{chore} commits are the most frequent (non-functional changes, including updating build scripts, dependencies, and continuous integration---31\%), followed by fix (bug fixes---27\%) and feat (implementation of new features---17\%)~\cite{zeng2025first}.

\begin{table}
    \small
    \caption{%
    Types of commits co-authored by Claude Code, for a random sample of 790 commits.
    Commit types are from the semantic commit messages convention.
    Commit classification performed by either parsing the first line of commit messages according to the Conventional Commits specification, version 1.0.0 (for the 513 commits that conform to it) or via manual review performed by two authors independently (277 commits).}
    \centering
    \begin{tabular}{l|r|r}
    \toprule
    \textbf{Type} & \textbf{Count} & \textbf{Ratio} (\%) \\
    \midrule
    \midrule
    feat     & 282 & 35.7\% \\
    fix      & 236 & 29.9\% \\
    docs     &  86 & 10.9\% \\
    refactor &  78 &  9.9\% \\
    chore    &  56 &  7.1\% \\
    test     &  43 &  5.4\% \\
    style    &   9 &  1.1\% \\
    \midrule
    \emph{Total} & 790 &        \\
    \bottomrule
    \end{tabular}
    \label{tab:claude-commit-types}
\end{table}

This is in stark contrast with our sample of Claude commits, as shown in \Cref{tab:claude-commit-types}.
Commits of types \textit{feat} and \textit{fix} are the most frequent in our sample, accounting for two thirds of the analyzed commits, while \textit{chore} commits only account for 7.1\%.
Notably, feature-introducing commits are about twice as common for agents as for humans.
This likely contributes to the pattern observed in \Cref{sec:contrib-size} that AI-assisted commits tend to modify more lines and touch more files than human-authored ones.
For the remaining categories, the percentages differ by two percent or less between the study of \citeauthor{zeng2025first} and ours, reflecting a similar distribution for these tasks between humans and agents.

Several factors might explain these differences.
As shown in \Cref{tab:file_adoption}, while agent adoption spans the whole spectrum, it is the highest in younger projects, where the implementation of new features might be prioritized over bug fixing and maintenance tasks such as \textit{chore}.
Besides, several \textit{chore} tasks are already handled well by automated bots (\eg dependabot for dependency management), so the cost of running agents might offset their benefits in this situation.
The low number of \textit{test} commits doesn't necessarily mean agents aren't being used for testing.
The Conventional Commit specification requires choosing a single category, so tests added alongside a new feature might be included in a \textit{feat} commit instead.
For example, commit \href{https://github.com/nizos/tdd-guard/commit/d8d58f61ed480c58ad1325a85badb1c1907235fa}{d8d58f61} in \texttt{nizos/tdd-guard} (\texttt{feat: add AnthropicModelClient with API key configuration}) introduces a new feature together with corresponding test cases and is classified as a \textit{feat} commit.
The proportion of standalone \textit{test} commits is similar to the proportion found by \citeauthor{zeng2025first}~\cite{zeng2025first}. 

Moreover, in line with the observation on contribution size, test commits, as well as other categories such as refactorings and documentation, despite being less common, can be quite large. For instance, commit \href{https://github.com/mamertofabian/bolt-to-github/commit/4a9dd8f7281f3b1f1f0cc3b2ed463f7e8b52b86d}{4a9dd8f} adds extensive unit tests for two components (two files added, +488 lines of code); commit \href{https://github.com/allenhutchison/obsidian-gemini/commit/3bcefd2707a3c913430e5001676e94b9880954bb}{3bcefd2} is a large refactoring that changes the structure of the code (and also adds some tests, 14 files changed; +912/-19 lines); commit \href{https://github.com/evmts/voltaire/commit/f6315ff2d8a5b3a360d33dc0c0689fccd2d610c1}{f6315ff} adds a high-level documentation of the structure of a module written in Zig, describing components, providing examples of use, with specific details on performance (one new file, 180 lines).

In summary, agents are used for a wide range of software development tasks, spanning the full set of categories defined by the Conventional Commits specification.
Compared with humans, however, agents are used more often to introduce new features and less often for long-term maintenance tasks such as dependency and continuous integration management.
These observations align with our previous findings about the age of agent-adopting projects and the size of the contributions attributed to them.


\section{Discussion}
\label{sec:discussion}
We first discuss the limitations of our study, focusing in particular on the ways our study could either under-estimate or over-estimate adoption; we conclude that an under-estimate is more likely. We then discuss the implications of this work for practitioners and researchers, keeping in mind that it is likely we are under-approximating coding agent adoption.

\subsection{Limitations}

\paragraph{Generalizability of our findings}
We took several precautions to maximize the external validity of our findings. We purposefully selected a large dataset of GitHub projects, selecting for projects that reach minimum size and activity requirements (5,000 lines of code; 100 commits). These minimum size and activity requirements are high enough to ensure we select projects that are more likely to reflect real programming activity, but low enough to increase diversity (\eg, including both small and large projects, projects with few and many contributors, that span many topics and programming languages). Further, after estimating adoption as a whole in RQ1, in RQ2 we analyzed our adoption metrics by relating them to other project characteristics (size; age; contributors; and activity indicators: commits, issues, pull requests) in order to better understand how adoption varied with these characteristics. However, we restricted these analyses to high-level metrics only, as carrying out similar analyses in the other RQs would have significantly lengthened the paper. 

Similarly, we opted to exclude forks to avoid including copies of seemingly large projects, but with possibly limited activity. Finally, to increase the relevance of our findings to real-world coding scenarios, we specifically looked at the adoption from repositories of specific, established organizations. 

Despite all of these mitigations, it is possible that our results do not translate well to more general use, and to industry in particular: we have reasons to think that industry use might be different than use in open-source\revise{}{, which we detail at the end of \secref{sub:under-estimate}; we think it is likely that the use in industry is higher than in open-source}. More generally, we have identified several reasons why our specific metrics could lead us to either \emph{over-estimate} the adoption and use of coding agents, or on the contrary, to \emph{under-estimate} it. We expand on these themes below. When relevant, we mention some of the perils of agent mining, that we identified and developed in another work \cite{agentminingpaper}. In particular, many of the reasons for under or over-approximating agent activity stem from the Peril of Partial Observability. 

\paragraph{Other limitations}
The commit ratio, for reasons explained earlier, is an imperfect metric. It is particularly the case at the individual project level, where the observability depends on the choices of individual contributors. This is why we refrain from studying projects in isolation. We assume that these kind of errors are evenly distributed when analyzing multiple projects, but we can not guarantee it. In particular, populations with a higher knowledge of coding agents may behave differently.

For the label analysis in RQ3, we are limited by the labels chosen by developers. A small minority of projects have such labels; moreover, many labels co-occur, which adds some degree of correlation between individual labels.

For the programming language analysis, we rely on GitHub Linguist. This is a proven solution, but in ideal scenarios, it uses both file names and file contents. This would not scale to our study, as we apply it to all committed files in 2025 for \TotalProjects projects; we limit GitHub linguist to file names and extensions, which makes it less precise. For instance, PHP is misclassified as one of its dialects, Hack. We tried to resolve the most important issues (\ie, Rust), but can not guarantee we found all the errors.

Pull requests that are ``squash merged'' condense multiple commits in a single one \cite{bird2009promises}. If parts of the commits in the PR was AI-assisted, we are forced to label the entire merge commit as AI-assisted. This may lead us to over-estimate the size of the AI contributions when the proportion of AI-assisted commits in the PR is small. Aside from analyzing the diff, which is difficult at the scale of the analysis, the other option left would be to ignore these commits, which we found to be a worse tradeoff.

\subsection{Reasons for over-estimating coding agent use}
We first focus on the reasons why we could over-estimate adoption.

\paragraph{Under-use of agents} A large proportion of projects have evidence of coding agent use at the file level, but none at the commit level: only \CommitUseAllFileUsersPercent of projects do; moreover, \figref{fig:files-vs-commits} shows that the correlation between the metrics is very weak. This a manifestation of the Peril of Partial Observability, and we can not know to which extent agents are used in these projects. It is very likely that, after early experiments, some of these projects are \emph{not} using coding agents, but we do not know the proportion.

\paragraph{Extent of the involvement of the agent} When we detect agentic activity, we can never be sure of the proportion of the activity that is due to the agent, due to the Peril of Partial Observability. In general, we can only observe the end result of the work (commits or pull requests). For a commit authored or co-authored by an agent, the involvement of the developer may range from the minimal (\eg ``vibe-coding'', or trusting the agent enough to give it extensive permissions and autonomy), to a very close oversight, closely reading the code, or even significantly editing it. 

Similarly, when we identify agent activity in a pull request for an agent that does not sign their commit (such as Codex), we assume that all the commits from the initial author stem from the agent. However it is possible that only part of the commits were done by the agent, and follow-up work was performed by the developer. The Peril of Diversity \cite{agentminingpaper} implies that there is a diversity of agent workflows, which makes tailoring heuristics for this difficult.

\paragraph{Size is not everything} In RQ5, we compare the size of the commits, finding that AI-assisted commits are larger in terms of lines added, lines deleted, and files involved. How this affects productivity is an open question, which we do not try to answer (we leave that to future work). For instance, the data is consistent with a high churn (lines added and quickly deleted), which might not translate to real productivity. Similarly, our analysis does not look at whether AI-assisted commits are more likely to be reverted (which we have seen on occasion), whether in other AI-assisted commits, or by humans. Computing churn in this way would require running origin analyses \cite{godfrey2005using} at scale (particularly in case of extensive refactorings \cite{hora2018assessing}), which is a significant endeavor.

\subsection{Reasons for under-estimating coding agent use}
\label{sub:under-estimate}
On the other hand, there are several reasons why we may \emph{under-estimate} the use of coding agents.

\paragraph{Absence of evidence is not evidence of absence} The flipside of the Peril of Partial Observability is that, for the projects in which we have no evidence of use (either at the commit level, or at the file level), there might very well be \emph{significant} use of coding agents. First, there is the simple fact that some agents simply do not co-author commits; this is an implementation choice in the harness. For instance, \codex does not sign it commits, partly because its workflow is more based on pull requests. Second, if a coding agent signs their commits or leaves traces in pull requests, there can be ways to disable this behavior. For instance, there is a boolean setting to toggle this in Claude Code; and there is a similar setting to set the branch prefix that Codex uses. This makes it very easy to make their activity undetectable. Anecdotally, we have seen developers go to considerable lengths to disable this behavior even without using these settings, such as including extensive guidance instructing Claude Code not to sign commits, or setting up a GitHub action to auto-remove labels inserted by Codex. Finally, developers whishing more oversight may integrate agents in their workflow, but prefer to commit manually, after checking the agent's contributions; we suspect this is a sizeable portion. 

In addition, we rely on the presence of files to scale our search for coding agent usage, but we have also found that as many projects, if not more, have visible coding agent activity only in commits. Once again, this depends on both the coding agent and the developer workflow. Some agents, such as Devin, work primarily using a cloud setup, where they clone the project and handle their state without committing to the repository. Other agents such as Lovable or Vercel v0 \footnote{Lovable: \url{https://lovable.dev/}; Vercel v0: \url{https://v0.app}} are dedicated to fast prototyping, starting from an installation in the cloud, and as such are under-represented in Github repositories. 

Some developers will avoid to have configuration files in their GitHub repositories, using specialized \dotfiles repositories. Indeed we have several hundred such \dotfiles repositories in our dataset (which we filter out to estimate overall adoption). To further improve our coverage of projects, we could check for the presence of agent configuration files in a \dotfiles repository, and link it to other repositories the developer contributes in. However, this would significantly complexify our mining pipeline, and it would add noise: not every developer has a public \dotfiles repository, and their agent use could be confined to a few of their repositories. 

\paragraph{Missing heuristics} There is a large number of coding agents (Peril of Multiplicity \cite{agentminingpaper}), which is increasing, and the agents themselves are evolving (Peril of Velocity). Keeping track of all the coding agents, and finding appropriate heuristics for all of them, is a significant endeavor (rendered more complex by the Peril of Diversity: each agent works differently). We tried to be as thorough as possible, but we have almost certainly missed some agents, or at least some heuristics for a given agent. This will lead us to under-estimate the use of these specific agents. Related to this is the fact that some agents can use the guidance of others, making the identification of individual agents difficult. This is why, for instance, we can not attribute with 100\% certainty all uses of \agentsmd to \codex.

\paragraph{Unmerged commits} Our analysis focuses on the commits that are merged in the main branch. There is more use of coding agents in unmerged Pull Request and branches, which we do not consider as we focus on the contributions that end up in the main version of the project. These PRs can be unmerged for a variety of reasons: \eg, they are very recent and have not been reviewed yet; or, the agent's work was subpar. We have occasionally seen pull requests that have been merged in non-conventional ways (\eg manually), which our study can not detect. In addition, unmerged PRs can represent proof of concept or prototyping work, which might be useful even if the actual code does not last. 

\paragraph{Things may be different in the industry} While it is well-known that studying open-source software is an imperfect proxy for industry software, it is particularly so for coding agents, due to their cost (Peril of Costs Shape Usage \cite{agentminingpaper}). Coding agents rely on long LLM inferences, over multiple turns, which is expensive. As a consequence, using coding agents most often requires a subscription, or metered API usage; free usage is limited to less powerful models and stricter rate limits. This means that industry developers may use coding agents more as part of a paid job (especially if it is subsidized by their employer), while open-source developers, who may not have such ease of access, may limit their usage. In addition, beyond the amount of usage, the \emph{kind} of usages made possible by less restrained use may be different (\eg, using multiple agents in parallel). Anecdotally, developers running high coding agent bills (hundreds of dollars per month or more) are not unheard of, while some are curious about the possibilities offered by spending even more \footnote{For instance, this Shopify executive wondering how a developer may spend ca. 10,000 dollars a month on coding agents \url{https://xcancel.com/fnthawar/status/1930367595670274058}}. Thus, it is possible that the usage in industry is higher than in open-source. We tried to look into this by studying specific organizations, finding that they had higher file-level adoption in RQ3 (a relative increase of \industryincrease). However, since this only based on public repositories, it likely does not reflect the internal usage.



\subsection{Putting it all together}
\label{sec:all_together}
Considering all the reasons for undercounting and overcounting AI adoption, we think it is more likely that we are undercounting it. First, we think that projects adopting coding agents at the file level that have no AI-assisted commits are more likely to be using coding agents, rather than having abandoned them. This is because it is very easy to configure the main coding agents not to leave traces in commits or PRs, and because, for now, we can not account for developers that commit manually yet still use coding agents. While this is only anecdotal, there seem to be a non negligeable number of projects that fit these cases \footnote{\label{invisible-projects}For some clear cut cases that our heuristics nevertheless would miss, see the following examples: \url{https://github.com/cloudflare/workers-oauth-provider}, \url{https://github.com/mitsuhiko/sloppy-xml-py}, \url{https://github.com/jackdoe/pico2-swd-riscv}}. While it is conceivable that projects with very limited and unchanging guidance have abandoned coding agents, it is less likely that projects with extensive and/or actively maintained guidance have done so.

We did a preliminary analysis of a related issue: after an initial usage, some projects experience a fall of their commit ratio to 0. A straightforward interpretation of this would be that such a project has abandoned coding agents. However, we found that on a sample of such projects (the 500 projects that had the largest commit ratio drop to zero), half of them had added or modified at least one agent guidance or configuration file after the commit ratio dropped to 0. It seems unlikely that a project that has abandoned coding agents would continue maintaining their coding agent guidance.

As another datapoint that fits with this narrative, consider \figref{fig:commit-adoption-by-topic}, in which some topics\revise{ have surprisingly low commit ratios (namely projects in the ``agents'', ``openai'', ``ai'', or ``llm'' categories)}{ such as \ghtopic{agents}, \ghtopic{llm}, or \ghtopic{openai}, have low to average commit ratios}, despite these topics showing very high file-level adoption in \figref{fig:file-adoption-by-topic}. Rather than having a lower than average level of adoption, we hypothesize it is more likely that contributors to these projects, having more knowledge of coding agents, are more likely to be aware of how to change their settings and disable the visibility of their activity.

Second, given the high ratio of file-level activity that has no commit level activity, we think there is a sizeable proportion of ``undetected use'' of coding agents: projects that have neither file-level nor commit-level agent activity, yet still use agents. For instance, the projects mentioned in this earlier footnote\footref{invisible-projects} fit this profile, making us believe that the true adoption is higher than our conservative estimate. To be clear, this is so far only a hypothesis; future work should define heuristics to detect coding agent activity in these (and more challenging cases) with sufficient precision. Only such a study could determine with more certainty where the true adoption lies between our conservative and our high estimate, and could provide a more precise estimate of the AI-assisted commit ratio in adopting projects.

Finally, the third way adoption could be higher is the difference between open-source and industry, for the reasons mentioned above. Taken together, we think that these three factors make it more likely that: 1) for a sizeable subset of projects where we have detected activity, the true commit ratio is significantly higher than what we can measure; 2) some projects we classify as non adopters are actually adopters; and 3) adoption in private repositories from industry (on GitHub and elsewhere) is likely higher than OSS. While on the other hand, there is certainly a portion of projects that do have abandoned coding agents, we think the balance weights towards higher, rather than lesser adoption.

\subsection{Implications}

\revise{}{We recall the main findings of our study (see \tabref{tab:questions-summary}), before discussing their implications.}

\begin{diffversion}{
\begin{table}
    \small
    \caption{%
    Main findings of our six research questions on the adoption of coding agents, based on analyzing \TotalProjects GitHub repositories.}
    \centering
    \begin{tabular}{l|l}
    \toprule
    \textbf{Research question} & \textbf{Summary}\\
    \midrule
    \midrule
    RQ1: Overall adoption & As of \studyenddate, \OverallAdoptionRatePercent to \HighEstimateOverallAdoptionRatePercent of our dataset adopted coding agents. \\
    RQ2:      & File-level adoption is higher for both \emph{younger} and \emph{larger} projects; \\
    Adoption and metrics & larger projects have more \emph{experimental} and less \emph{pervasive} commit-level adoption. \\
    RQ3:     &  Adoption is \industryincrease larger in the top 20 organizations; \\
    Contexts of adoption & it is broad in terms of languages and topics. \\
    RQ4: Evolution  &  Adoption is increasing over time, with a last acceleration in early 2026. \\
    RQ5: Contribution size    &  AI-assisted commits are \emph{larger} than human-authored commits. \\
    RQ6: Contribution types     &  Most AI-assisted commits are \emph{feature additions} or \emph{bug fixes}. \\
    \bottomrule
    \end{tabular}
    \label{tab:questions-summary}
\end{table}}
\end{diffversion}

\paragraph{The adoption of coding agents is fast, and broad.} According to RQ1, our conservative estimate of coding agents is \OverallAdoptionRatePercent, as of \studyenddate while our high estimate is \HighEstimateOverallAdoptionRatePercent. This is a technology that saw the light of day in 2024; it was confidential in 2024 and before the spring of 2025, when increase in model capabilities and the release of coding agents by major actors coincided with a drastic increase in adoption rate. 

Moreover, if there are variations in adoption intensity in terms of project characteristics, project topic, or programming language, we can see that the adoption is \emph{broad}: 
\begin{itemize}
    \item Adoption is not especially concentrated on smaller projects (be it measured in lines of code, contributors or commits). File-level adoption rather suggests the opposite, and commit-level adoption, even if less strong on such larger projects, is far from insignificant (``Pervasive'' adopters are decreasing with size, but ``Consistent'' adopters are stable). Moreover, important organizations have high adoption of coding agents. There is, however, a bias towards younger projects.
    \item Likewise, adoption is not especially concentrated on particular types of projects. Some categories of projects may be more or less popular than the average, but there does not appear to be many categories that have near-zero adoption. This holds for both file-level and commit-level adoption. At the file level, some categories of projects are seeing much larger adoption, particularly in topics close to AI (a sort of ``dogfooding'' effect?), however, at the commit level even these categories see comparable usage to the rest.
    \item Finally, adoption is not especially concentrated on the small set of the most popular programming languages. Some less common languages (\eg, Dart, Ruby, Swift, PHP) see similar commit-level adoption to major languages. However, data is limited for the truly uncommon languages. 
\end{itemize}


\paragraph{Coding agents are a step change from LLM-based code completion}
Looking at the results of RQ5 and RQ6, we can infer that coding agents go beyond traditional LLM-based code completion. While LLM-based code completion helped developers write lines or blocks of code, the scope of coding agents is larger. This can be seen just by the size of AI contributions, which is clearly larger than human contributions: the median lines of code in an AI-assisted commit is triple that of human-authored ones; the proportion of very large AI-assisted commits (500 or more lines of code, or even more than 1000) is twice as frequent. AI-assisted commits are also larger in terms of deleted lines, and involved files. In addition, the type of contributions is telling: the most common type of contributions that we see in RQ6 are the implementation of \emph{features}, indicating that coding agents are used in tasks spanning a relatively wide scope.

Note that, if, as we suspect, we are missing a sizeable proportion of AI-assisted commits, it is likely that the difference between the human and AI-assisted commits is \emph{larger}, since those undetected commits are at the moment classified as human-authored commits.

\paragraph{Performance Outlook}
If, as the adoption curve in RQ4 suggest, this trend continues, we can expect coding agents to become extremely common in 2026 or 2027. 
This is even more the case, if, as we hypothesize, we are under-estimating, rather than over-estimating, the adoption of coding agents. Related to this is the increase of capabilities of the underlying models: much as stronger models were a contributing factor to the adoption of coding agents in 2025, similar increases could drive the adoption further in 2026 and beyond. The Anthropic economic index \cite{anthropic2025economicindex} finds that a growing number of tasks are delegated to LLMs, pointing at increased autonomy. Similarly, recent work~\cite{kwa2025measuring} looks at 50\% and 80\%-task-completion time horizon for agents (the time humans typically take to complete tasks that AI models can complete with 50\%/80\% success rate). They observe that, historically, the task completion time has been doubling every 7 months. This increased autonomy, due to evolving model capabilities, is consistent with the way coding agents work in the ideal cases, and is consistent with the increased adoption we observe in \Cref{fig:overall-adoption}.

\paragraph{The use of coding agents in practice should be studied extensively}
Taken together, the previous three observations come to an important implication. Coding agents, making large contributions are adopted quickly, broadly and sometimes very extensively (a small but non-negligeable proportion of projects have a very high ratio of AI-assisted commits). We have reasons to expect this trend to continue further. Therefore, it is \emph{extremely important} to study their impact in practice, in order to provide actionable advice to practitioners. 


Since the emergent of practical coding agents is very new, the body of work so far is very limited (see \secref{sec:background}). Given the speed of the phenomenon, we argue that the research community should study this with urgency; we hope that the initial strides we made in this field will be helpful for others \footnote{we will share all the datasets and analyses that we made to facilitate this}. While many aspects can be studied, we have identified a few we wish to highlight:
\begin{itemize}
    \item While the large contributions of coding agents can be seen as an indicator of productivity, the large amount of deleted code is an indicator of churn, which has been traditionally associated with defects \cite{nagappan2005use}. It is thus important to investigate to which extent coding agents cause high code churn, and what its consequences are. There is already evidence that AI-assisted development in general contributes to high code churn \cite{hardinggitclear2024, hardinggitclear2025}. In addition, recent work points at emerging quality issues associated with the use of coding agents \cite{he2025speed}.
    \item On the other hand, perhaps such a high churn is not always problematic. Coding agents, by making code extremely cheap to write, can enable scenarios that were not possible before (e.g., prototyping multiple variants of a solution to a problem, before making a more informed decision). In essence, coding agents greatly facilitate the creation of disposable, exploratory code, which may have a very short, but very useful, life. Characterizing these scenarios would be valuable future work.
    \item More generally, we think it is valuable to study the minority of extreme adopters of coding agents (those that we identify as having very high commit ratios). These adopters may give us early signals that could be useful to the broader community. They might act as ``canaries in a coal mine'', or cautionary tales for prospective users of coding agents. On the other hand, they might help discover new usages of coding agents that may be useful to the community at large.
    \item An essential component of coding agents is their guidance. As we have seen in RQ2, the amount of guidance used, and its update frequency, varies by several orders of magnitude, from the most basic to the most sophisticated. Detailed studies of this guidance, what it contains, and how it is used, will be essential to better understand how it impacts the behavior of coding agents. There is very early work in classifying the type of guidance \cite{mohsenimofidi2025context}.
    \item Last but not least, underlying all of this is the identification of coding agent activity that we are not able to detect at the moment. There is considerable future work in devising heuristics to detect this activity, and strengthen the body of work that will be based on this.
\end{itemize}

\paragraph{A call for clear attribution} To study the use of coding agents in the best conditions, proper identification is essential. While we think detecting agent activity is an important avenue for future research, an even more effective solution would be to reliably document it. This is why we call on the practitioners to document, rather than hide, their use of coding agents. We also call on providers of coding agents to continue their standardization efforts, while defining mechanisms to identify individual coding agents.

\section{Conclusion}

Coding agents based on LLMs appeared in 2024, and took off in 2025. This work presented a study of the adoption of coding agents in GitHub projects. We find that, as of \studyenddate, between \OverallAdoptionRatePercent and \HighEstimateOverallAdoptionRatePercent of projects in a large dataset of \TotalProjects GitHub projects show traces of use of coding agents. We find that this adoption was fast (it happened mainly from March to October 2025), broad (it includes all kinds of projects, in a variety of programming languages and organizations), and is still increasing. At the commit level, we find that a sizeable proportion of projects exhibit ``Consistent'' (5 to 20\%) or even ``Pervasive'' use (more than 20\%); a small, but non-negligible proportion of projects show extreme use. Moreover, we find that coding agents are used to implement features and bugs fixes, and do so via large contributions (larger than humans). These findings have implications both for the practice, which is seeing a very rapid transition towards coding agents, and for researchers, which should study this phenomenon in order to provide advice to practictioners during this transition.


\begin{acks}
This study received financial support from the French State (Investments for the Future programme, IdEx université de Bordeaux).
\end{acks}














\bibliographystyle{ACM-Reference-Format}
\bibliography{biblio}

@inproceedings{zeng2025first,
  title={A First Look at Conventional Commits Classification},
  author={Zeng, Qunhong and Zhang, Yuxia and Qiu, Zhiqing and Liu, Hui},
  booktitle={Proceedings of the IEEE/ACM 47th International Conference on Software Engineering},
  pages={2277--2289},
  year={2025}
}

@misc{hardinggitclear2025,
title = {AI Copilot Code Quality: Evaluating 2024's Increased Defect Rate via Code Quality Metrics},
url = {https://www.gitclear.com/ai_assistant_code_quality_2025_research},
author = {Harding, William},
year = {2025},
note = {Accessed on October 13th, 2025}
}

@misc{hardinggitclear2024,
title = {Coding on Copilot: 2023 Data Suggests Downward Pressure on Code Quality},
url = {https://www.gitclear.com/coding_on_copilot_data_shows_ais_downward_pressure_on_code_quality},
author = {Harding, William and Kloster, Matthew},
year = {2024},
note = {Accessed on 03 24, 2024}
}

@misc{li2025aidev,
  author       = {Li, Hao and Zhang, Haoxiang and Hassan, Ahmed E.},
  title        = {{AIDev}: Studying {AI} Coding Agents on {GitHub}},
  howpublished = {\url{https://doi.org/10.5281/zenodo.16919051}},
  year         = {2025},
  note         = {Accessed 2025-10-21}
}

@article{bouzenia2025understanding,
  title={Understanding Software Engineering Agents: A Study of Thought-Action-Result Trajectories},
  author={Bouzenia, Islem and Pradel, Michael},
  journal={arXiv preprint arXiv:2506.18824},
  year={2025}
}

@inproceedings{bird2009promises,
  title={The promises and perils of mining git},
  author={Bird, Christian and Rigby, Peter C and Barr, Earl T and Hamilton, David J and German, Daniel M and Devanbu, Prem},
  booktitle={2009 6th IEEE International Working Conference on Mining Software Repositories},
  pages={1--10},
  year={2009},
  organization={IEEE}
}

@techreport{anthropic2025economicindex,
  author       = {Anthropic},
  title        = {Anthropic Economic Index: Uneven Geographic and Enterprise AI Adoption},
  institution  = {Anthropic PBC},
  year         = {2025},
  month        = sep,
  note         = {Report, 15 Sept 2025},
  url          = {https://assets.anthropic.com/m/218c82b858610fac/original/Economic-Index.pdf}
}

@misc{Osmani2025_AI_Assisted_SE_Productivity,
  author       = {Osmani, Addy},
  title        = {The reality of AI-Assisted software engineering productivity},
  howpublished = {Substack blog post},
  month        = {Aug},
  year         = {2025},
  note         = {“What the data really shows about AI coding tools in 2025.”},
  url          = {https://addyo.substack.com/p/the-reality-of-ai-assisted-software-engineering-productivity}
}

@misc{Tufte2004_SparklineTheoryAndPractice,
  author       = {Tufte, Edward R.},
  title        = {Sparkline Theory and Practice},
  howpublished = {Online article},
  month        = {May},
  day          = {27},
  year         = {2004},
  note         = {“A small, intense, word-sized graphic with typographic resolution.”},
  url          = {https://www.edwardtufte.com/notebook/sparkline-theory-and-practice-edward-tufte/},
  urldate      = {2025-11-13}
}

@article{lex2014upset,
  title        = {UpSet: Visualization of intersecting sets},
  author       = {Lex, Alexander and Gehlenborg, Nils and Strobelt, Hendrik and Vuillemot, Romain and Pfister, Hanspeter},
  journal      = {IEEE Transactions on Visualization and Computer Graphics},
  volume       = {20},
  number       = {12},
  pages        = {1983--1992},
  year         = {2014},
  publisher    = {IEEE},
  doi          = {10.1109/TVCG.2014.2346248}
}

@inproceedings{izadi2024language,
  title={Language models for code completion: A practical evaluation},
  author={Izadi, Maliheh and Katzy, Jonathan and Van Dam, Tim and Otten, Marc and Popescu, Razvan Mihai and Van Deursen, Arie},
  booktitle={Proceedings of the IEEE/ACM 46th International Conference on Software Engineering},
  pages={1--13},
  year={2024}
}

@inproceedings{swebench2024,
  author       = {Carlos E. Jimenez and
                  John Yang and
                  Alexander Wettig and
                  Shunyu Yao and
                  Kexin Pei and
                  Ofir Press and
                  Karthik R. Narasimhan},
  title        = {SWE-bench: Can Language Models Resolve Real-world Github Issues?},
  booktitle    = {The Twelfth International Conference on Learning Representations,
                  {ICLR} 2024, Vienna, Austria, May 7-11, 2024},
  publisher    = {OpenReview.net},
  year         = {2024},
  bibsource    = {dblp computer science bibliography, https://dblp.org}
}

@article{wang2025solved,
  title={Are" Solved Issues" in SWE-bench Really Solved Correctly? An Empirical Study},
  author={Wang, You and Pradel, Michael and Liu, Zhongxin},
  journal={arXiv preprint arXiv:2503.15223},
  year={2025}
}

@misc{chowdhury2024swebenchverified,
  title={Introducing {SWE}-bench Verified},
  author={Chowdhury, Neil and Aung, James and Shern, Chan Jun and Jaffe, Oliver and Sherburn, Dane and Starace, Giulio and Mays, Evan and Dias, Rachel and Aljubeh, Marwan and Glaese, Mia and Jimenez, Carlos E. and Yang, John and Ho, Leyton and Patwardhan, Tejal and Liu, Kevin and Madry, Aleksander},
  year={2024},
  url={https://openai.com/index/introducing-swe-bench-verified/},
}

@inproceedings{zhang2024autocoderover,
  title={Autocoderover: Autonomous program improvement},
  author={Zhang, Yuntong and Ruan, Haifeng and Fan, Zhiyu and Roychoudhury, Abhik},
  booktitle={Proceedings of the 33rd ACM SIGSOFT International Symposium on Software Testing and Analysis},
  pages={1592--1604},
  year={2024}
}

@article{xia2024agentless,
  title={Agentless: Demystifying llm-based software engineering agents},
  author={Xia, Chunqiu Steven and Deng, Yinlin and Dunn, Soren and Zhang, Lingming},
  journal={arXiv preprint arXiv:2407.01489},
  year={2024}
}

@article{kwa2025measuring,
  title={Measuring ai ability to complete long tasks},
  author={Kwa, Thomas and West, Ben and Becker, Joel and Deng, Amy and Garcia, Katharyn and Hasin, Max and Jawhar, Sami and Kinniment, Megan and Rush, Nate and Von Arx, Sydney and others},
  journal={arXiv preprint arXiv:2503.14499},
  year={2025}
}

@article{xiao2025self_admitted_ai,
  author       = {Tao Xiao and
                  Youmei Fan and
                  Fabio Calefato and
                  Christoph Treude and
                  Raula Gaikovina Kula and
                  Hideaki Hata and
                  Sebastian Baltes},
  title        = {Self-Admitted GenAI Usage in Open-Source Software},
  journal      = {CoRR},
  volume       = {abs/2507.10422},
  year         = {2025},
  url          = {https://doi.org/10.48550/arXiv.2507.10422},
  doi          = {10.48550/ARXIV.2507.10422},
  eprinttype    = {arXiv},
  eprint       = {2507.10422},
  bibsource    = {dblp computer science bibliography, https://dblp.org}
}

@article{stol2018abc,
  title={The ABC of software engineering research},
  author={Stol, Klaas-Jan and Fitzgerald, Brian},
  journal={ACM Transactions on Software Engineering and Methodology (TOSEM)},
  volume={27},
  number={3},
  pages={1--51},
  year={2018},
  publisher={ACM New York, NY, USA}
}

@article{kumar2025sharp,
  title={Sharp Tools: How Developers Wield Agentic AI in Real Software Engineering Tasks},
  author={Kumar, Aayush and Bajpai, Yasharth and Gulwani, Sumit and Soares, Gustavo and Murphy-Hill, Emerson},
  journal={arXiv e-prints},
  pages={arXiv--2506},
  year={2025}
}

@article{becker2025measuring,
  title={Measuring the Impact of Early-2025 AI on Experienced Open-Source Developer Productivity},
  author={Becker, Joel and Rush, Nate and Barnes, Elizabeth and Rein, David},
  journal={arXiv preprint arXiv:2507.09089},
  year={2025}
}

@inproceedings{dabic2021sampling,
  title={Sampling projects in github for MSR studies},
  author={Dabic, Ozren and Aghajani, Emad and Bavota, Gabriele},
  booktitle={2021 IEEE/ACM 18th International Conference on Mining Software Repositories (MSR)},
  pages={560--564},
  year={2021},
  organization={IEEE}
}

@inproceedings{DiCosmo2017,
  author = {Roberto Di Cosmo and Stefano Zacchiroli},
  title = {Software Heritage: Why and How to Preserve Software Source Code},
  booktitle = {Proceedings of the 14th International Conference on Digital Preservation (iPRES 2017)},
  year = {2017},
  pages = {1--10},
  publisher = {ACM},
  doi = {10.1145/nnnnnnn.nnnnnnn},
  url = {https://doi.org/10.1145/nnnnnnn.nnnnnnn},
}

@inproceedings{vaithilingam2022expectation,
  title={Expectation vs. experience: Evaluating the usability of code generation tools powered by large language models},
  author={Vaithilingam, Priyan and Zhang, Tianyi and Glassman, Elena L},
  booktitle={Chi conference on human factors in computing systems extended abstracts},
  pages={1--7},
  year={2022}
}

@inproceedings{imai2022github,
  title={Is github copilot a substitute for human pair-programming? an empirical study},
  author={Imai, Saki},
  booktitle={Proceedings of the ACM/IEEE 44th International Conference on Software Engineering: Companion Proceedings},
  pages={319--321},
  year={2022}
}

@inproceedings{sandoval2023lost,
  title={Lost at c: A user study on the security implications of large language model code assistants},
  author={Sandoval, Gustavo and Pearce, Hammond and Nys, Teo and Karri, Ramesh and Garg, Siddharth and Dolan-Gavitt, Brendan},
  booktitle={32nd USENIX Security Symposium (USENIX Security 23)},
  pages={2205--2222},
  year={2023}
}

@inproceedings{perry2023users,
  title={Do users write more insecure code with AI assistants?},
  author={Perry, Neil and Srivastava, Megha and Kumar, Deepak and Boneh, Dan},
  booktitle={Proceedings of the 2023 ACM SIGSAC Conference on Computer and Communications Security},
  pages={2785--2799},
  year={2023}
}

@article{peng2023impact,
  title={The impact of ai on developer productivity: Evidence from github copilot},
  author={Peng, Sida and Kalliamvakou, Eirini and Cihon, Peter and Demirer, Mert},
  journal={arXiv preprint arXiv:2302.06590},
  year={2023}
}

@article{barke2023grounded,
  title={Grounded copilot: How programmers interact with code-generating models},
  author={Barke, Shraddha and James, Michael B and Polikarpova, Nadia},
  journal={Proceedings of the ACM on Programming Languages},
  volume={7},
  number={OOPSLA1},
  pages={85--111},
  year={2023},
  publisher={ACM New York, NY, USA}
}

@article{mozannar2022reading,
  title={Reading between the lines: Modeling user behavior and costs in AI-assisted programming},
  author={Mozannar, Hussein and Bansal, Gagan and Fourney, Adam and Horvitz, Eric},
  journal={arXiv preprint arXiv:2210.14306},
  year={2022}
}

@article{wang2023investigating,
  title={Investigating and designing for trust in ai-powered code generation tools},
  author={Wang, Ruotong and Cheng, Ruijia and Ford, Denae and Zimmermann, Thomas},
  journal={arXiv preprint arXiv:2305.11248},
  year={2023}
}

@article{godfrey2005using,
  title={Using origin analysis to detect merging and splitting of source code entities},
  author={Godfrey, Michael W and Zou, Lijie},
  journal={IEEE Transactions on Software Engineering},
  volume={31},
  number={2},
  pages={166--181},
  year={2005},
  publisher={IEEE}
}

@inproceedings{hora2018assessing,
  title={Assessing the threat of untracked changes in software evolution},
  author={Hora, Andre and Silva, Danilo and Valente, Marco Tulio and Robbes, Romain},
  booktitle={Proceedings of the 40th International Conference on Software Engineering},
  pages={1102--1113},
  year={2018}
}

@inproceedings{liang2024large,
  title={A large-scale survey on the usability of ai programming assistants: Successes and challenges},
  author={Liang, Jenny T and Yang, Chenyang and Myers, Brad A},
  booktitle={Proceedings of the 46th IEEE/ACM International Conference on Software Engineering},
  pages={1--13},
  year={2024}
}

@inproceedings{ziegler2022productivity,
  title={Productivity assessment of neural code completion},
  author={Ziegler, Albert and Kalliamvakou, Eirini and Li, X Alice and Rice, Andrew and Rifkin, Devon and Simister, Shawn and Sittampalam, Ganesh and Aftandilian, Edward},
  booktitle={Proceedings of the 6th ACM SIGPLAN International Symposium on Machine Programming},
  pages={21--29},
  year={2022}
}

@article{tufano2024unveiling,
  title={Unveiling ChatGPT's Usage in Open Source Projects: A Mining-based Study},
  author={Tufano, Rosalia and Mastropaolo, Antonio and Pepe, Federica and Dabi{\'c}, Ozren and Di Penta, Massimiliano and Bavota, Gabriele},
  journal={arXiv preprint arXiv:2402.16480},
  year={2024}
}

@article{murali2023codecompose,
  title={CodeCompose: A large-scale industrial deployment of AI-assisted code authoring},
  author={Murali, Vijayaraghavan and Maddila, Chandra and Ahmad, Imad and Bolin, Michael and Cheng, Daniel and Ghorbani, Negar and Fernandez, Renuka and Nagappan, Nachiappan},
  journal={arXiv preprint arXiv:2305.12050},
  year={2023}
}

@book{cochran1977sampling,
  title     = {Sampling Techniques},
  author    = {Cochran, William G.},
  year      = {1977},
  edition   = {3rd},
  publisher = {John Wiley \& Sons},
  address   = {New York},
  isbn      = {978-0471162407}
}

@book{triola2006elementary,
  title={Elementary statistics},
  author={Triola, Mario F and Goodman, William Martin and Law, Richard and Labute, Gerry},
  year={2006},
  publisher={Pearson/Addison-Wesley Reading, MA}
}

@article{he2025speed,
  title={Speed at the Cost of Quality? The Impact of LLM Agent Assistance on Software Development},
  author={He, Hao and Miller, Courtney and Agarwal, Shyam and K{\"a}stner, Christian and Vasilescu, Bogdan},
  journal={arXiv preprint arXiv:2511.04427},
  year={2025}
}

@article{mohsenimofidi2025context,
  title={Context Engineering for AI Agents in Open-Source Software},
  author={Mohsenimofidi, Seyedmoein and Galster, Matthias and Treude, Christoph and Baltes, Sebastian},
  journal={arXiv preprint arXiv:2510.21413},
  year={2025}
}

@inproceedings{nagappan2005use,
  title={Use of relative code churn measures to predict system defect density},
  author={Nagappan, Nachiappan and Ball, Thomas},
  booktitle={Proceedings of the 27th international conference on Software engineering},
  pages={284--292},
  year={2005}
}

@article{asare2023user,
  title={A User-centered Security Evaluation of Copilot},
  author={Asare, Owura and Nagappan, Meiyappan and Asokan, N},
  journal={arXiv preprint arXiv:2308.06587},
  year={2023}
}

@article{rupert2012simultaneous,
  title={Simultaneous statistical inference},
  author={Rupert Jr, G and others},
  year={2012},
  publisher={Springer Science \& Business Media}
}

@article{goodman1965simultaneous,
  title={On simultaneous confidence intervals for multinomial proportions},
  author={Goodman, Leo A},
  journal={Technometrics},
  volume={7},
  number={2},
  pages={247--254},
  year={1965},
  publisher={Taylor \& Francis}
}

@article{agentminingpaper,
  title={Promises, Perils, and (Timely) Heuristics for Mining Coding Agent Activity},
  author={Robbes, Romain and Matricon, Théo and Degueule, Thomas and Hora, Andre and Zacchiroli, Stefano},
  journal={Accepted for publication at MSR 2026},
  year={2025}
}

\appendix
\section{Selected coding agents}

\tabref{tab:coding-agents} shows the list of coding agents we include in this study.

\begin{table}
    \tiny
    \caption{List of coding agents with their URLs}
    \centering
    \begin{tabular}{l|l}
    \toprule
    \textbf{Name} & \textbf{URL}\\
    \midrule
    Aider & \url{https://aider.chat/}\\
    Alibaba Lingma & \url{https://www.alibabacloud.com/en/product/lingma}\\
    AmazonQ & \url{https://aws.amazon.com/q/}\\
    Amp & \url{https://sourcegraph.com/amp}\\
    Atlassian Rovodev & \url{https://www.atlassian.com/software/rovo-dev}\\
    Augment Code & \url{https://www.augmentcode.com/}\\
    Azad & \url{https://www.azad.bot}\\
    Baidu Comate & \url{https://comate.baidu.com/}\\
    Brokk & \url{https://brokk.ai/}\\
    Charlie & \url{https://www.charlielabs.ai}\\
    ChatGPT & \url{https://chatgpt.com/}\\
    Claude Code & \url{https://www.claude.com/product/claude-code}\\
    Cline & \url{https://cline.bot/}\\
    Codebuddy & \url{https://www.codebuddy.ai/}\\
    Codegen & \url{https://codegen.com/}\\
    Coderabbit & \url{https://coderabbit.ai/}\\
    Codex & \url{https://openai.com/codex}\\
    Continue & \url{https://www.continue.dev/}\\
    Copilot & \url{https://github.com/features/copilot}\\
    CopilotSwe & \url{https://github.blog/changelog/2025-09-22-copilot-swe-model-rolling-out-to-visual-studio-code-insiders/}\\
    Crush & \url{https://github.com/charmbracelet/crush}\\
    Cursor & \url{https://cursor.com/}\\
    Deepsource & \url{https://deepsource.com/}\\
    Devin & \url{https://devin.ai/}\\
    Factory & \url{https://factory.ai/}\\
    Factory Droid & \url{https://factory.ai}\\
    Fly & \url{https://fly.io/}\\
    GPT-Engineer & \url{https://github.com/AntonOsika/gpt-engineer}\\
    Gemini & \url{https://gemini.google.com/}\\
    Generic & \url{https://agents.md}\\
    Goose & \url{https://github.com/block/goose}\\
    Gru & \url{https://gru.ai/}\\
    Jules & \url{https://jules.google/}\\
    Junie & \url{https://www.jetbrains.com/junie/}\\
    Kilo Code & \url{https://kilocode.ai/}\\
    Kiro & \url{https://kiro.dev/}\\
    Langchain Open SWE & \url{https://github.com/langchain-ai/open-swe}\\
    Letta Code & \url{https://www.letta.com}\\
    Microsoft Amplifier & \url{https://github.com/microsoft/amplifier}\\
    Mistral Vibe & \url{https://docs.mistral.ai/mistral-vibe/introduction}\\
    Ona & \url{https://ona.com/}\\
    OpenHands & \url{https://github.com/All-Hands-AI/OpenHands}\\
    Opencode & \url{https://opencode.site}\\
    Phoenix & \url{https://phoenix.new/}\\
    Pi & \url{https://pi.dev/}\\
    Plandex & \url{https://plandex.ai/}\\
    Qodo & \url{https://qodo.ai}\\
    Qwen Coder & \url{https://github.com/QwenLM/qwen-code}\\
    Roo Code & \url{https://github.com/RooCodeInc/Roo-Code}\\
    Rulesync & \url{https://github.com/dyoshikawa/rulesync/tree/main}\\
    Sentry Seer & \url{https://sentry.io/product/seer}\\
    Serena & \url{https://github.com/oraios/serena}\\
    Sketch & \url{https://www.sketch.com/}\\
    Sourcery & \url{https://sourcery.ai/}\\
    SpecKit & \url{https://github.com/github/spec-kit}\\
    Sweep & \url{https://sweep.dev/}\\
    Taskmaster & \url{https://github.com/eyaltoledano/claude-task-master}\\
    Tessl & \url{https://tessl.io}\\
    Trae & \url{https://www.trae.ai/}\\
    Verdent & \url{https://verdent.ai}\\
    Warp & \url{https://www.warp.dev/}\\
    Windsurf & \url{https://windsurf.com/}\\
    Zed & \url{https://zed.dev/}\\
    \bottomrule
    \end{tabular}
    \label{tab:coding-agents}
\end{table}

\end{document}